\documentclass[aps,prl,twocolumn,superscriptaddress,longbibliography]{revtex4-2}
\usepackage{graphicx,comment}
\usepackage{bm}
\usepackage{microtype}
\usepackage{amsmath}
\usepackage{amssymb}
\usepackage{amsthm}
\usepackage[english]{babel}
\usepackage[table]{xcolor}
\usepackage{array}
\usepackage{multirow}
\usepackage[caption=false]{subfig}
\usepackage{physics}
\usepackage[normalem]{ulem}
\usepackage[verbose]{placeins}
\usepackage{dsfont}
\usepackage[scr=boondoxo,frak=boondox,scrscaled=1.05]{mathalfa}
\usepackage{tikz}
\usetikzlibrary{calc,arrows.meta}
\definecolor{wblue}{HTML}{2E5FA3}
\definecolor{wpurple}{HTML}{7B52A8}
\definecolor{worange}{HTML}{E07B29}
\tikzset{
  wbar/.style={draw=wblue,fill=wblue!40,line width=0.4pt},
  wdbar/.style={draw=black!35,fill=black!8,line width=0.4pt,dash pattern=on 1.2pt off 0.8pt},
  wax/.style={draw=black!70,line width=0.5pt},
  wleaf/.style={circle,draw=wblue,fill=wblue!20,line width=0.5pt,inner sep=1.5pt},
  wgate/.style={circle,draw=wpurple,fill=wpurple!25,line width=0.5pt,inner sep=1.9pt},
  wroot/.style={circle,draw=worange,fill=worange!35,line width=0.6pt,inner sep=2.3pt},
  wpsi/.style={circle,draw=worange,fill=worange!35,line width=0.6pt,minimum size=0.36cm,inner sep=0pt,font=\footnotesize},
  wwire/.style={line width=0.4pt,draw=black!40},
  wimp/.style={-{Stealth[length=2.4mm,width=1.9mm]},line width=1.2pt,draw=black!55},
  wlab/.style={font=\footnotesize},
  wsf/.style={font=\fontfamily{phv}\selectfont\footnotesize},
}
\usepackage{hyperref}
\hypersetup{breaklinks=true, colorlinks=true, linkcolor=violet,citecolor=violet, urlcolor=violet}
\renewcommand{\bra}[1]{\langle#1|}
\renewcommand{\ket}[1]{|#1\rangle}
\renewcommand{\braket}[2]{\langle#1|#2\rangle}

\renewcommand{\vec}{\boldsymbol}

\newcommand{\wz}{{\mathrm{W}^{Z}_\alpha}}

\newcommand{\w}{{{\mathrm{W}_\alpha}}}
\newtheoremstyle{prlthm}
  {0pt}{0pt}            % no extra space above/below (try \smallskipamount if cramped)
  {\itshape}            % body font
  {\parindent}          % indent like an ordinary paragraph
  {\bfseries\itshape}   % head font
  {}                   % head punctuation
  { }                   % space after head
  {\thmname{#1}\thmnumber{ #2}.\thmnote{ {\normalfont(#3)}}}
\theoremstyle{prlthm}
\newtheorem{theorem}{Theorem}
\newtheorem{proposition}{Proposition}

\begin{document}
\raggedbottom
%\title{Walshness: an intrinsic complexity metric for neural quantum states}
%\title{Walshness: an intrinsic learnability metric for quantum states}
\title{Walshness: an intrinsic neural-network representability metric for quantum states}

\author{Nisarga Paul}
\email{npaul@caltech.edu}
\affiliation{Department of Physics and Institute for Quantum Information and Matter,
California Institute of Technology, Pasadena, California 91125, USA}

\author{Kehuang Chen}
\affiliation{Department of Physics, Tsinghua University, Beijing 100084, China}

\author{Jessica K. Jiang}
\affiliation{Department of Physics and Institute for Quantum Information and Matter,
California Institute of Technology, Pasadena, California 91125, USA}

\author{Haimeng Zhao}
\affiliation{Department of Physics and Institute for Quantum Information and Matter,
California Institute of Technology, Pasadena, California 91125, USA}

\author{Di Luo}
\email{diluo@tsinghua.edu.cn}
\affiliation{Department of Physics, Tsinghua University, Beijing 100084, China}
\affiliation{Institute of Advanced Study, Tsinghua University, Beijing 100084, China}

\begin{abstract}
Neural quantum states (NQS) have emerged as powerful representations of quantum states with rapidly expanding applications across quantum many-body physics. Yet our understanding of when neural networks can efficiently represent physical quantum states remains limited, in part due to the nonlinear parameterization of NQS, the intricate sign structure of quantum states, and the sensitive dependence on basis choice. We introduce \textit{Walshness}, a complexity metric of a quantum state which quantifies whether it admits a compact neural-network representation. We rigorously prove that a general quantum many-body state admits an efficient NQS representation if it has bounded Walshness, and vice versa, for various NQS architectures.
We provide an algorithm for finding the optimal basis by minimizing the energy of a corresponding classical spin model. 
This allows us to empirically improve the learnability of various physical quantum states, including ground states of the transverse-field Ising model and mixed-field toric code, often reducing infidelity by orders of magnitude, and generalizes the basis choice given by the well-known Marshall sign rule in frustrated antiferromagnets. Our results provide an intrinsic complexity metric for quantum states with provable connections to their neural-network representability.
\end{abstract}

\maketitle

\textbf{\textit{Introduction.}}
Efficiently exploring and understanding the exponentially large Hilbert space of quantum systems is the central challenge in quantum many-body physics.
This curse of dimensionality has been proven to resist the attack of conventional classical computers and, in general, even future quantum computers \cite{kitaev2002classical}.
Fortunately, nature is not structureless, and physical quantum states always possess some structure that can be exploited \cite{poulinQuantumSimulationTimedependent2011}.
This has motivated the development of many celebrated classical simulation methods, such as density functional theory \cite{hohenberg1964inhomogeneous,kohn1965self}, tensor networks \cite{white1992density,schollwock2011density,cirac2021matrix}, quantum Monte Carlo \cite{Foulkes2001,becca2017quantum}, and neural network quantum states \cite{carleo2017solving}, along with quantum simulation methods with more general applicability~\cite{feynman1986quantum,lloyd1996universal,georgescu2014quantum}.

Beyond furnishing solutions of concrete quantum systems, these methods also allow us to better understand the structure and resources needed to efficiently describe and manipulate quantum many-body states.
For example, the applicability and limitations of tensor network and stabilizer methods are dictated by entanglement and magic \cite{vidal2003efficient,eisert2010colloquium,bravyi2005universal,aaronson2004improved,leone2022stabilizer}, respectively.
These properties can then be measured with the entanglement entropy and stabilizer R\'enyi entropy, which themselves turned out to be useful metrics for characterizing various quantum phenomena, including quantum criticality~\cite{calabrese2004entanglement}, topological phases \cite{levin2006detecting,kitaev2006topological}, measurement-induced phase transitions \cite{fisher2023random}, and spacetime geometry in holographic duality \cite{cao2025gravitational}.

Neural quantum states (NQS), a newer classical method, have been shown to be competitive in understanding quantum many-body systems \cite{carleo2017solving,wu2024variational}.
As ground state variational ans\"atze, NQS have already achieved cutting-edge results across condensed matter, quantum chemistry, and nuclear physics~\cite{luo2019backflow,cassella2023discovering,chen2024empowering,rende2024simple,teng2025solving,geier2025self,pfau2020ab,yang2023deep,adams2021variational,gnech2024distilling,fore2025investigating} and have further expanded into time dynamics, open quantum systems, and beyond. In many instances, NQS has demonstrated state-of-the-art performance in quantum many-body simulations~\cite{giuliani2023learning,PhysRevLett.134.079701,kufel2025approximately,deng2017quantum}. 
Nonetheless, despite many previous attempts to understand the limitations of NQS \cite{szabo2020neural,westerhout2020generalization,kol2024neural}, it remains unclear whether a quantifiable, intrinsic notion of complexity exists that predicts the difficulty of learning a given quantum state with NQS --- analogous to the role entanglement plays for tensor networks.

We first note that any such metric characterizing the neural network representability of quantum states should be basis dependent.
This is because an NQS parameterizes a quantum state in a given basis $\ket{\vec s}$ as $\ket{\psi} = \sum_{\vec s}\psi_{\vec \theta}(\vec s)\ket{\vec s}$, where $\psi_{\vec \theta}$ is a neural network with weights $\vec \theta$.
When we examine the same quantum state under a local change of basis $U_{\vec s,\vec s'}$, the wavefunction components in the new basis can change drastically.
In practice, a neural network that can efficiently represent $\psi(\vec s)$ may struggle to represent the amplitudes $\sum_{\vec s'} U_{\vec s,\vec s'} \psi(\vec s')$ in the new basis.
This phenomenon can already be seen even in trivial product states. Despite several recent empirical studies \cite{cortes2025basis,kovzic2025exploring,doschl2025towards}, a systematic theory of basis dependence in NQS has so far been lacking.
This motivates the following questions: Is there a \textit{basis-dependent} metric for the neural-network-representability of a wavefunction~\footnote{We note there are three aspects of `learnability' in neural network representations of quantum states: representability, optimization, and sampling. In this work, we focus on establishing theoretical aspects of representability and empirically establishing aspects of optimization.}? Can such a metric be used to define and compute an \textit{optimal} local basis and signify physical phenomena? 

In this work, we answer these questions in the affirmative by proposing a new quantity, \textit{Walshness}, which captures the amplitude and sign structure complexity in a quantum state for any given basis. We also define an intrinsic version, obtained by minimizing over all local basis choices. Using Walshness, we establish a precise two-sided characterization of the representability of a quantum state with NQS: a quantum state with bounded Walshness has a local basis in which it admits a compact NQS representation at fixed accuracy; conversely, a wavefunction with a compact NQS representation must have a suitably small Walshness. This answers an open question in Ref.~\cite{yang2024can}. 

\par 
Walshness provides not only a conceptual lens for diagnosing quantum states different from entanglement and magic, but also a practical tool for NQS modeling. The importance of sign structure for neural quantum states has long been recognized, reminiscent of its deep significance in quantum Monte Carlo~\cite{henelius2000sign,troyer2005computational,levy2021mitigating}, and has recently been sharpened using quantities derived from the discrete Fourier transform in Refs.~\cite{schurov2025learning,doschl2025towards,cortes2025basis,kumar2026unlearnable} and particularly Ref.~\cite{wang2026expressibility}. We use Walshness as a criterion for selecting an optimal local basis for NQS representation. Training with an infidelity loss, we find that the Walshness-minimizing basis outperforms the computational basis for a variety of quantum states. Moreover, Walshness recasts the search for an optimal sign rule as the minimization of an associated classical spin model, allowing us to recover and generalize the well-known Marshall sign rule in frustrated antiferromagnets.

\par 

In the following, we introduce Walshness, provide an intuitive account of its main idea and underlying theory, compute it in simple cases, rigorously establish its connection to NQS learnability, and demonstrate it with numerical experiments. In the End Matter, we provide a guide for NQS practitioners to computing Walshness and using it to find an optimal basis for the neural representation of a quantum state. \par 
\begin{figure}
    \centering   \includegraphics[width=\linewidth]{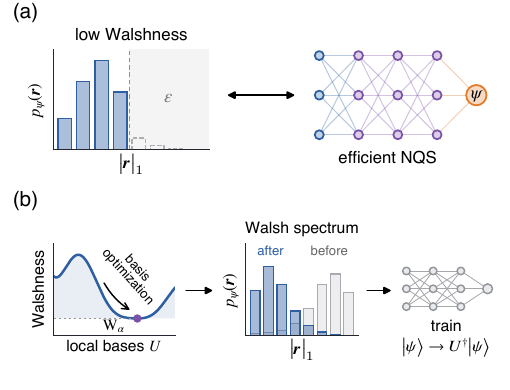}
\caption{
\textbf{Walshness.} (a) For a quantum state $\ket\psi$ on $n$ qubits, low Walshness implies the existence of an efficient (i.e. compact) neural quantum state (NQS) representation in some basis at fixed accuracy. Conversely, a compact NQS has a correspondingly low Walshness determined by depth. (b) We provide an algorithm for finding Walshness-minimizing bases and numerically show they ease NQS training.}
\label{fig:1}
\end{figure}

\textbf{\textit{Main definitions and results.}} Let $\ket\psi \in (\mathbb{C}^2)^{\otimes n}$ be a quantum state on $n$ qubits (generalization to other degrees of freedom is possible) and let $X_i,Y_i,Z_i$ denote the Pauli operators acting on qubit $i \in [n]$. Let $\ket{\vec s}, \vec s \in \{0,1\}^n$, denote a $Z$-basis state. Motivated by the importance of the sign structure in characterizing the complexity of a general wavefunction, we consider the Walsh-Hadamard transform~\cite{nielsen2000quantum} (the discrete Fourier transform over $\mathbb Z_2^n$) of the wavefunction $\psi(\vec s)$:
\begin{equation}
\label{eq:wht}
\widehat\psi(\vec r) = 2^{-n}\sum_{\vec s} (-1)^{\vec r\cdot \vec s} \psi(\vec s), \quad \vec r\in \{0,1\}^n, 
\end{equation}
where $\ket{\psi} = \sum_{\vec s}\psi(\vec s)\ket{\vec s} \neq 0$ is a general, possibly unnormalized quantum state. We define the \textit{Walsh spectrum}
\begin{equation}
    p_\psi(\vec r) = |\widehat\psi(\vec r)|^2 /\sum_{\vec r} |\widehat\psi(\vec r)|^2
\end{equation}
as a classical distribution satisfying $p_\psi\geq 0, \sum_{\vec r} p_\psi(\vec r) = 1$. This allows us to introduce the \textit{fixed-basis Walshness} in the $Z$ basis as
\begin{equation}    
\label{eq:fbwalshness}
\wz\left[\ket\psi\right] =  \frac{1}{\alpha} \log \mathbb{E}_{\vec r\sim p_\psi}\!\left[ e^{\alpha |\vec r|_1}\right], \quad \alpha>0.
\end{equation}
$\alpha$-Walshness satisfies $0\leq \wz[\ket\psi]\leq n$ and depends on the distribution of Hamming weights $|\vec r|_1$ in $p_\psi(\vec r)$. Intuitively, it quantifies how many discrete Fourier modes are present in the $\vec s$-basis wavefunction, with more oscillatory Fourier modes contributing more. Alternatively, the Walsh spectrum can be regarded as the outcome distribution of measurements in the $X$ basis. Thus $\alpha \wz$ is the cumulant-generating function of $|\vec r|_1$, which counts the number of $X=-1$ spins, in this outcome distribution.
\par
By design, $\wz$ is sensitive to the choice of basis $\vec s$. To capture an intrinsic, local-basis-independent complexity, we consider the minimal $\alpha$-Walshness under single-site unitaries $\mathcal{U} = \left\{\otimes_{i=1}^nU_i \big\vert U_i \in \mathrm{SU}(2) \right\}$:
\begin{equation}
\label{eq:walshness}  
\w[\ket\psi] = \min_{U\in \mathcal{U}} \wz[U\ket\psi],
\end{equation}
which we call the \emph{intrinsic Walshness}, or simply \emph{Walshness},
as sketched in Fig.~\ref{fig:1}. 
Walshness is a highly nonlinear functional of $\ket \psi$.
Nevertheless, we show in Proposition~\ref{prop:spin} that one can recast the optimization problem in Eq.~\eqref{eq:walshness} as minimizing the energy of a classical spin model, with one spin for each $U_i$.\par

Several useful properties of Walshness are as follows~\cite{supp}. Walshness is invariant under rescaling $\ket\psi$ and permuting the $n$ qubits, and, by design, under single-site unitaries $U\in\mathcal{U}$. It is additive under tensor products, $\w[\ket\psi\otimes \ket\phi] = \w[\ket\psi]+\w[\ket\phi]$, and it vanishes if and only if $\ket\psi$ is a product state. By tuning the parameter $\alpha>0$, we obtain a family that is nondecreasing in $\alpha$~\cite{supp}. In the limit of $\alpha\to 0$, $\wz$ reduces to the mean Hamming weight $\mathbb{E}_{\vec r \sim p_{\psi}}[|\vec r|_1]$ of the Walsh spectrum, while as $\alpha\to\infty$ it is the largest Hamming weight in the support of $p_\psi$. \par 

For several common families of quantum states, we can analytically calculate $Z$-basis Walshness~\cite{supp}. GHZ states $\ket{\psi_{\mathrm{GHZ}}} = (\ket{0^n}+\ket{1^n})/\sqrt{2}$ satisfy $\wz[\ket{\psi_{\mathrm{GHZ}}}] \approx n g_\alpha$ up to corrections exponentially small in $n$, where $g_\alpha \equiv \frac{1}{\alpha} \log \frac{1+e^\alpha}{2}$. Meanwhile, they have intrinsic Walshness $\w[\ket{\psi_{\mathrm{GHZ}}}] = n/2 < ng_\alpha$ for $n$ even. Haar-random states and random stabilizer states satisfy $\wz = ng_\alpha + \tilde o(1)$, where $\tilde o(1)$ denotes an additive correction tending to zero in probability at large $n$. Numerical results further confirm this scaling in the Supplemental Material~\cite{supp}.\par 

The intrinsic Walshness is invariant under single-site unitaries, similar to entanglement. 
For any bipartition $L:R$ of the qubits, the entanglement entropy is upper bounded by the intrinsic Walshness as~\cite{supp}
\begin{equation}
	S_{L:R} \leq \frac{n}{2}\,h_2\left(\frac{\w}{n}\right)
\end{equation}
whenever $\w\leq n/2$, where $h_2(x) = -x\log x-(1-x)\log(1-x)$ is the binary entropy. Consequently, extensive entanglement can only exist in the presence of extensive $\w$. In contrast, low entanglement states can have extensive $\w$ and therefore these are distinct metrics, as illustrated by GHZ states $\ket{\psi_{\mathrm{GHZ}}}$ having constant entanglement across any cut but extensive $\w$.\par

\par

Walshness captures a notion of complexity which connects to NQS learnability.
We formalize this connection in the Supplemental Material~\cite{supp} and rigorously establish the following theorem connecting the Walshness and the NQS learnability of quantum states.

\begin{theorem}[MLP representability; informal]\label{thm:mlp}
A wavefunction with fixed-basis Walshness $\omega=\wz[\ket\psi]$ admits an MLP representation of $\psi(\vec s)$ of width $n^{O\left(\omega\right)}$ and depth $O(\log(\omega))$ to any constant infidelity. Conversely, if $\mathrm{W}_0^Z[\ket\psi]=\Omega(n)$ or $\mathrm{W}_{\alpha}^Z[\ket\psi]\geq n-O(1)$ then any MLP representation with sufficiently small constant infidelity must have a depth $\Omega(\log n)$, or $\Omega\left(\frac{\log n}{\log\log n}\right)$, for polynomial or standard analytic nonlinearities, respectively.
\end{theorem}
\par 
Intuitively, this theorem establishes that low Walshness implies the existence of a basis in which $\ket\psi$ has an efficient MLP representation.
Conversely, a shallow MLP-NQS can only represent states with low Walshness [Fig.~\ref{fig:1}]. Above, we have suppressed dependence on various constants, kept $\alpha$ and infidelity $\varepsilon$ fixed, and used asymptotic notation with respect to $n$.

\par 
We now lay out the key proof ideas of the theorem above. We first show how small $\wz[\ket \psi]$ allows for a compact MLP representation. Let $\ket\psi$ be normalized and define a truncated expansion 
\begin{equation}
\label{eq:psitilde}
    \widetilde\psi(\vec s)
    =
    \sum_{|\vec r|_1\leq q}
    \widehat\psi(\vec r)(-1)^{\vec r\cdot\vec s},
\end{equation}
where $q< n$ is a cutoff to be chosen. We will construct an NQS to represent the truncated state $\ket{\widetilde \psi} = \sum_{\vec s}\widetilde{\psi}(\vec s)\ket{\vec s}$, which has fidelity $|\langle \psi|\widetilde\psi\rangle|^2/\langle\widetilde\psi|\widetilde\psi\rangle = \sum_{|\vec r|_1\leq q}p_\psi(\vec r)$. With Walshness $\omega = \wz[\ket\psi]$ the infidelity is bounded by
\begin{equation}
    \sum_{|\vec r|_1>q}p_\psi(\vec r)
    \leq
    e^{-\alpha q}
    \sum_{\vec r}e^{\alpha|\vec r|_1}p_\psi(\vec r)
    =
    e^{-\alpha(q-\omega)}.
\end{equation}
Thus, choosing the cutoff
$q=\lceil\omega+\alpha^{-1}\log(1/\varepsilon)\rceil$
ensures infidelity at most $\varepsilon$. 

We now construct an explicit MLP representing the truncated wavefunction $\widetilde\psi(\vec s)$ with cutoff $q$. According to Eq.~\eqref{eq:psitilde}, it is a linear combination of functions of $\vec s$ with the form $(-1)^{\vec r\cdot\vec s}$ and coefficients $\widehat\psi(\vec r)$. We first construct a small MLP realizing the former. The identity $xy=[\sigma(x+y)-\sigma(x-y)]/4$ with $\sigma(x) = x^2$ allows us to express multiplication using two
quadratic nonlinearities. Multiplying the inputs in pairs, then multiplying the resulting products in pairs, and so on, allows us to express $(-1)^{\vec r\cdot\vec s}=\prod_{i:r_i=1}(-1)^{s_i}$ as an MLP with quadratic nonlinearities using $O(\log q)$ layers. Finally, there are $\sum_{k=0}^q \binom nk = n^{O(q)}$ terms in Eq.~\eqref{eq:psitilde}. Evaluating in parallel and summing with coefficients $\widehat\psi(\vec r)$ therefore gives the claimed width and depth at the desired accuracy $\varepsilon$. In particular, for constant $\omega,\alpha,\varepsilon$, the neural network has polynomial width and constant depth. \par 

To establish the converse, we make essential use of properties of the Walsh-Hadamard transform [Eq.~\eqref{eq:wht}]. In particular, we prove bounds on the largest weight $|\vec r|_1$ appearing in the Walsh-Hadamard transform for an MLP-NQS of depth $d$ recursively. Applying a nonlinearity $\sigma$ to any function only multiplies this largest weight by a constant factor $c_{\sigma}$ depending on the nonlinearity.
For instance, we show a polynomial nonlinearity with degree $p$ has $c_\sigma=p$. Meanwhile, linear combinations of functions with maximum weight at most $q$ again have a largest weight $q$. Recursively applying these properties, we conclude that an MLP of depth $d$ has Walshness at most $c_\sigma^{d}$. Hence, a target wavefunction with $\Omega(n)$ Walshness requires a depth $\Omega(\log n)$ for constant $c_\sigma$. These conclusions remain robust to a small infidelity under the conditions stated, and for more generic analytic nonlinearities such as $\tanh$, the effective $c_\sigma$ scales with depth $d$ and results in the slightly weaker bound~\cite{supp}.
\par 
An analogous characterization holds for restricted Boltzmann
machines (RBMs), whose wavefunctions naturally admit a multiplicative factorization through an exponential parametrization.
The relevant quantity is \textit{multiplicative Walshness},
which assesses truncated approximations of the form in Eq.~\eqref{eq:psitilde} for $\log\psi(\vec s)$. 
\par 
\begin{theorem}[RBM representability; informal]
\label{thm:rbm}
For fixed $\alpha>0$, a wavefunction
$\psi=e^K$ with bounded multiplicative Walshness
$\mathrm W_\alpha^*(K)$ admits an RBM representation with
$\mathrm{poly}(n)$ hidden units and bounded
connectivity, at fixed infidelity $0<\varepsilon<1$.
Conversely, every such RBM representation admits such a $K$ with
bounded $\mathrm W_\alpha^*(K)$.
\end{theorem}
\par 
The definition and construction are detailed in the End Matter. The formal theorem and proof are left to~\cite{supp}.

\par
We have thus established a property with a two-sided guarantee: low Walshness implies a compact neural network representation, and vice versa. In the following, we apply this to quantum state learning, provide an algorithm for finding bases minimizing Walshness, and numerically show that they ease the training of NQS in practice.
\par

\begin{figure}
\centering
\includegraphics[width=\linewidth]{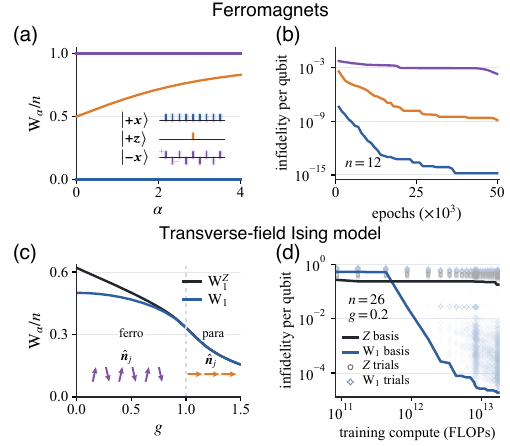}
\caption{\textbf{NQS learning.} (a) Fixed-basis Walshness density of $\ket{\vec{+x}},\ket{\vec{+z}},$ and $\ket{\vec{-x}}$ ferromagnets. All three have zero intrinsic Walshness. (b) Loss during a common training budget for MLP-NQS fits of these states, using infidelity as loss. Ordering of losses follows that of $\wz$. (c) Fixed-basis Walshness density and intrinsic Walshness density at $\alpha=1$ for the TFIM ground state. Insets show $\mathrm{W}_\alpha$-minimizing bases $(\hat n_{j,x},\hat n_{j,z})$ at $g=0.2$ and $1.2$.
(d) Minimum infidelity per qubit for learning the $g=0.2$ TFIM ground state over many trials (with varying MLP architectures, hyperparameters, and epochs~\cite{supp}) for Walshness-minimizing (blue) and computational (black) bases. These data demonstrate the advantage of the Walshness-minimizing basis in contexts of training NQS.}
\label{fig:quantumstatelearning}
\end{figure}

\textbf{\textit{NQS learning.}} As illustrative examples, consider the ferromagnets $\left|\vec{+x}\right\rangle =\ket{+}^n$, $\left|\vec{+z}\right\rangle = \ket{0}^n$, and $\left|\vec{-x}\right\rangle =\ket{-}^n$, where $X\ket{\pm} = \pm \ket{\pm}$.  Although they are related by single-qubit rotations, neural networks have dramatically different difficulties in representing their wavefunctions in the $Z$-basis. By a direct calculation, 
\begin{equation}
\label{eq:ferromagnets}
\wz[\left|\vec{+x}\right\rangle] = 0, \,\,\, \wz[\left|\vec{+z}\right\rangle] = ng_\alpha,\,\,\, \wz[\left|\vec{-x}\right\rangle] = n,
\end{equation}
as shown in Fig.~\ref{fig:quantumstatelearning}a. However, each has $\mathrm{W}_\alpha = 0$. Eq.~\eqref{eq:ferromagnets} aligns with our expectations as follows. Representing the constant wavefunction of $\left|\vec{+x}\right\rangle$ is trivial for any neural network. In contrast, representing $\left|\vec{-x}\right\rangle$ amounts to learning the $n$-bit \textit{parity function}, a notoriously difficult problem in learning theory with exponential training hardness~\cite{o2014analysis,shoshani2025hardness}. Finally, $\left|\vec{+z}\right\rangle$ has a single-neuron representation~\footnote{For example $\delta_{\vec 0,\vec s} = \mathrm{ReLU}(1-|\vec s|_1)$, where $\mathrm{ReLU}(x)= \max(0,x)$ is the rectified linear unit.}, and $\wz$ places its complexity in between the previous cases.\par
To corroborate these expectations, in Fig.~\ref{fig:quantumstatelearning}b we plot the infidelity during $5\cdot 10^4$ epochs of Adam optimization when training MLP-NQS to represent $\ket{\vec{+x}},\ket{\vec{+z}},\ket{\vec{-x}}$ with $\tanh$ nonlinearities, two hidden layers of width six, and infidelity as the loss function. (We present numerical details in~\cite{supp}.) The ordering of the final infidelity confirms that learning $\left|\vec{+x}\right\rangle$ is by far easiest, $\left|\vec{+z}\right\rangle$ is intermediate, and $\left|\vec{-x}\right\rangle$ is hardest. \par

\par 
Next, we ask whether choosing a basis with minimal Walshness enables more efficient training of NQS more generally. We consider the ground state of the $n$-qubit transverse-field Ising model (TFIM), $H_{\mathrm{TFIM}} = -\sum_{i} Z_iZ_{i+1} - g\sum_i X_i$. The TFIM has a quantum critical point separating a ferromagnetic phase ($g<1$) from a paramagnetic phase ($g>1$). We analytically calculate the Walshness $\wz[\ket\psi]$ of its ground state using its free-fermion solution~\cite{supp} and obtain
\begin{equation}
\label{eq:tfimW0}
\wz[\ket\psi] = n\left(\frac{1}{2} + \frac{1}{2\pi\alpha}\int_0^\pi \dd k\,\log\rho(\alpha,k) +o(1)\right)
\end{equation}
where $\rho(\alpha,k) = \cosh\alpha - \frac{g-\cos k}{\varepsilon_k}\sinh\alpha$ and $\varepsilon_k = \sqrt{1+g^2-2g\cos k}$ is the free-fermion dispersion. As $g\to 0$ or $\infty$, the ground state $\ket\psi$ approaches $\ket{\psi_{\mathrm{GHZ}}}$ or $\ket{\vec{+x}}$ and its Walshness indeed attains the GHZ value or zero. 
At the phase transition point $g=1$, its Walshness is continuous with a logarithmically diverging derivative, $\partial_g(\wz/n) \simeq (2\pi\alpha)^{-1}\tanh\alpha\,\log|g-1|$.

To calculate the intrinsic Walshness $\w$, we need to perform a minimization over all local bases according to Eq.~\eqref{eq:walshness}. The following proposition, proved in the End Matter, shows that this optimization problem can be mapped into a classical spin problem with concrete physical interpretation. \par

\begin{proposition}[Walshness minimization as a classical spin problem]\label{prop:spin}
For any $n$-qubit state $\ket\psi$ and $\alpha>0$, parameterize single-qubit unitaries $U_j\in \mathrm{SU}(2)$ with 3D unit vectors $\hat{\vec n}_j\in \mathbb R^3$ as $U_j^\dagger X_j U_j = \hat{\vec n}_j\cdot\vec\sigma_j$, where $\vec\sigma_j = (X_j,Y_j,Z_j)$.
Let $A_\alpha(\hat{\vec n}) = \cosh(\alpha/2)\,I - \sinh(\alpha/2)\,\hat{\vec n}\cdot\vec\sigma$ and $E[\{\hat{\vec n}_j\}]= \langle \prod_{j=1}^n A_\alpha(\hat{\vec n}_j)\rangle$, where $\langle\,\cdot\,\rangle$ denotes expectation in $\ket\psi$. 
Then, the intrinsic Walshness satisfies
\begin{equation}
\label{eq:spinmodel}
\w[\ket\psi] = \frac{n}{2} + \frac{1}{\alpha}\log \min_{\{\hat{\vec n}_j\}} E[\{\hat{\vec n}_j\}].
\end{equation}
In other words, minimizing Walshness amounts to minimizing the energy $E[\{\hat{\vec n}_j\}]$ of a classical spin model, with couplings given by the multi-spin correlations $\langle \sigma_{j_1}\cdots\sigma_{j_k}\rangle$.
\end{proposition}
From this mapping, we see that the intrinsic Walshness is a highly nonlinear functional of $\ket\psi$ and therefore the Walshness-minimizing basis can exhibit spontaneous ordering of its own. We apply this to the TFIM, plotting ground state Walshness in Fig.~\ref{fig:quantumstatelearning}c, and find $\w\approx \wz$ for $g$ large, while $\w$ and $\wz$ differ as $g\to 0$.
Moreover, the Walshness-minimizing basis for $g<1$ (inset) exhibits a period-two staggering that spontaneously breaks the translation symmetry of the TFIM. \par

\par 
Next, we compare the training of NQS on the ground state of the TFIM in the $Z$- and Walshness-minimizing bases. We train several MLP-NQS on the ferromagnetic ground state at $g=0.2$ in both bases and find that the Walshness-minimizing basis can lead to several orders of magnitude lower infidelity at comparable training compute [Fig.~\ref{fig:quantumstatelearning}d]. The advantage of using the Walshness-minimizing basis is most prominent for small $g$ in the ferromagnetic phase and gradually fades as $g\to 1$ and the two bases coincide.
In the End Matter, we provide additional numerical experiments for the mixed-field toric code, again finding that training in the Walshness-minimizing basis can lead to substantially lower infidelity compared to the computational basis when $\w < \wz$. \par 

\textbf{\textit{Applications in frustrated antiferromagnets.} } In the context of frustrated antiferromagnets, the Marshall sign rule provides a familiar example of a local basis change simplifying a many-body wavefunction. For an
unfrustrated bipartite antiferromagnet,
amplitudes can be made nonnegative, hence amenable to quantum Monte Carlo simulation, by a $\pi$ rotation about $\hat{\vec z}$
on one sublattice~\cite{auerbach2012interacting}. For a sublattice $A$, this transformation is $M_A=\prod_{j\in A}Z_j$ and yields the transformed wavefunction $\bra{\vec s}M_A\ket\psi =(-1)^{\sum_{j\in A}s_j}\psi(\vec s)$.  
\par 

\begin{figure}
    \centering
    \includegraphics[width=\linewidth]{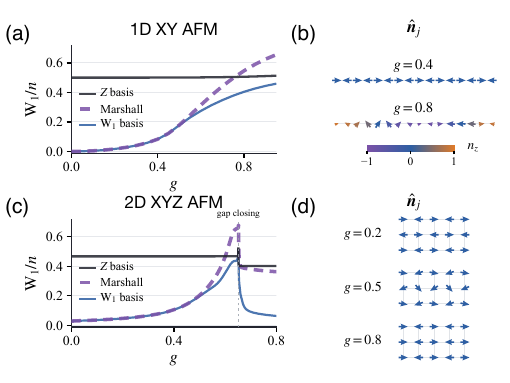}
    \caption{\textbf{Walshness and the Marshall sign rule.}     (a,c) Walshness $\mathrm W_1^Z[U\ket\psi]$
    in the computational (black), Marshall (purple dashed),
    and numerically optimized (blue) bases.
    (a,b) Results for $H_{\mathrm{XY}}$
    [Eq.~\eqref{eq:xy_af_main}], with varying anisotropy $g$, in the symmetry-broken ground state.
    (c,d) Results for $H_{\mathrm{XYZ}}$
    [Eq.~\eqref{eq:j1j2_xyz_main}], with varying frustration $g$.
    Within the exact ground state manifold, we select the state
    maximizing $\langle X\rangle$ on a corner site of the finite lattice.
    (b,d) Optimized basis axes at the indicated couplings.
    Arrows show the $xy$ projections of
    $\hat{\vec n}_j$, defined by
    $U_j^\dagger X_jU_j=\hat{\vec n}_j\cdot\vec\sigma_j$.}
\label{fig:frustratedAFM}
\end{figure}

We demonstrate that Walshness minimization can independently reproduce the Marshall sign rule and provide a prescription for generalizing it to other settings. We illustrate this with two antiferromagnetic models. First, in an XY antiferromagnet,
\begin{equation}
\label{eq:xy_af_main}
    H_{\mathrm{XY}}
    =    \sum_j\left(X_jX_{j+1}+gY_jY_{j+1}\right),
\end{equation}
we numerically minimize the Walshness of the symmetry-broken ground state over single-site rotations. At small $g>0$, we find the Marshall sign rule emerges from Walshness minimization with no input other than the wavefunction. Moreover, at larger $g$, the Walshness-minimizing basis departs from the Marshall sign rule and yields a lower Walshness [Fig.~\ref{fig:frustratedAFM}a-b], suggesting its wider applicability than the simple Marshall sign rule.
\par 
Second, we consider the 2D XYZ antiferromagnet
\begin{equation}
   \label{eq:j1j2_xyz_main}
    H_{\mathrm{XYZ}}
    =
    \sum_{\langle ij\rangle}K_{ij}
    +g\sum_{\langle\langle ij\rangle\rangle}K_{ij}
\end{equation}
on a square lattice in the frustrated regime $g>0$, with $K_{ij}=X_iX_j+0.3Y_iY_j+0.1Z_iZ_j$ and sums running over nearest- and next-nearest-neighbor bonds [Fig.~\ref{fig:frustratedAFM}c-d]. We observe that the Walshness-minimizing basis recovers the Marshall sign rule on a checkerboard sublattice at weak frustration $g=0.2$ and a stripe sublattice at $g=0.8$. Dominant correlations are also checkerboard- and stripe-like in these respective regimes, with a crossover or transition at $g\approx 0.65$. At $g=0.5$, Walshness minimization produces a canted basis outside the purview of the Marshall sign rule. Together, these results suggest that Walshness can recover and generalize standard heuristics for basis optimization. 
\par 
\textbf{\textit{Outlook.}} Motivated by neural quantum states, we have introduced a new measure of wavefunction complexity, \textit{Walshness}, and laid out its strong connection to representability by small neural networks. This is analogous to the role of entanglement for tensor networks. Walshness is intrinsic, highly nonlinear, and distinct from entanglement, raising the possibility that it may provide new insight into various physical phenomena such as quantum phases, criticality, dynamics, and topological order.\par 

Our work opens up many interesting future directions. First, enlarging the allowed basis transformations from single-site rotations to finite-depth local circuits would define a hierarchy of increasingly intrinsic notions of Walshness. Second, natural next steps include generalizing our results to other NQS architectures, such as autoregressive models, and to fermionic degrees of freedom. On the practical side, extending our work from quantum state learning to variational ground-state calculations would widen applicability. Finally, as demonstrated by the recovery and generalization of the Marshall sign rule, Walshness may also find use in alleviating the sign problem in quantum Monte Carlo, where identifying the appropriate basis is likewise of central importance. 

\begin{acknowledgments}
\textit{Acknowledgements.} We thank David Dai, Liang Fu, Joel Moore, Nils Niggemann, Xiaoliang Qi, Zack Weinstein, and especially Taige Wang for helpful discussions. NP, JKJ, and HZ acknowledge support from the Institute for Quantum Information and Matter, an NSF Physics Frontiers Center (NSF Grant PHY-2317110), and the AI for Quantum Matter program at KITP. This research was supported in part by grant NSF PHY-2309135 to the Kavli Institute for Theoretical Physics (KITP). NP acknowledges AI assistance (Codex, GPT $\leq 5.6$) in coding and checking theoretical results. 
JKJ. is supported by the U.S. Department of Energy, Office of Science, Office of Advanced Scientific Computing Research, Department of Energy Computational Science Graduate Fellowship under Award Number(s) DE-SC0025528. JKJ used the Lawrencium computational cluster resource provided by the IT Division at the Lawrence Berkeley National Laboratory (Supported by the Director, Office of Science, Office of Basic Energy Sciences, of the U.S. Department of Energy under Contract No. DE-AC02-05CH11231). \par 

This report was prepared as an account of work sponsored by an agency of the
United States Government. Neither the United States Government nor any agency thereof, nor
any of their employees, makes any warranty, express or implied, or assumes any legal liability
or responsibility for the accuracy, completeness, or usefulness of any information, apparatus,
product, or process disclosed, or represents that its use would not infringe privately owned
rights. Reference herein to any specific commercial product, process, or service by trade name,
trademark, manufacturer, or otherwise does not necessarily constitute or imply its
endorsement, recommendation, or favoring by the United States Government or any agency
thereof. The views and opinions of authors expressed herein do not necessarily state or reflect
those of the United States Government or any agency thereof.
\end{acknowledgments}

\bibliography{bib, nqs}
\clearpage
\section*{End Matter}
\appendix

\section{Proof of Prop.~\ref{prop:spin} and computing Walshness}
We first prove Prop.~\ref{prop:spin} and then describe a practical algorithm to compute Walshness. We use the notation of the main text, with $\langle O\rangle \equiv \bra{\psi}O\ket\psi/\braket{\psi}{\psi}$. We begin with the identity
\begin{equation}
\label{eq:W0operator}
\wz[\ket\psi]= \frac{1}{\alpha} \log \Big\langle \prod_{j=1}^n \left(c_{\alpha+} I - c_{\alpha-}X_j\right)\Big\rangle,
\end{equation}
where $c_{\alpha\pm} \equiv (e^\alpha\pm 1)/2$. To see this, note that the Walsh-Hadamard transform [Eq.~\eqref{eq:wht}] is, up to normalization, the $Z$-basis amplitude of $H^{\otimes n}\ket\psi$, where $H$ is the Hadamard, and hence $p_\psi(\vec r) = |\bra{\vec r}H^{\otimes n}\ket\psi|^2/\braket{\psi}{\psi}$. Since $[ \sum_j (1-Z_j)/2]\ket{\vec r}=|\vec r|_1 \ket{\vec r}$, we have 
\begin{multline}
\mathbb{E}_{\vec r\sim p_\psi}\big[e^{\alpha|\vec r|_1}\big]= \Big\langle H^{\otimes n}\prod_j e^{\alpha(1-Z_j)/2}H^{\otimes n}\Big\rangle  \\
= \Big\langle\prod_j e^{\alpha(1-X_j)/2}\Big\rangle.
\end{multline}
Using $e^{\alpha(1-X)/2} = c_{\alpha+}I - c_{\alpha-}X$ yields Eq.~\eqref{eq:W0operator}. Next, for $U = \prod_j U_j\in\mathcal U$, applying Eq.~\eqref{eq:W0operator} to $U\ket\psi$ and using $U_j^\dagger X_j U_j = \hat{\vec n}_j\cdot\vec\sigma_j$ yields
\begin{equation}\label{eq:walshness_spin_formulation}
\wz[U\ket\psi] = \frac{1}{\alpha}\log\Big\langle \prod_{j=1}^n \left(c_{\alpha+}I - c_{\alpha-}\hat{\vec n}_j\cdot\vec\sigma_j\right)\Big\rangle.
\end{equation}
Finally, $c_{\alpha+}I - c_{\alpha-}\hat{\vec n}\cdot\vec\sigma = e^{\alpha/2}A_\alpha(\hat{\vec n})$, the prefactor $e^{n\alpha/2}$ contributes $n/2$, and minimizing over $\hat{\vec n}_j\in S^2$ gives Eq.~\eqref{eq:spinmodel}. Expanding $\prod_j A_\alpha(\hat{\vec n}_j)$ yields terms proportional to various $k$-point correlators $\langle\prod_{j\in S}\sigma_j^{a_j}\rangle$ for $|S|=k\leq n$, which completes the proof.
\par 
We now discuss minimizing $E[\hat{\vec n}_1,\ldots,\hat{\vec n}_n]$ in Eq.~\eqref{eq:spinmodel}. Although the couplings involve many spins, $E$ is linear in each $\hat{\vec n}_j$ separately, which permits an exact single-site update: (1) freeze every spin except $\hat{\vec n}_j$, (2) define an environment state $\ket{\phi_{-j}}
=
\prod_{k\neq j}
A_\alpha(\hat{\boldsymbol n}_k)\ket{\psi}$ and a vector $\boldsymbol v_j
=
\bra{\psi}
\boldsymbol\sigma_j
\ket{\phi_{-j}}$, (3) update $\hat{\boldsymbol n}_j\leftarrow \boldsymbol v_j / \|\boldsymbol v_j\|$ whenever $\vec v_j \neq \vec 0$, (4) sweep across $j$ and repeat until convergence. In practice, we have found rapid convergence using a tolerance of $10^{-10}$, with agreement across many random initializations. We retain the smallest value found, though we note that local minima are possible in this nonconvex optimization~\cite{supp}.\par 

This procedure produces (a numerical upper bound on) $\w[\ket\psi]$ and a set of unit vectors $\{\hat{\vec n}_j\}$, from which we choose $U=\prod_j U_j$ satisfying
$U_j^\dagger X_jU_j=\hat{\vec n}_j\cdot\vec\sigma_j$. Whenever
$\wz[U\ket\psi]< \wz[\ket\psi]$,
these axes identify a basis with lower fixed-basis Walshness.
To use this basis for learning, we fit an MLP to the rotated
amplitudes: $\psi_{\vec \theta}(\vec s)
    \approx
    \bra{\vec s}U \ket\psi$ with $\ket{\psi_{\vec \theta}}
    =
    \sum_{\vec s}\psi_{\vec\theta}(\vec s)\ket{\vec s}.$ Empirically, we find that this preprocessing improves learning with shallow MLP architectures, as reported in the main text for the transverse-field Ising model and in the following section for the mixed-field toric code. \par

For variational ground-state calculations, a basis selected
from any known \textit{approximation} of the ground state can similarly be used
to define the ansatz $U$. The energy
in a Hamiltonian $H$ is $E_{\vec \theta}=\frac{\bra{\psi_{\vec \theta}}UHU^\dagger\ket{\psi_{\vec \theta}}
    }{\langle\psi_{\vec \theta}|\psi_{\vec\theta}\rangle}$, so one can train the NQS with the rotated Hamiltonian $UHU^\dagger$. Because $U$ is a product
of single-site unitaries, this transformation preserves the
support of every Hamiltonian term. Another possibility is to compute the exact ground state of $H$ at small system sizes, find the Walshness-minimizing basis, and extrapolate to larger system sizes. We leave a detailed study of Walshness in the context of variational energy minimization to future work.
\par 
Finally, we note that, in our expression for Walshness in Eq.~(\ref{eq:walshness_spin_formulation}), the expectation is that of a spin operator that is highly nonlocal, and hence computing Walshness with Monte Carlo methods may be challenging.
It would be useful for future work to devise a method to efficiently compute Walshness-minimizing bases for large system sizes, where NQS methods are expected to provide the greatest advantage.

\section{Mixed-field toric code}

We consider the mixed-field $\mathbb Z_2$ toric code (MFTC)
\begin{equation}
    H_{\mathrm{TC}}
    =
    -\sum_s A_s
    -\sum_p B_p
    -h_X\sum_i X_i
    -h_Z\sum_i Z_i,
\end{equation}
on an $L\times L$ square lattice with periodic boundary
conditions, with $n=2L^2$ qubits on its edges,
$A_s=\prod_{i\ni s}X_i$ and
$B_p=\prod_{i\in\partial p}Z_i$~\cite{kitaev2003fault}. At $h_X=h_Z=0$, there exists a closed-form expression for the fixed-basis Walshness for any logical sector:
\begin{equation}\label{eq:tc_walshness_density}
\lim_{n\rightarrow\infty}\frac{\wz[\ket{\psi}]}{n}
=\frac12+\frac{\Phi(\alpha/2)-\ln 2}{2\alpha}~,
\end{equation}
where $\Phi(x)$ is the proportional to the free energy density of the 2D Ising model in the thermodynamic limit. Interestingly, we find that tuning $\alpha$ across $\ln(1+\sqrt{2})$ encounters the phase transition of the 2D Ising model~\cite{supp}.

\par
\begin{figure}[t]
    \centering
    \includegraphics[width=\columnwidth]{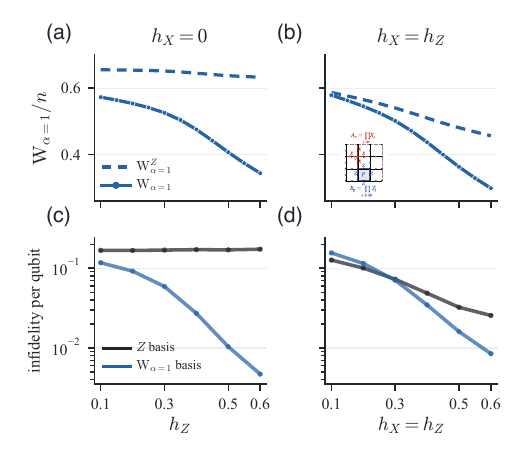}
\caption{\textbf{Mixed-field toric code.} (a,b) $Z$-basis (dashed) and numerically minimized
(solid) Walshness at $\alpha=1$, along $h_X=0$ and
$h_X=h_Z$ for the $L=3$ torus [$n=18$, inset in (b)].
(c,d) Median best infidelity per qubit during
$2\cdot 10^4$ Adam epochs for a one-hidden-layer $\tanh$
MLP of width $63$, aggregated over $36$ runs per field point
and basis.}    \label{fig:mftc_summary}
\end{figure}

For the MFTC, we obtain ground states by
exact diagonalization (ED) along $h_X=0$ and the
self-dual line $h_X=h_Z$. For each ground state, we compare three product bases: the $Z$ basis and the
Walshness-minimizing bases at $\alpha=1$ and $\alpha=2$ ($\mathrm W_1$ and $\mathrm W_2$ bases).
Fig.~\ref{fig:mftc_summary}a-b compares the fixed-basis and
intrinsic Walshness along these two cuts~\cite{supp}.
Along $h_X=0$, the optimized bases substantially reduce
Walshness. The line
$h_X=h_Z$ exhibits a more modest reduction. Interestingly, the self-dual line $h_X=h_Z$ yields the largest intrinsic Walshness (at these system sizes, at constant $\sqrt{h_X^2+h_Z^2}$).

\par
We test learnability using real one-hidden-layer $\tanh$ MLPs
(widths $42$ and $63$), trained on the target amplitudes in each
basis using Adam for $2\cdot 10^4$ epochs with minibatches of
$8192$ samples. For each architecture, field point, and basis, we aggregate $36$ training runs and report the median best infidelity per qubit, $1-\mathrm{F}^{1/n}$.
We note that the learning trends in
Fig.~\ref{fig:mftc_summary}c-d broadly track the Walshness trends: the clearest advantage
appears along $h_X=0$, where basis optimization produces
the larger reduction in Walshness, and a more modest advantage appears along $h_X=h_Z\gtrsim 0.3$.

\section{Restricted Boltzmann machines (RBMs)}
\label{sec:rbm-endmatter}

We explain Thm.~\ref{thm:rbm}, the analog of Thm.~\ref{thm:mlp} for RBMs. An RBM comprises $n$
visible spins $x_i=(-1)^{s_i}$ and $M$ hidden spins
$h_j=\pm1$ with hidden-visible couplings:
\begin{align}
\label{eq:psiRBMdef}
\psi(\vec s)
&=
\sum_{\vec h}
e^{\sum_i a_i x_i+\sum_j b_j h_j+\sum_{ij}W_{ji}x_i h_j}
\nonumber\\
&=
e^{\sum_i a_i x_i}
\prod_{j=1}^{M}
2\cosh\!\left(b_j+\sum_i W_{ji}x_i\right),
\end{align}
with biases $a_i,b_j$ and weights $W_{ji}$. This form motivates applying the Walsh-Hadamard transformation to
$\log\psi$. Assume $\psi(\vec s)\neq0$ for every $\vec s$, and choose a
logarithm $K$ with $e^K=\psi$ (with the choice of $K$ part of
the representation). Let
$K_{\le D}$ denote the truncated expansion [Eq.~\eqref{eq:psitilde}] of $K$ at cutoff $D$. To compute how accurately such truncations recover $\psi$, define
\begin{equation}
E_D^*(K)
=
\min_{0\le j\le D}
\left[1-\operatorname{F}\!\left(e^K,e^{K_{\le j}}\right)\right],
\end{equation}
where $\operatorname{F}$ is the normalized fidelity. The decreases $E_{D-1}^*-E_D^*$ measure the
improvement at each degree and sum to one, with
$E_{-1}^*=1$ and $E_n^*=0$, allowing us to define a probability distribution $p_D^* = E_{D-1}^* - E_D^*$ as the multiplicative analog of the Walsh spectrum and
\begin{equation}
\mathrm W_\alpha^*(K)
=
\frac{1}{\alpha}\log
\sum_{D=0}^{n}
e^{\alpha D}
p_D^*(K)
\end{equation}
as the \textit{multiplicative Walshness}. $\mathrm W_\alpha^*$ is small when low-order truncations of $K$ suffice to approximate $\psi$.
\par 
The idea of Thm.~\ref{thm:rbm} is to truncate the Walsh-Hadamard expansion of
$\log\psi$ and assign at most one hidden unit to each term. Each hidden unit connects only to the bounded number of spins appearing in the given term. Small $W^*_\alpha$ enables accurate truncation at low $D$,
and only polynomially many hidden units are needed.
Conversely, assuming bounded hidden-visible connectivity in each factor in Eq.~\eqref{eq:psiRBMdef} can be shown to strongly limit $W_\alpha^*(K)$. We give the full construction and proof details in the Supplemental Material~\cite{supp}.

\par
We test quantum state learning with RBMs in the TFIM ($n=10$, $g=1.5$) under the global basis rotation protocol of Ref.~\cite{cortes2025basis}, using
fully connected complex RBMs with $M/n=1,2,3$ and
$1000$ steps of stochastic reconfiguration~\cite{supp}.
Since $\psi>0$, $K=\log\psi$
is real. We find that as multiplicative Walshness increases, energy
error and infidelity broadly increase, even as RBM size is increased. While this is a test at modest sizes, multiplicative Walshness demonstrates agreement with training ease for RBMs as well as rigorous two-sided representability guarantees. Our main results thus extend to RBMs with the above considerations.

\begin{figure}[h]
    \centering   \includegraphics[width=\linewidth]{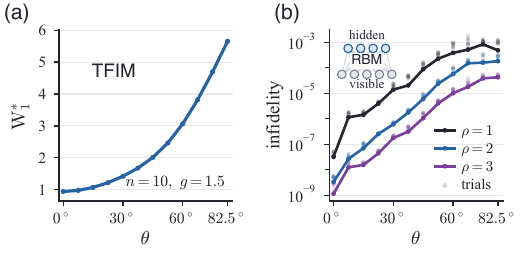}
{\caption{\textbf{Multiplicative Walshness and RBM learning.}
(a) $\mathrm W_\alpha^*$ for $\alpha=1$ and (b) infidelity versus global basis
rotation angle $\theta$ in the TFIM.
Curves show the lowest-energy retained trial after
$1000$ SR steps, with hidden-unit density $\rho=M/n$. Data in part reproduced from \cite{cortes2025basis} (see~\cite{supp}). Increase in $\mathrm{W}_{1}^*$ tracks increase in infidelity across $\theta$. 
}}
    \label{fig:RBMs}
\end{figure}

\par

\end{document}

\title{\Large Supplemental Material:\\
\large
Walshness: an intrinsic neural-network representability metric for quantum states}

\author{Nisarga Paul}
\email{npaul@caltech.edu}
\affiliation{Department of Physics and Institute for Quantum Information and Matter,
California Institute of Technology, Pasadena, California 91125, USA}

\author{Kehuang Chen}
\affiliation{Department of Physics, Tsinghua University, Beijing 100084, China}

\author{Jessica K. Jiang}
\affiliation{Department of Physics and Institute for Quantum Information and Matter,
California Institute of Technology, Pasadena, California 91125, USA}

\author{Haimeng Zhao}
\affiliation{Department of Physics and Institute for Quantum Information and Matter,
California Institute of Technology, Pasadena, California 91125, USA}

\author{Di Luo}
\email{diluo@tsinghua.edu.cn}
\affiliation{Department of Physics, Tsinghua University, Beijing 100084, China}
\affiliation{Institute of Advanced Study, Tsinghua University, Beijing 100084, China}
\maketitle
\onecolumngrid
\tableofcontents

\section{Overview}

This Supplemental Material is organized as follows.
Section~\ref{sec:defns} develops the definitions and properties of
fixed-basis and intrinsic Walshness. We establish their basic
structural properties, analyze examples, and derive a bound relating Walshness to bipartite
entanglement. We also formulate the minimization over local bases
as a classical spin optimization problem and describe the numerical
minimization used in this work.

Section~\ref{sec:theorems} provides the precise statements and proofs
connecting Walshness to neural network representability. We
derive bounds on fixed-basis Walshness for multilayer perceptrons
(MLPs) with polynomial or analytic nonlinearities, making explicit relevant architectural assumptions. Conversely, we show how low
fixed-basis Walshness permits an approximate MLP representation by
truncating the Walsh spectrum and constructing a network that
evaluates the retained terms. These results quantify the relation
between Walshness, network depth and size, and approximation accuracy.
\par 
Section~\ref{sec:RBM} introduces multiplicative Walshness for
wavefunctions represented through their logarithmic amplitudes.
We relate this quantity to representability by restricted Boltzmann machines (RBMs). This section proceeds as an RBM analog of the previous section.
\par 

Section~\ref{sec:casestudies} presents additional analytical results
and numerical details for various quantum state families: ferromagnetic product
states, stabilizer states, the transverse-field Ising model, and the
mixed-field toric code. We describe
the machine learning details and examine the structure of
Walshness-minimizing bases. The learning experiments complement our theory by affirming that reductions in Walshness translate into easier NQS training.

\par 
\vspace{0.5cm}

\section{Walshness: definitions and properties}
\label{sec:defns}

\subsection{Definitions}
\begin{figure}[t]
\centering
\resizebox{0.95\textwidth}{!}{%
\begin{tikzpicture}[
  x=1cm,y=1cm,
  bar/.style={draw=wblue,fill=wblue!40,line width=0.5pt},
  sbar/.style={draw=wblue,fill=wblue!40,line width=0.4pt},
  ax/.style={draw=black!70,line width=0.6pt},
  sax/.style={draw=black!50,line width=0.4pt},
  plab/.style={font=\small},
  tlab/.style={font=\tiny},
  spin/.style={->,>=stealth,line width=1.4pt,draw=wpurple}
]
 
% =========== (a) the three examples, as a table ===========
\fill[wblue!12] (-2.3,4.15) rectangle (4.1,4.7);
\fill[black!5]  (-2.3,2.85) rectangle (4.1,4.1);
\fill[black!5]  (-2.3,0.25) rectangle (4.1,1.5);
\draw[draw=black!35,line width=0.5pt] (-2.3,4.13) -- (4.1,4.13);
% header
\node[plab] at (-1.4,4.42) {$\psi(\vec s)$};
\node[plab] at (0.55,4.42) {$p(\vec r)$};
\node[plab] at (2.35,4.42) {$\wz$};
\node[plab] at (3.6,4.42)  {$\w$};
% row 1, constant
\node[plab] at (-1.4,3.5) {$\mathrm{const}$};
\draw[sbar] (-0.35,3.25) rectangle (-0.19,3.97);
\draw[sax] (-0.4,3.25) -- (1.5,3.25);
\draw[sax] (-0.4,3.25) -- (-0.4,4.05);
\node[tlab,below=1.5pt] at (0.55,3.25) {$|\vec r|_1$};
\node[plab] at (2.35,3.5) {$0$};
\node[plab] at (3.6,3.5)  {$0$};
% row 2, basis state
\node[plab] at (-1.4,2.2) {$\delta_{\vec s,\vec s_0}$};
\foreach \i/\h in {0/0.04,1/0.14,2/0.34,3/0.58,4/0.72,5/0.58,6/0.34,7/0.14,8/0.04}{
  \draw[sbar] ({0.2*\i-0.35},1.95) rectangle ({0.2*\i-0.19},{1.95+\h});}
\draw[sax] (-0.4,1.95) -- (1.5,1.95);
\draw[sax] (-0.4,1.95) -- (-0.4,2.75);
\node[tlab,below=1.5pt] at (0.55,1.95) {$|\vec r|_1$};
\node[plab] at (2.35,2.2) {$\in(\tfrac n2,\,n)$};
\node[plab] at (3.6,2.2)  {$0$};
% row 3, parity
\node[plab] at (-1.4,0.9) {$(-1)^{|\vec s|_1}$};
\draw[sbar] (1.25,0.65) rectangle (1.41,1.37);
\draw[sax] (-0.4,0.65) -- (1.5,0.65);
\draw[sax] (-0.4,0.65) -- (-0.4,1.45);
\node[tlab,below=1.5pt] at (0.55,0.65) {$|\vec r|_1$};
\node[plab] at (2.35,0.9) {$n$};
\node[plab] at (3.6,0.9)  {$0$};
\node[plab] at (-2.6,5.15) {(a)};
 
\foreach \i/\h in {0/0.35,1/0.9,2/1.5,3/1.05,4/1.3,5/0.7,6/0.45,7/0.18}{
  \draw[bar] ({0.36*\i+6.06},1.3) rectangle ({0.36*\i+6.3},{1.3+\h});}
\draw[ax] (6.0,1.3) -- (9.0,1.3);
\draw[ax] (6.0,1.3) -- (6.0,3.4);
\draw[line width=1pt,draw=worange]
  (6.06,1.37) .. controls (7.3,1.48) and (8.0,1.8) .. (8.9,3.05);
\node[plab,worange] at (8.35,3.28) {$e^{\alpha|\vec r|_1}$};
\node[plab,rotate=90] at (5.55,2.3) {$p(\vec r)$};
\node[plab,below=3pt] at (7.5,1.3) {$|\vec r|_1$};
\node[plab] at (6.75,3.75) {$\wz$};
\node[plab] at (5.55,5.15) {(b)};
 
% --- the single-site rotations ---
\draw[line width=0.5pt,draw=black!35] (9.35,2.7) -- (11.5,2.7);
\foreach \x in {9.5,9.95,10.4,10.85,11.3}{\fill[black!45] (\x,2.7) circle (1.2pt);}
\foreach \x/\a in {9.5/115, 9.95/-90, 10.4/65, 10.85/-140, 11.3/15}{
  \draw[spin] (\x,2.7) ++(\a+180:0.17) -- ++(\a:0.34);}
\node[plab] at (10.45,3.35) {$\hat{\vec n}_j$};
\draw[->,>=stealth,line width=0.8pt,draw=black!55] (9.85,2.0) -- (11.05,2.0)
  node[midway,below=4pt,plab] {$\min\limits_{\mathcal{U}}$};
\node[font=\tiny\itshape,align=center] at (10.45,0.98)
  {classical spin\\[0.5pt] optimization};
 
% --- spectrum after the minimization ---
\foreach \i/\h in {0/1.6,1/0.95,2/0.55,3/0.3,4/0.12,5/0.08,6/0.03,7/0.02}{
  \draw[bar] ({0.36*\i+12.06},1.3) rectangle ({0.36*\i+12.3},{1.3+\h});}
\draw[ax] (12.0,1.3) -- (15.0,1.3);
\draw[ax] (12.0,1.3) -- (12.0,3.4);
\node[plab,below=3pt] at (13.5,1.3) {$|\vec r|_1$};
\node[plab] at (13.5,3.75) {$\w$};
\end{tikzpicture}}
\caption{
\textbf{Fixed-basis and intrinsic Walshness.}
(a) Walsh spectra, fixed-basis Walshness, and (intrinsic) Walshness for the three ferromagnets $\ket{\vec{+x}},\ket{\vec{+z}},\ket{\vec{-x}}$, which have constant, delta-function, or parity-like wavefunctions, respectively. Fixed-basis Walshness ranges from $0$ to $n$, but all have vanishing Walshness.
(b) Schematic: fixed-basis Walshness exponentially weights high-degree components of the Walsh spectrum. It can also be regarded as a cumulant-generating function of the Walsh spectrum. Minimizing over single-site rotations ${\hat{\vec n}_j}$ shifts spectral weight toward low $|\vec r|_1$ and defines the Walshness $\w$. The minimization maps to a classical spin optimization problem.}
\label{fig:walshness-defs}
\end{figure}
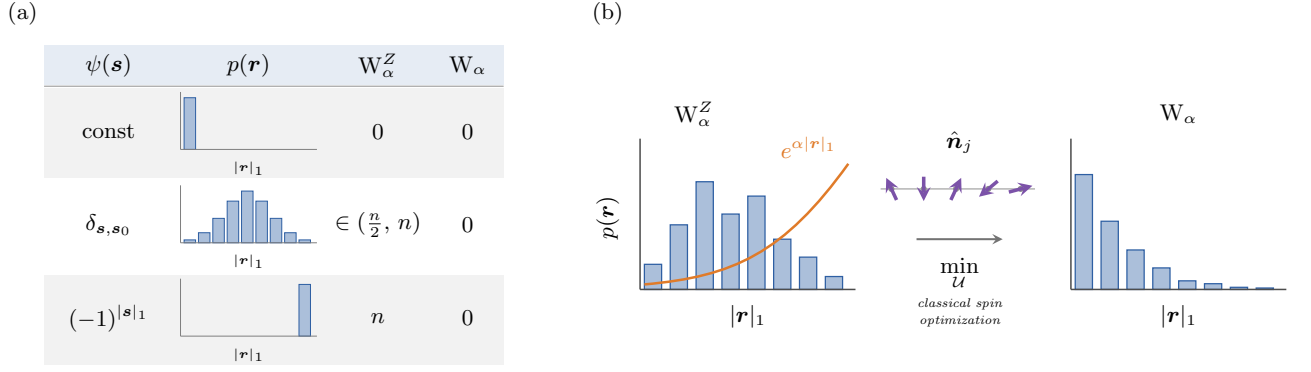
We collect definitions in this section. We will primarily be concerned with pure quantum states defined on $n$ qubits,
\begin{equation}
    \ket{\psi} = \sum_{\vec s\in\{0,1\}^n} \psi(\vec s)\ket{\vec s},\qquad \psi: \{0,1\}^n\mapsto \C,
\end{equation}
where $\ket{\vec s}$ is a tensor product state in the Pauli $Z$ basis, $Z\ket{0} = \ket{0}, Z\ket{1} = -\ket{1}$. We will not require $\langle \psi |\psi\rangle = 1$ though, as crucially, our definitions will be invariant under rescaling $\psi$. Hence our definitions are appropriate for quantum states. We will assume $\ket{\psi} \neq 0$. Define, for $\vec r,\vec s \in \{0,1\}^n$, the \textbf{\textit{Walsh-Hadamard transform}}~\cite{nielsen2000quantum}
\begin{equation}
    \widehat\psi(\vec r) \coloneqq 2^{-n}\sum_{\vec s} (-1)^{\vec r\cdot \vec s} \psi(\vec s). 
\end{equation}
Equivalently, $\widehat \psi(\vec r)\sim \langle \vec r|H^{\otimes n}\ket\psi$ where $H$ is the Hadamard. Further define the \textbf{\textit{Walsh spectrum}}
\begin{equation}
    p(\vec r) \coloneqq \frac{|\widehat\psi(\vec r)|^2}{\sum_{\vec r} |\widehat\psi(\vec r)|^2}
\end{equation}
which is a classical distribution, $p(\vec r)\geq 0, \sum_{\vec r} p(\vec r) = 1$. Next, define the \textbf{\textit{fixed-basis Walshness}} 
\begin{equation}   \wz \coloneqq  \frac{1}{\alpha} \log \mathbb{E}_p\left[ e^{\alpha |\vec r|_1}\right]=\frac{1}{\alpha} \log\left(\sum_{\vec r} e^{\alpha |\vec r|_1} p(\vec r)\right),
\end{equation}
where $\alpha >0$ is a free parameter and $|\vec r|_1=\sum_{i=1}^n r_i \in \{0,\ldots, n\}$ is the Hamming weight of $\vec r$. Fixed-basis Walshness is invariant under rescaling $\psi$. Intuitively, $\wz$ is a measure of the weight distribution of $p(\vec r)$ ordered by Hamming weight. For instance, if $\psi(\vec s) = (-1)^{\sum_i s_i}$ is the parity function, then $p(\vec r) = \delta_{\vec r, \vec j}$ where $\vec j = 111\cdots$ is the string of 1s, which has maximal Hamming weight. Consequently, 
\begin{equation}
    \wz = \frac{1}{\alpha} \log \sum_{\vec r} \delta_{\vec r,\vec j} e^{\alpha |\vec r|_1} = \frac{1}{\alpha}\log e^{\alpha n} =n.\tag{parity}
\end{equation}
Conversely, if $\psi(\vec s) = $ constant, then 
\begin{equation}
    \wz = 0. \tag{constant}
\end{equation}
These are two extremes. In general, we have
\begin{equation}
    \wz \in [0,n].
\end{equation}
Another illustrative example is when $\psi(\vec s) = \delta_{\vec s,\vec s_0}$ is localized at some basis state $\ket{\vec s_0}$. In this case, it is simple to show that $\widehat\psi(\vec r) = 2^{-n}(-1)^{\vec r\cdot \vec s_0}$, so $p(\vec r)$ is the uniform distribution, and 
\begin{equation}
    \wz = \frac{1}{\alpha}\log \left[2^{-n}\sum_{k=0}^n \binom{n}{k} e^{\alpha k} \right]=\frac{n}{\alpha}\log \frac{1+e^\alpha}{2}\tag{$\delta$-function}
\end{equation}
which monotonically interpolates between the mean Hamming weight $n/2$ and the maximum $n$ as $\alpha\to \infty$. We regard $\alpha$ as a parameter with no \textit{a priori} preferred value; indeed, different values of $\alpha$ may prove useful across different applications. In this regard, we may draw an analogy with the index of the R\'enyi entropy in quantum information.\par 

We will note another convenient expression for the fixed-basis Walshness:

\begin{equation}
\label{eq:fbWalternate}
    \wz = \frac{1}{\alpha} \log \left\langle \prod_{j=1}^n \left(c_{\alpha+} I - c_{\alpha-}X_j\right)\right\rangle,\qquad c_{\alpha\pm} \coloneqq \frac{e^\alpha\pm 1}{2}
\end{equation}
where $\left\langle O\right\rangle \coloneqq \frac{\left \langle \psi | O |\psi\right\rangle}{\left\langle \psi|\psi\right\rangle}$. 
This expression is more convenient for use in numerical calculations. We leave the proof of Eq.~\eqref{eq:fbWalternate} to the end of this section. 
\par 

So far, $\wz$ depends explicitly and sensitively on the choice of basis. That is, $\wz(\ket\psi) \neq \wz(U\ket\psi)$ for a generic local unitary $U$. This is by design, and finding the change of basis $U$ that minimizes $\wz$ is a useful objective. It is also desirable to have a basis-independent version of $\wz$, and for this we will explicitly minimize over some set of local unitaries. In general, such a minimization is a highly nontrivial analytical or numerical task, and hence in this work we content ourselves with the simplest such set: single-site unitaries
\begin{equation}
    \mathcal{U} \coloneqq \left\{\prod_{j=1}^n U_j \big\vert U_j \in \mathrm{SU}(2) \right\}.
\end{equation}
We then define the \textbf{\textit{intrinsic Walshness}}, or simply \textbf{\textit{Walshness}}, as 
\begin{equation}
    \w \coloneqq \min_{U\in \mathcal{U}} \wz(U \ket\psi). 
\end{equation}
We note that all three of our examples (parity, constant, and $\delta$-function) from earlier have \textit{vanishing} intrinsic Walshness. Indeed, this is the case for all product states. Note that parity, constant, and $\delta$-function correspond to the product states $\otimes_j \ket{-}, \otimes_j \ket{+}_j,$ and $\otimes_j \ket{s_{0,j}}_j$ respectively where $\ket{\pm}$ are $X$ eigenstates. All product states are single-qubit rotations away from $\otimes_j\ket{+}$, which is the constant function, and hence has vanishing Walshness.

\par 
Following Eq.~\eqref{eq:fbWalternate}, a convenient, alternate form for $\w$ is given by 
\begin{equation}
    \label{eq:Walternate}
    \w = \frac{1}{\alpha} \min_{\hat{\vec n}_j \in S^2} \log \left\langle \prod_{j=1}^n \left(c_{\alpha+}I - c_{\alpha -} \hat{\vec n}_j \cdot \vec \sigma_j\right) \right\rangle, 
\end{equation}
where the $\hat{\vec n}_j$ are unit vectors defined by $U_j^\dagger X_j U_j = \hat{\vec n}_j \cdot \vec \sigma_j$, where $\vec \sigma_j = (X_j,Y_j,Z_j)$ are Paulis. Hence the $\hat{\vec n}_j$ parameterize single-site unitaries $U_1 \cdots U_n \in \mathcal{U}$. Moreover, each factor has eigenvalues $1,e^\alpha$, so the expectation is
positive and the logarithm is increasing on its range, allowing the minimization to be
moved inside the logarithm: 
\begin{equation}
    \label{eq:Walternate2}
    \w = \frac{1}{\alpha}  \log \min_{\hat{\vec n}_j \in S^2}\left\langle \prod_{j=1}^n \left(c_{\alpha+}I - c_{\alpha -} \hat{\vec n}_j \cdot \vec \sigma_j\right) \right\rangle.
\end{equation}
The argument of the minimization in Eq.~\eqref{eq:Walternate2} has an appealing interpretation as the energy of a classical $O(3)$ spin model. To make this more explicit, define
\begin{equation}
A_\alpha(\hat{\boldsymbol n})
=
\cosh\left(\frac{\alpha}{2}\right)I
-
\sinh\left(\frac{\alpha}{2}\right)
\hat{\boldsymbol n}\cdot\boldsymbol\sigma,
\end{equation}
so we have $c_{\alpha+}I-c_{\alpha-}\hat{\boldsymbol n}\cdot\boldsymbol\sigma
=
e^{\alpha/2}A_\alpha(\hat{\boldsymbol n})$, and further define
\begin{equation}
E[\{\hat{\vec n}_j\}]
=
\left\langle
\prod_{j=1}^n
A_\alpha(\hat{\boldsymbol n}_j)
\right\rangle.
\end{equation}
The Walshness is then
\begin{equation}\label{eq:Walternate3}
\w 
=
\frac n2+
\frac{1}{\alpha}
\log
\min_{\hat{\vec n}_j\in S^2}E[\{\hat{\vec n}_j\}].
\end{equation}
Thus, it is sufficient to minimize the scalar objective $E[\{\hat{\vec n}_j\}]$, which could be thought of as the energy functional of a classical spin model whose couplings are determined by $\psi$. Finally, it is convenient to define the \textbf{\textit{(fixed-basis) Walshness density}} by dividing by $n$: 
\begin{equation}
    \wdd = \frac1n \w \in [0,1],\qquad \wzd = \frac1n \wz \in [0,1].
\end{equation}
We detail our approach to calculating $\w$ numerically, or in some cases analytically, in a later section. \par 

We now prove Eqs.~\eqref{eq:fbWalternate} and \eqref{eq:Walternate}. Both follow from a
single identity, Eq.~\eqref{eq:rotatedW} below, which expresses the fixed-basis Walshness
of $U\ket\psi$ for arbitrary $U \in \mathcal{U}$. Let $U_j^\dagger X_j U_j = \hat{\vec n}_j \cdot \vec \sigma_j$ and, noting that
$(\hat{\vec n}_j\cdot \vec \sigma_j)^2 = I$, define the projectors
$P_j^{\pm} \coloneqq \frac12 (I\pm \hat{\vec n}_j \cdot \vec \sigma_j)$ onto the $\pm 1$
eigenstates of $\hat{\vec n}_j \cdot \vec \sigma_j$. We may then write
\begin{equation}
\label{eq:projectors}
    c_{\alpha+}I-c_{\alpha-}\hat{\vec n}_j \cdot \vec \sigma_j
    = (c_{\alpha+}-c_{\alpha-})P_j^+ + (c_{\alpha+}+c_{\alpha-})P_j^-
    = P_j^+ +e^\alpha P_j^- = e^{\alpha P_j^-},
\end{equation}
where the second equality used $c_{\alpha+}-c_{\alpha-} = 1$ and
$c_{\alpha+}+c_{\alpha-}=e^\alpha$, and the third holds because $(P_j^-)^2 =P_j^-$
truncates the exponential series to $e^{\alpha P_j^-} = I+(e^\alpha-1)P_j^-
= P_j^+ + e^\alpha P_j^-$. Factors on distinct sites commute, so
\begin{equation}
    \prod_{j=1}^n (c_{\alpha+}I-c_{\alpha-}\hat{\vec n}_j \cdot \vec \sigma_j)
    = \exp\Big(\alpha\sum_j P_j^-\Big)
    =\sum_{\vec u\in \{0,1\}^n} e^{\alpha|\vec u|_1}
      \ket{\vec u}_{\hat{\vec n}}\!\bra{\vec u}_{\hat{\vec n}},
\end{equation}
where $\ket{\vec u}_{\hat{\vec n}} \coloneqq \bigotimes_j \ket{u_j}_{\hat{\vec n}_j}$ is
the joint eigenbasis of the $\hat{\vec n}_j\cdot\vec \sigma_j$ and $u_j \in \{0,1\}$ is the
eigenvalue of $P_j^-$.

\par 
Since $\hat{\vec n}_j \cdot \vec \sigma_j = U_j^\dagger X_j U_j$, the eigenstates of
$\hat{\vec n}_j \cdot \vec \sigma_j$ are $U_j^\dagger$ applied to those of $X_j$. The
latter are $\ket{+} = H\ket 0$ and $\ket - = H\ket 1$ (where $H$ is the Hadamard) with $u_j = 0$ and $u_j = 1$
respectively. Hence $\ket{\vec u}_{\hat{\vec n}} = U^\dagger H^{\otimes n}\ket{\vec u}$ where $\ket{\vec u}$ is in the standard computational basis. Writing $\ket{\psi_U} \coloneqq U\ket\psi$, we obtain
\begin{equation}
    _{\hat{\vec n}}\langle\vec u|\psi\rangle = \bra{\vec u} H^{\otimes n}\ket{\psi_U}
    = 2^{n/2}\, \widehat{\psi_U}(\vec u),
\end{equation}
and therefore
\begin{equation}
    \left\langle \prod_{j=1}^n \left(c_{\alpha+}I - c_{\alpha-}
    \hat{\vec n}_j \cdot \vec \sigma_j\right)\right\rangle
    = \frac{\sum_{\vec u} e^{\alpha |\vec u|_1}|\widehat{\psi_U}(\vec u)|^2}
           {\sum_{\vec u} |\widehat{\psi_U}(\vec u)|^2}
    = \mathbb{E}_{p_U}\!\left[e^{\alpha |\vec u|_1}\right],
\end{equation}
where $p_U$ is the Walsh spectrum of $\ket{\psi_U}$.
Taking the logarithm,
\begin{equation}
    \label{eq:rotatedW}
    \wz[U\ket\psi] = \frac{1}{\alpha} \log \left\langle
    \prod_{j=1}^n \left(c_{\alpha+}I - c_{\alpha-}\hat{\vec n}_j \cdot \vec \sigma_j
    \right)\right\rangle.
\end{equation}
Eq.~\eqref{eq:fbWalternate} is the case $U = I$, for which $\hat{\vec n}_j\cdot\vec\sigma_j
= X_j$. Eq.~\eqref{eq:Walternate} follows by minimizing Eq.~\eqref{eq:rotatedW} over
$\mathcal{U}$. Since $\hat{\vec n}_j$ sweeps $S^2$ as $U_j$ sweeps
$\mathrm{SU}(2)$, the minimization over $\mathcal{U}$ is a minimization over
$(S^2)^n$.

\subsection{Properties of Walshness}
\label{sec:properties}
In this section, we collect various properties, theoretical results, and potentially useful facts about (fixed-basis) Walshness. 
\begin{fact}
    $\wz[\ket\psi]$ and $\w[\ket\psi]$ are nonnegative, lie in the range $[0,n]$, and are invariant under permutations of the $n$ qubits. Moreover, $\w[\ket\psi]$ is invariant under single-site unitaries. 
\end{fact}
\begin{proof}
    These follow straightforwardly from the definitions. 
\end{proof}

\begin{fact}
    $\w[\ket\psi] = 0$ if and only if $\ket\psi$ is a product state.
\end{fact}
\begin{proof}
    If $\ket\psi$ is a product state, then it is a single-site unitary rotation away from $\ket{\vec{+x}}$, the state with constant wavefunction, and hence $\w[\ket\psi] = 0$. Conversely if $\w[\ket\psi] = 0$ then there is a single-site unitary $U$ such that $\ket{\tilde \psi} = U\ket\psi$ and $\wz[\ket{\tilde \psi}] = 0$. Let $\ket{\tilde \psi} = \sum_{\vec s} \tilde\psi(\vec s)\ket{\vec s}$ and let $p(\vec r)$ be the Walsh spectrum of $\ket{\tilde \psi}$. We must have $\mathbb{E}[e^{\alpha |\vec r|_1}] = 1$, which implies $p(\vec r)$ has support only on $|\vec r|_1 = 0$. This implies $p(\vec r)$ vanishes unless $\vec r = 0^n$, which in turn implies $\tilde\psi(\vec s)$ is constant. This implies $\ket{\tilde \psi} = \ket{\vec{+x}}$, so in particular $\ket\psi = U^\dagger \ket{\vec{+x}}$ is a product state. 
\end{proof}
\begin{fact}[Tensor products]
Fixed-basis Walshness and intrinsic Walshness are additive under tensor products:
\begin{equation}
\wz[\ket\psi\otimes \ket\phi] = \wz[\ket\psi]+\wz[\ket\phi],\quad \w[\ket\psi\otimes \ket\phi] = \w[\ket\psi]+\w[\ket\phi].
\end{equation}
\end{fact}
\begin{proof}
    We prove this for fixed-basis Walshness $\wz$; it then follows trivially for intrinsic Walshness. Let $\ket\Psi \equiv \ket\psi \otimes \ket \phi \in \mathcal{H}_1\otimes \mathcal{H}_2$, where $\mathcal{H}_i$ is the Hilbert space of $n_i$ qubits. Let $N = n_1+n_2$. Using Eq.~\eqref{eq:fbWalternate} and factorizing, we have 
\begin{equation}
    \begin{aligned}     \wz\left[\ket\Psi\right] &= \frac{1}{\alpha} \log \left\langle \Psi \left| \otimes_{j=1}^{N} (c_{\alpha+}I-c_{\alpha -}X_j)\right|\Psi\right\rangle \\
        &= \frac{1}{\alpha} \log \left\langle\psi \left| \otimes_{j=1}^{n_1} (c_{\alpha+}I-c_{\alpha -}X_j) \right| \psi \right\rangle +\frac{1}{\alpha}\log \left\langle\phi \left|\otimes_{j=1}^{n_2} (c_{\alpha+}I-c_{\alpha -}X_j) \right|\phi \right\rangle \\
        &= \wz\left[\ket\phi\right]+ \wz\left[\ket\psi\right]
    \end{aligned}
\end{equation}
as was to be shown. 
\end{proof}
An immediate corollary is that tensoring a trivial product state $\ket{\phi_0}$ does not increase Walshness. This behavior with respect to tensoring naturally raises the question of whether there are nice properties under addition. In general we cannot expect subadditivity, since the superposition of states with zero Walshness can have nonzero Walshness. However, one additivity property is the following. 
\begin{fact}[Superpositions]
For $m$ orthonormal states $\ket{\psi_1},\ldots, \ket{\psi_m}$, $\langle\psi_i|\psi_j\rangle = \delta_{ij}$, we have 
\begin{equation}
\label{eq:additivity}
\wz\left[\sum_{j=1}^m \ket{\psi_j}\right] \leq \max_j \left\{ \wz[\ket{\psi_j}]\right\} + \frac{1}{\alpha}\log m. 
\end{equation}
\end{fact}
\begin{proof}
Define the Hermitian, positive operator $A_\alpha = \otimes_{j=1}^n (c_{\alpha+}I-c_{\alpha-}X_j)$, which is related to the fixed-basis Walshness via $e^{\alpha \wz[\ket\psi]} = \frac{\langle \psi | A_\alpha |\psi\rangle}{\langle\psi|\psi\rangle}$ for general states $\ket\psi$. We introduce the norm $    \|\ket\psi\|_\alpha = \sqrt{\langle\psi|A_\alpha|\psi\rangle} = \|\ket\psi\| e^{\alpha \wz[\ket\psi]/2}.$ Let $M = \max_j \left\{ \wz[\ket{\psi_j}]\right\}$. The triangle inequality for $\|\cdot\|_\alpha$ then yields 
\begin{equation}
    \left\| \sum_{j=1}^m \ket{\psi_i}\right\|_\alpha  \leq \sum_{j=1}^m \left\|\ket{\psi_j}\right\|_\alpha \leq e^{\alpha M/2} \sum_{j=1}^m \|\ket{\psi_j}\| = me^{\alpha M/2}.
\end{equation}
Consequently, $ \frac{\left\| \sum_{j=1}^m \ket{\psi_i}\right\|_\alpha }{\left\| \sum_{j=1}^m \ket{\psi_i}\right\|} \leq \sqrt{m} e^{\alpha M/2}.$ Squaring, we obtain $ e^{\alpha \wz[\sum_{j=1}^m\ket{\psi_j}]}\leq m e^{\alpha M},$ which directly yields Eq.~\eqref{eq:additivity}. 
\end{proof}

\begin{fact}[Monotonicity in $\alpha$] For any state $\ket\psi$ and $0<\alpha_1<\alpha_2$, we have the following: 
\begin{equation} \label{eq:convexity}   \mathrm{W}_{\alpha_1}^Z[\ket\psi] \leq \mathrm{W}_{\alpha_2}^Z[\ket\psi],\qquad \mathrm{W}_{\alpha_1}[\ket\psi] \leq \mathrm{W}_{\alpha_2}[\ket\psi].
\end{equation}
In other words, (fixed-basis) Walshness is nondecreasing in $\alpha$. 
\end{fact}
\begin{proof}
Fix a basis and let $p(\vec r)$ be the Walsh spectrum of $\ket\psi$. Moreover, let $K = |\vec r|_1$, let $Z(\alpha) = \mathbb{E}_p[e^{\alpha K}]$, and let $F(\alpha) = \log Z(\alpha)$. Define also the ``tilted" distributions $p_\alpha(\vec r) = \frac{e^{\alpha|\vec r|_1}p(\vec r)}{Z(\alpha)}.$ Differentiating, we have $F'(\alpha) = \frac{\mathbb{E}_p[Ke^{\alpha K}]}{\mathbb{E}_p[e^{\alpha K}]} = \mathbb{E}_{p_\alpha}[K]$. 
Differentiating once more, we have $F''(\alpha) = \mathbb{E}_{p_\alpha}[K^2]-\mathbb{E}_{p_\alpha}[K]^2 = \mathrm{Var}_{p_\alpha}[K]\geq 0$. Therefore, $F$ is convex. Now let $0< \alpha_1<\alpha_2$ and set $\lambda = \frac{\alpha_1}{\alpha_2} \in (0,1)$. Since $\alpha_1 = \lambda\alpha_2 + (1-\lambda)\cdot 0$, convexity of $F$ together with $F(0) = 0$ imply 
\begin{equation}
    F(\alpha_1) \leq \lambda F(\alpha_2) + (1-\lambda)F(0) = \frac{\alpha_1}{\alpha_2}F(\alpha_2). 
\end{equation}
Dividing by $\alpha_1$, we obtain 
\begin{equation}
    \mathrm{W}_{\alpha_1}^Z[\ket\psi] = \frac{F(\alpha_1)}{\alpha_1} \leq \frac{F(\alpha_2)}{\alpha_2} = \mathrm{W}_{\alpha_2}^Z[\ket\psi]
\end{equation}
as was to be shown. Next, replacing $\ket\psi$ with $U\ket\psi$ and taking the minimum over $U$ implies $\mathrm{W}_{\alpha_1}[\ket\psi] \leq \mathrm{W}_{\alpha_2}[\ket\psi]$. 
\end{proof}

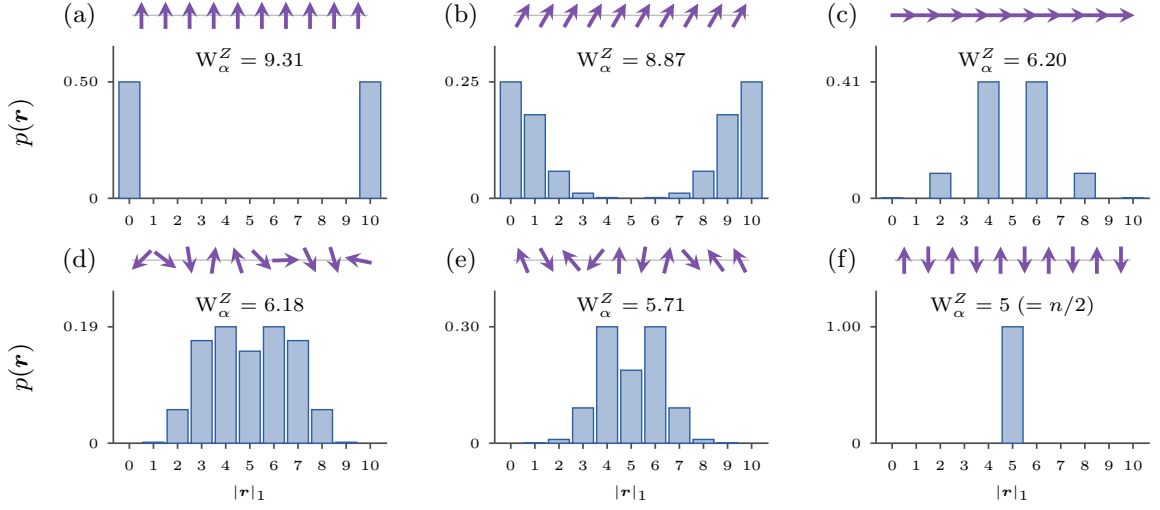
\begin{figure}[t]
\centering
\resizebox{0.85\textwidth}{!}{%
\begin{tikzpicture}[
  x=1cm,y=1cm,
  bar/.style={draw=wblue,fill=wblue!40,line width=0.5pt},
  sbar/.style={draw=wblue,fill=wblue!40,line width=0.4pt},
  ax/.style={draw=black!70,line width=0.6pt},
  plab/.style={font=\small}, tlab/.style={font=\tiny},
  spin/.style={->,>=stealth,line width=1.35pt,draw=wpurple},
  sspin/.style={->,>=stealth,line width=1.35pt,draw=wblue}]
\draw[ax] (0.000,3.050) -- (3.410,3.050);
\draw[ax] (0.000,3.050) -- (0.000,4.950);
\draw[bar] (0.070,3.050) rectangle (0.330,4.500);
\draw[bar] (3.070,3.050) rectangle (3.330,4.500);
\draw[ax] (0.200,3.050) -- ++(0,-0.07);
\node[tlab,below=2pt] at (0.200,3.020) {$0$};
\draw[ax] (0.500,3.050) -- ++(0,-0.07);
\node[tlab,below=2pt] at (0.500,3.020) {$1$};
\draw[ax] (0.800,3.050) -- ++(0,-0.07);
\node[tlab,below=2pt] at (0.800,3.020) {$2$};
\draw[ax] (1.100,3.050) -- ++(0,-0.07);
\node[tlab,below=2pt] at (1.100,3.020) {$3$};
\draw[ax] (1.400,3.050) -- ++(0,-0.07);
\node[tlab,below=2pt] at (1.400,3.020) {$4$};
\draw[ax] (1.700,3.050) -- ++(0,-0.07);
\node[tlab,below=2pt] at (1.700,3.020) {$5$};
\draw[ax] (2.000,3.050) -- ++(0,-0.07);
\node[tlab,below=2pt] at (2.000,3.020) {$6$};
\draw[ax] (2.300,3.050) -- ++(0,-0.07);
\node[tlab,below=2pt] at (2.300,3.020) {$7$};
\draw[ax] (2.600,3.050) -- ++(0,-0.07);
\node[tlab,below=2pt] at (2.600,3.020) {$8$};
\draw[ax] (2.900,3.050) -- ++(0,-0.07);
\node[tlab,below=2pt] at (2.900,3.020) {$9$};
\draw[ax] (3.200,3.050) -- ++(0,-0.07);
\node[tlab,below=2pt] at (3.200,3.020) {$10$};
\draw[ax] (0.000,3.050) -- ++(-0.07,0);
\node[tlab,left=2pt] at (0.000,3.050) {$0$};
\draw[ax] (0.000,4.500) -- ++(-0.07,0);
\node[tlab,left=2pt] at (0.000,4.500) {$0.50$};
\node[font=\scriptsize] at (1.700,4.770) {$\wz = 9.31$};
\node[plab] at (-0.420,5.330) {(a)};
\draw[line width=0.4pt,draw=black!30] (0.230,5.330) -- (3.170,5.330);
\draw[spin] (0.350,5.330) ++(270.0:0.170) -- ++(90.0:0.340);
\draw[spin] (0.650,5.330) ++(270.0:0.170) -- ++(90.0:0.340);
\draw[spin] (0.950,5.330) ++(270.0:0.170) -- ++(90.0:0.340);
\draw[spin] (1.250,5.330) ++(270.0:0.170) -- ++(90.0:0.340);
\draw[spin] (1.550,5.330) ++(270.0:0.170) -- ++(90.0:0.340);
\draw[spin] (1.850,5.330) ++(270.0:0.170) -- ++(90.0:0.340);
\draw[spin] (2.150,5.330) ++(270.0:0.170) -- ++(90.0:0.340);
\draw[spin] (2.450,5.330) ++(270.0:0.170) -- ++(90.0:0.340);
\draw[spin] (2.750,5.330) ++(270.0:0.170) -- ++(90.0:0.340);
\draw[spin] (3.050,5.330) ++(270.0:0.170) -- ++(90.0:0.340);
\draw[ax] (4.750,3.050) -- (8.160,3.050);
\draw[ax] (4.750,3.050) -- (4.750,4.950);
\draw[bar] (4.820,3.050) rectangle (5.080,4.500);
\draw[bar] (5.120,3.050) rectangle (5.380,4.091);
\draw[bar] (5.420,3.050) rectangle (5.680,3.387);
\draw[bar] (5.720,3.050) rectangle (5.980,3.114);
\draw[bar] (6.020,3.050) rectangle (6.280,3.059);
\draw[bar] (6.620,3.050) rectangle (6.880,3.059);
\draw[bar] (6.920,3.050) rectangle (7.180,3.114);
\draw[bar] (7.220,3.050) rectangle (7.480,3.387);
\draw[bar] (7.520,3.050) rectangle (7.780,4.091);
\draw[bar] (7.820,3.050) rectangle (8.080,4.500);
\draw[ax] (4.950,3.050) -- ++(0,-0.07);
\node[tlab,below=2pt] at (4.950,3.020) {$0$};
\draw[ax] (5.250,3.050) -- ++(0,-0.07);
\node[tlab,below=2pt] at (5.250,3.020) {$1$};
\draw[ax] (5.550,3.050) -- ++(0,-0.07);
\node[tlab,below=2pt] at (5.550,3.020) {$2$};
\draw[ax] (5.850,3.050) -- ++(0,-0.07);
\node[tlab,below=2pt] at (5.850,3.020) {$3$};
\draw[ax] (6.150,3.050) -- ++(0,-0.07);
\node[tlab,below=2pt] at (6.150,3.020) {$4$};
\draw[ax] (6.450,3.050) -- ++(0,-0.07);
\node[tlab,below=2pt] at (6.450,3.020) {$5$};
\draw[ax] (6.750,3.050) -- ++(0,-0.07);
\node[tlab,below=2pt] at (6.750,3.020) {$6$};
\draw[ax] (7.050,3.050) -- ++(0,-0.07);
\node[tlab,below=2pt] at (7.050,3.020) {$7$};
\draw[ax] (7.350,3.050) -- ++(0,-0.07);
\node[tlab,below=2pt] at (7.350,3.020) {$8$};
\draw[ax] (7.650,3.050) -- ++(0,-0.07);
\node[tlab,below=2pt] at (7.650,3.020) {$9$};
\draw[ax] (7.950,3.050) -- ++(0,-0.07);
\node[tlab,below=2pt] at (7.950,3.020) {$10$};
\draw[ax] (4.750,3.050) -- ++(-0.07,0);
\node[tlab,left=2pt] at (4.750,3.050) {$0$};
\draw[ax] (4.750,4.500) -- ++(-0.07,0);
\node[tlab,left=2pt] at (4.750,4.500) {$0.25$};
\node[font=\scriptsize] at (6.450,4.770) {$\wz = 8.87$};
\node[plab] at (4.330,5.330) {(b)};
\draw[line width=0.4pt,draw=black!30] (4.980,5.330) -- (7.920,5.330);
\draw[spin] (5.100,5.330) ++(240.0:0.170) -- ++(60.0:0.340);
\draw[spin] (5.400,5.330) ++(240.0:0.170) -- ++(60.0:0.340);
\draw[spin] (5.700,5.330) ++(240.0:0.170) -- ++(60.0:0.340);
\draw[spin] (6.000,5.330) ++(240.0:0.170) -- ++(60.0:0.340);
\draw[spin] (6.300,5.330) ++(240.0:0.170) -- ++(60.0:0.340);
\draw[spin] (6.600,5.330) ++(240.0:0.170) -- ++(60.0:0.340);
\draw[spin] (6.900,5.330) ++(240.0:0.170) -- ++(60.0:0.340);
\draw[spin] (7.200,5.330) ++(240.0:0.170) -- ++(60.0:0.340);
\draw[spin] (7.500,5.330) ++(240.0:0.170) -- ++(60.0:0.340);
\draw[spin] (7.800,5.330) ++(240.0:0.170) -- ++(60.0:0.340);
\draw[ax] (9.500,3.050) -- (12.910,3.050);
\draw[ax] (9.500,3.050) -- (9.500,4.950);
\draw[bar] (9.570,3.050) rectangle (9.830,3.057);
\draw[bar] (10.170,3.050) rectangle (10.430,3.361);
\draw[bar] (10.770,3.050) rectangle (11.030,4.500);
\draw[bar] (11.370,3.050) rectangle (11.630,4.500);
\draw[bar] (11.970,3.050) rectangle (12.230,3.361);
\draw[bar] (12.570,3.050) rectangle (12.830,3.057);
\draw[ax] (9.700,3.050) -- ++(0,-0.07);
\node[tlab,below=2pt] at (9.700,3.020) {$0$};
\draw[ax] (10.000,3.050) -- ++(0,-0.07);
\node[tlab,below=2pt] at (10.000,3.020) {$1$};
\draw[ax] (10.300,3.050) -- ++(0,-0.07);
\node[tlab,below=2pt] at (10.300,3.020) {$2$};
\draw[ax] (10.600,3.050) -- ++(0,-0.07);
\node[tlab,below=2pt] at (10.600,3.020) {$3$};
\draw[ax] (10.900,3.050) -- ++(0,-0.07);
\node[tlab,below=2pt] at (10.900,3.020) {$4$};
\draw[ax] (11.200,3.050) -- ++(0,-0.07);
\node[tlab,below=2pt] at (11.200,3.020) {$5$};
\draw[ax] (11.500,3.050) -- ++(0,-0.07);
\node[tlab,below=2pt] at (11.500,3.020) {$6$};
\draw[ax] (11.800,3.050) -- ++(0,-0.07);
\node[tlab,below=2pt] at (11.800,3.020) {$7$};
\draw[ax] (12.100,3.050) -- ++(0,-0.07);
\node[tlab,below=2pt] at (12.100,3.020) {$8$};
\draw[ax] (12.400,3.050) -- ++(0,-0.07);
\node[tlab,below=2pt] at (12.400,3.020) {$9$};
\draw[ax] (12.700,3.050) -- ++(0,-0.07);
\node[tlab,below=2pt] at (12.700,3.020) {$10$};
\draw[ax] (9.500,3.050) -- ++(-0.07,0);
\node[tlab,left=2pt] at (9.500,3.050) {$0$};
\draw[ax] (9.500,4.500) -- ++(-0.07,0);
\node[tlab,left=2pt] at (9.500,4.500) {$0.41$};
\node[font=\scriptsize] at (11.200,4.770) {$\wz = 6.20$};
\node[plab] at (9.080,5.330) {(c)};
\draw[line width=0.4pt,draw=black!30] (9.730,5.330) -- (12.670,5.330);
\draw[spin] (9.850,5.330) ++(180.0:0.170) -- ++(0.0:0.340);
\draw[spin] (10.150,5.330) ++(180.0:0.170) -- ++(0.0:0.340);
\draw[spin] (10.450,5.330) ++(180.0:0.170) -- ++(0.0:0.340);
\draw[spin] (10.750,5.330) ++(180.0:0.170) -- ++(0.0:0.340);
\draw[spin] (11.050,5.330) ++(180.0:0.170) -- ++(0.0:0.340);
\draw[spin] (11.350,5.330) ++(180.0:0.170) -- ++(0.0:0.340);
\draw[spin] (11.650,5.330) ++(180.0:0.170) -- ++(0.0:0.340);
\draw[spin] (11.950,5.330) ++(180.0:0.170) -- ++(0.0:0.340);
\draw[spin] (12.250,5.330) ++(180.0:0.170) -- ++(0.0:0.340);
\draw[spin] (12.550,5.330) ++(180.0:0.170) -- ++(0.0:0.340);
\draw[ax] (0.000,0.000) -- (3.410,0.000);
\draw[ax] (0.000,0.000) -- (0.000,1.900);
\draw[bar] (0.370,0.000) rectangle (0.630,0.012);
\draw[bar] (0.670,0.000) rectangle (0.930,0.418);
\draw[bar] (0.970,0.000) rectangle (1.230,1.278);
\draw[bar] (1.270,0.000) rectangle (1.530,1.450);
\draw[bar] (1.570,0.000) rectangle (1.830,1.146);
\draw[bar] (1.870,0.000) rectangle (2.130,1.450);
\draw[bar] (2.170,0.000) rectangle (2.430,1.278);
\draw[bar] (2.470,0.000) rectangle (2.730,0.418);
\draw[bar] (2.770,0.000) rectangle (3.030,0.012);
\draw[ax] (0.200,0.000) -- ++(0,-0.07);
\node[tlab,below=2pt] at (0.200,-0.030) {$0$};
\draw[ax] (0.500,0.000) -- ++(0,-0.07);
\node[tlab,below=2pt] at (0.500,-0.030) {$1$};
\draw[ax] (0.800,0.000) -- ++(0,-0.07);
\node[tlab,below=2pt] at (0.800,-0.030) {$2$};
\draw[ax] (1.100,0.000) -- ++(0,-0.07);
\node[tlab,below=2pt] at (1.100,-0.030) {$3$};
\draw[ax] (1.400,0.000) -- ++(0,-0.07);
\node[tlab,below=2pt] at (1.400,-0.030) {$4$};
\draw[ax] (1.700,0.000) -- ++(0,-0.07);
\node[tlab,below=2pt] at (1.700,-0.030) {$5$};
\draw[ax] (2.000,0.000) -- ++(0,-0.07);
\node[tlab,below=2pt] at (2.000,-0.030) {$6$};
\draw[ax] (2.300,0.000) -- ++(0,-0.07);
\node[tlab,below=2pt] at (2.300,-0.030) {$7$};
\draw[ax] (2.600,0.000) -- ++(0,-0.07);
\node[tlab,below=2pt] at (2.600,-0.030) {$8$};
\draw[ax] (2.900,0.000) -- ++(0,-0.07);
\node[tlab,below=2pt] at (2.900,-0.030) {$9$};
\draw[ax] (3.200,0.000) -- ++(0,-0.07);
\node[tlab,below=2pt] at (3.200,-0.030) {$10$};
\draw[ax] (0.000,0.000) -- ++(-0.07,0);
\node[tlab,left=2pt] at (0.000,0.000) {$0$};
\draw[ax] (0.000,1.450) -- ++(-0.07,0);
\node[tlab,left=2pt] at (0.000,1.450) {$0.19$};
\node[font=\scriptsize] at (1.700,1.720) {$\wz = 6.18$};
\node[plab] at (-0.420,2.280) {(d)};
\node[tlab,below=2pt] at (1.700,-0.340) {$|\vec r|_1$};
\draw[line width=0.4pt,draw=black!30] (0.230,2.280) -- (3.170,2.280);
\draw[spin] (0.350,2.280) ++(45.0:0.170) -- ++(-135.0:0.340);
\draw[spin] (0.650,2.280) ++(143.0:0.170) -- ++(-37.0:0.340);
\draw[spin] (0.950,2.280) ++(99.2:0.170) -- ++(-80.8:0.340);
\draw[spin] (1.250,2.280) ++(261.1:0.170) -- ++(81.1:0.340);
\draw[spin] (1.550,2.280) ++(288.1:0.170) -- ++(108.1:0.340);
\draw[spin] (1.850,2.280) ++(134.5:0.170) -- ++(-45.5:0.340);
\draw[spin] (2.150,2.280) ++(181.9:0.170) -- ++(1.9:0.340);
\draw[spin] (2.450,2.280) ++(115.6:0.170) -- ++(-64.4:0.340);
\draw[spin] (2.750,2.280) ++(106.9:0.170) -- ++(-73.1:0.340);
\draw[spin] (3.050,2.280) ++(348.5:0.170) -- ++(168.5:0.340);
\draw[ax] (4.750,0.000) -- (8.160,0.000);
\draw[ax] (4.750,0.000) -- (4.750,1.900);
\draw[bar] (5.120,0.000) rectangle (5.380,0.002);
\draw[bar] (5.420,0.000) rectangle (5.680,0.047);
\draw[bar] (5.720,0.000) rectangle (5.980,0.440);
\draw[bar] (6.020,0.000) rectangle (6.280,1.450);
\draw[bar] (6.320,0.000) rectangle (6.580,0.909);
\draw[bar] (6.620,0.000) rectangle (6.880,1.450);
\draw[bar] (6.920,0.000) rectangle (7.180,0.440);
\draw[bar] (7.220,0.000) rectangle (7.480,0.047);
\draw[bar] (7.520,0.000) rectangle (7.780,0.002);
\draw[ax] (4.950,0.000) -- ++(0,-0.07);
\node[tlab,below=2pt] at (4.950,-0.030) {$0$};
\draw[ax] (5.250,0.000) -- ++(0,-0.07);
\node[tlab,below=2pt] at (5.250,-0.030) {$1$};
\draw[ax] (5.550,0.000) -- ++(0,-0.07);
\node[tlab,below=2pt] at (5.550,-0.030) {$2$};
\draw[ax] (5.850,0.000) -- ++(0,-0.07);
\node[tlab,below=2pt] at (5.850,-0.030) {$3$};
\draw[ax] (6.150,0.000) -- ++(0,-0.07);
\node[tlab,below=2pt] at (6.150,-0.030) {$4$};
\draw[ax] (6.450,0.000) -- ++(0,-0.07);
\node[tlab,below=2pt] at (6.450,-0.030) {$5$};
\draw[ax] (6.750,0.000) -- ++(0,-0.07);
\node[tlab,below=2pt] at (6.750,-0.030) {$6$};
\draw[ax] (7.050,0.000) -- ++(0,-0.07);
\node[tlab,below=2pt] at (7.050,-0.030) {$7$};
\draw[ax] (7.350,0.000) -- ++(0,-0.07);
\node[tlab,below=2pt] at (7.350,-0.030) {$8$};
\draw[ax] (7.650,0.000) -- ++(0,-0.07);
\node[tlab,below=2pt] at (7.650,-0.030) {$9$};
\draw[ax] (7.950,0.000) -- ++(0,-0.07);
\node[tlab,below=2pt] at (7.950,-0.030) {$10$};
\draw[ax] (4.750,0.000) -- ++(-0.07,0);
\node[tlab,left=2pt] at (4.750,0.000) {$0$};
\draw[ax] (4.750,1.450) -- ++(-0.07,0);
\node[tlab,left=2pt] at (4.750,1.450) {$0.30$};
\node[font=\scriptsize] at (6.450,1.720) {$\wz = 5.71$};
\node[plab] at (4.330,2.280) {(e)};
\node[tlab,below=2pt] at (6.450,-0.340) {$|\vec r|_1$};
\draw[line width=0.4pt,draw=black!30] (4.980,2.280) -- (7.920,2.280);
\draw[spin] (5.100,2.280) ++(292.1:0.170) -- ++(112.1:0.340);
\draw[spin] (5.400,2.280) ++(119.0:0.170) -- ++(-61.0:0.340);
\draw[spin] (5.700,2.280) ++(310.3:0.170) -- ++(130.3:0.340);
\draw[spin] (6.000,2.280) ++(51.7:0.170) -- ++(-128.3:0.340);
\draw[spin] (6.300,2.280) ++(269.0:0.170) -- ++(89.0:0.340);
\draw[spin] (6.600,2.280) ++(80.8:0.170) -- ++(-99.2:0.340);
\draw[spin] (6.900,2.280) ++(256.0:0.170) -- ++(76.0:0.340);
\draw[spin] (7.200,2.280) ++(131.5:0.170) -- ++(-48.5:0.340);
\draw[spin] (7.500,2.280) ++(306.1:0.170) -- ++(126.1:0.340);
\draw[spin] (7.800,2.280) ++(296.6:0.170) -- ++(116.6:0.340);
\draw[ax] (9.500,0.000) -- (12.910,0.000);
\draw[ax] (9.500,0.000) -- (9.500,1.900);
\draw[bar] (11.070,0.000) rectangle (11.330,1.450);
\draw[ax] (9.700,0.000) -- ++(0,-0.07);
\node[tlab,below=2pt] at (9.700,-0.030) {$0$};
\draw[ax] (10.000,0.000) -- ++(0,-0.07);
\node[tlab,below=2pt] at (10.000,-0.030) {$1$};
\draw[ax] (10.300,0.000) -- ++(0,-0.07);
\node[tlab,below=2pt] at (10.300,-0.030) {$2$};
\draw[ax] (10.600,0.000) -- ++(0,-0.07);
\node[tlab,below=2pt] at (10.600,-0.030) {$3$};
\draw[ax] (10.900,0.000) -- ++(0,-0.07);
\node[tlab,below=2pt] at (10.900,-0.030) {$4$};
\draw[ax] (11.200,0.000) -- ++(0,-0.07);
\node[tlab,below=2pt] at (11.200,-0.030) {$5$};
\draw[ax] (11.500,0.000) -- ++(0,-0.07);
\node[tlab,below=2pt] at (11.500,-0.030) {$6$};
\draw[ax] (11.800,0.000) -- ++(0,-0.07);
\node[tlab,below=2pt] at (11.800,-0.030) {$7$};
\draw[ax] (12.100,0.000) -- ++(0,-0.07);
\node[tlab,below=2pt] at (12.100,-0.030) {$8$};
\draw[ax] (12.400,0.000) -- ++(0,-0.07);
\node[tlab,below=2pt] at (12.400,-0.030) {$9$};
\draw[ax] (12.700,0.000) -- ++(0,-0.07);
\node[tlab,below=2pt] at (12.700,-0.030) {$10$};
\draw[ax] (9.500,0.000) -- ++(-0.07,0);
\node[tlab,left=2pt] at (9.500,0.000) {$0$};
\draw[ax] (9.500,1.450) -- ++(-0.07,0);
\node[tlab,left=2pt] at (9.500,1.450) {$1.00$};
\node[font=\scriptsize] at (11.200,1.720) {$\wz = 5\;(=n/2)$};
\node[plab] at (9.080,2.280) {(f)};
\node[tlab,below=2pt] at (11.200,-0.340) {$|\vec r|_1$};
\draw[line width=0.4pt,draw=black!30] (9.730,2.280) -- (12.670,2.280);
\draw[spin] (9.850,2.280) ++(270.0:0.170) -- ++(90.0:0.340);
\draw[spin] (10.150,2.280) ++(90.0:0.170) -- ++(-90.0:0.340);
\draw[spin] (10.450,2.280) ++(270.0:0.170) -- ++(90.0:0.340);
\draw[spin] (10.750,2.280) ++(90.0:0.170) -- ++(-90.0:0.340);
\draw[spin] (11.050,2.280) ++(270.0:0.170) -- ++(90.0:0.340);
\draw[spin] (11.350,2.280) ++(90.0:0.170) -- ++(-90.0:0.340);
\draw[spin] (11.650,2.280) ++(270.0:0.170) -- ++(90.0:0.340);
\draw[spin] (11.950,2.280) ++(90.0:0.170) -- ++(-90.0:0.340);
\draw[spin] (12.250,2.280) ++(270.0:0.170) -- ++(90.0:0.340);
\draw[spin] (12.550,2.280) ++(90.0:0.170) -- ++(-90.0:0.340);
\node[plab,rotate=90] at (-1.15,4.000) {$p(\vec r)$};
\node[plab,rotate=90] at (-1.15,0.950) {$p(\vec r)$};
\end{tikzpicture}}
\caption{\textbf{Walsh spectrum of the GHZ state.} Walsh spectrum $p(\vec r)$ after a single-site rotation for the $n=10$ GHZ state. Local unitary defined by the axes $\hat{\vec n}_j$, depicted above each
panel. Panels are ordered by decreasing
$\wz$: (a) all axes $+Z$, (b) a uniform tilt, (c) the $X$ basis in which $\ket{\psi_{\mathrm{GHZ}}}= \frac{1}{\sqrt{2}}(\ket{0^n}+\ket{1^n})$,
(d),(e) random axes, (f) half the axes flipped to $-Z$. No local basis can reach below $\w[\ket{\psi_{\mathrm{GHZ}}}] = n/2$, which
(f) attains.}
\label{fig:ghz-bases}
\end{figure}

\begin{fact}[GHZ states] Fix $n\geq 2$. Let $\ket{\psi_{\mathrm{GHZ}}} = \frac{\ket{0^n}+\ket{1^n}}{\sqrt{2}}$ be the GHZ (or ``cat") state on $n$ qubits. Then
\begin{equation}  
\label{eq:GHZWalshness}
\wz[\ket{\psi_{\mathrm{GHZ}}}] = \frac{n}{\alpha}\log\frac{1+e^\alpha}{2} + o(1) 
\end{equation}
where the $o(1)$ correction decays exponentially in $n$. For $n$ even, we also have
\begin{equation}   
\label{eq:GHZconstant}
\w[\ket{\psi_{\mathrm{GHZ}}}] = \frac{n}{2}.
\end{equation}
\end{fact}
\begin{proof}
The computational-basis wavefunction is $\psi(\vec s) = \frac{1}{\sqrt{2}}(\delta_{\vec s,0^n} + \delta_{\vec s,1^n})$ and the Walsh transform thereof is $\widehat\psi(\vec r) = \frac{2^{-n}}{\sqrt{2}}\left[1+(-1)^{|\vec r|_1}\right]$. It follows straightforwardly that the Walsh spectrum is given by 
\begin{equation}
    p(\vec r) = \begin{cases}
        2^{1-n} & |\vec r|_1 \,\,\,\text{even}\\
        0 & \text{else}
    \end{cases}.
\end{equation}
In other words, the Walsh spectrum is uniform over even-parity bitstrings. Using the identity $    (1-x)^n +(1+x)^n = 2\sum_{\substack{k=0\\k \,\text{even}}}^n \binom{n}{k}x^k
$, we find  
\begin{equation}
    e^{\alpha \wz[\ket{\psi_{\mathrm{GHZ}}}]} = 2^{1-n}\sum_{\substack{\vec r\\|\vec r|_1\,\text{even}}} e^{\alpha |\vec r|_1}  = 2^{-n}[(1+e^\alpha)^n+(1-e^\alpha)^n]
\end{equation}
from which it follows 
\begin{equation}
    \begin{aligned}
\wz[\ket{\psi_{\mathrm{GHZ}}}] &= \frac{1}{\alpha} \log \left[2^{-n}[(1+e^\alpha)^n+(1-e^\alpha)^n]\right]\\
        &= \frac{n}{\alpha} \log \frac{1+e^\alpha}{2} + \frac{1}{\alpha}\log \left[1+ \left(-\tanh\frac{\alpha}{2}\right)^n\right]\\
        &= \frac{n}{\alpha} \log \frac{1+e^\alpha}{2} + o(1)
    \end{aligned}
\end{equation}
where in the last step we used $\left|\tanh\frac{\alpha}{2}\right|<1$. This proves Eq.~\eqref{eq:GHZWalshness}. Next, we aim to prove Eq.~\eqref{eq:GHZconstant}. As in Eq.~\eqref{eq:projectors}, let $U_j^\dagger X_j U_j = \hat{\vec n}_j \cdot \vec \sigma_j$ and define projectors
$P_j^{\pm} \coloneqq \frac12 (I\pm \hat{\vec n}_j \cdot \vec \sigma_j)$. Write $U = \bigotimes_{j=1}^n U_j$ and $K = \sum_{j=1}^n P_j^-$, so that $\wz[U\ket{\psi_{\mathrm{GHZ}}}] = \frac{1}{\alpha}\log\mathbb{E}[e^{\alpha K}]$, noting that $K$ implicitly depends on $U$ via the $P_j^-$. Independently of the unitary $U$, however, we claim that 
\begin{equation}
    \mathbb{E}[K] = n/2. 
\end{equation}
To see this, we note that $\mathbb{E}[K] = \sum_{j=1}^n \mathbb{E}[P_j^-] = \sum_{j=1}^n \mathbb{P}(P_j^- = 1) =  \sum_{j=1}^n\Tr[P_j^- \rho_j^{\mathrm{GHZ}}]$, where $\rho_j^{\mathrm{GHZ}}$ is the one-site reduced density matrix (RDM) of $U\ket{\psi_{\mathrm{GHZ}}}$. It is simple to see that $\rho_j^{\mathrm{GHZ}} = \mathrm{Tr}_{\bar j} \ket{\psi_{\mathrm{GHZ}}}\!\bra{\psi_{\mathrm{GHZ}}} = \frac12 (\ket{0}\!\bra{0} + \ket{1}\!\bra{1}) = I/2$, and hence $\mathrm{Tr}[P_j^- \rho_j^{\mathrm{GHZ}}] = \frac12$, from which $\mathbb{E}[K] = n/2$ follows. \par 

Next, since $x\mapsto e^{\alpha x}$ is convex, Jensen's inequality implies 
\begin{equation}
    \mathbb{E}[e^{\alpha K}]\geq e^{\alpha \mathbb{E}[K]} = e^{\alpha n/2}. 
\end{equation}
Therefore, in every local basis, $\wz = \frac{1}{\alpha}\log\mathbb{E}[e^{\alpha K}] \geq n/2$. Hence $\w[\ket{\psi_{\mathrm{GHZ}}}] \geq n/2$. We now show that this lower bound is tight, i.e. it can be achieved in some local basis. \par 

A Walshness minimizing basis for the GHZ state is the following: \textit{choose $n/2$ local axes to be $+\hat{\vec z}$, and choose the other $n/2$ local axes to be $-\hat{\vec z}$.} In other words, choose any subset $A$ of $n/2$ qubits, define its complement to be $B$, and take the unitary 
\begin{equation}
    U = U_A U_B,\qquad U_A = \prod_{j\in A} H_j X_j,\quad U_B = \prod_{j\in B} H_j
\end{equation}
where $H_j$ is the Hadamard. Note that $(HX)^\dagger X (HX) = -Z$ and $H^\dagger X H = Z$. Using $H\ket{0} = \ket{+}$ and $H\ket{1} = \ket{-}$, we have $U\ket{\psi_{\mathrm{GHZ}}} = \frac{1}{\sqrt{2}}\left(\ket{-}_A\ket{+}_B + \ket{+}_A\ket{-}_B\right)$ and, moreover, $H^{\otimes n} U\ket{\psi_{\mathrm{GHZ}}}  = \frac{1}{\sqrt{2}}\left(\ket{1_A0_B}+\ket{0_A1_B}\right)$ in obvious notation. The Walsh spectrum in this basis is thus 
\begin{equation}
    p(\vec r) = |\langle \vec r |H^{\otimes n}U\ket{\psi_{\mathrm{GHZ}}} |^2 =  \frac12 \delta_{\vec r,1_A0_B} + \frac12\delta_{\vec r,0_A1_B}. 
\end{equation}
Note that both $1_A0_B$ and $0_A1_B$ have Hamming weight $n/2$. Therefore $W_\alpha^Z[U\ket{\psi_{\mathrm{GHZ}}}] = \frac{1}{\alpha}\log e^{\alpha n/2} = n/2$. Combined with $\w[\ket{\psi_{\mathrm{GHZ}}}] \geq n/2$, it follows that
\begin{equation}
\w[\ket{\psi_{\mathrm{GHZ}}}] = n/2,
\end{equation}
completing the proof. 
\end{proof}
For odd $n$, a similar result holds with a small additive correction. Though straightforward, we omit it from here.

\begin{fact}[Walshness and entanglement]
    Let $[n]=L\sqcup R$ be a bipartition of the qubits, and let $\ell = \min\{|L|,|R|\}$. Let $S_{L:R}[\ket\psi]$ denote the bipartite entanglement entropy of $\ket\psi$ between $L$ and $R$. Then for every $\alpha>0$, we have 
\begin{equation}
\label{eq:SLRequation}
    S_{L:R}[\ket\psi] \leq \min\left( \ell\log 2, \frac{n}{2} h_2\left[\min\left\{\frac1n \w[\ket\psi],\frac12\right\}\right]\right),
\end{equation}
where $h_2(x) = -x\log x-(1-x)\log(1-x)$ is the binary entropy. In particular, $\w[\ket\psi] = o(n)$ implies $S_{L:R}[\ket\psi] = o(n)$ for any $L:R$. Thus, subextensive Walshness rules out volume-law entanglement.
\end{fact}
\begin{proof}
Without loss of generality, we may assume $\ket\psi$ is normalized. Moreover, suppose $\ket\psi$ is already in the local basis which minimizes fixed-basis Walshness (or else replace $\ket\psi\mapsto U\ket\psi$ with the appropriate $U$, which does not change $S_{L:R}$). Let $p(\vec r)$ denote the Walsh spectrum of $\ket\psi$, and further write $p(\vec r_L,\vec r_R) = p(\vec r)$ for a decomposition $\vec r = (\vec r_L,\vec r_R)$ of the bitstring $\vec r$ into left and right components. We also define the left and right marginals of the Walsh spectrum as 
\begin{equation}
    q_L(\vec r_L) = \sum_{\vec r_R}p(\vec r_L,\vec r_R),\quad q_R(\vec r_R) = \sum_{\vec r_L} p(\vec r_L,\vec r_R).
\end{equation}
Let us also introduce the concept of the reduced density matrix in the Walsh-Hadamard basis. Since $\widehat\psi(\vec r) \sim \langle \vec r |H^{\otimes n}|\psi\rangle$, where $H$ is the Hadamard, it is convenient to define the density matrix in the Walsh-Hadamard basis as $\rho = H^{\otimes n}\ket\psi\!\bra\psi H^{\otimes n}$ and to note that the diagonals are precisely the Walsh spectrum, assuming $\ket\psi$ is normalized: $p(\vec r) = \langle \vec r|\rho|\vec r\rangle$. We then define the reduced density matrices in the Walsh-Hadamard basis as
\begin{equation}
    \rho_L = \Tr_R \rho,\qquad \rho_R = \Tr_L\rho. 
\end{equation}
The diagonals thereof satisfy $\langle \vec r_L|\rho_L|\vec r_L\rangle = q_L(\vec r_L)$ and similarly for $R$. Next, we use the fact that dephasing $\rho_L$ (setting all off-diagonals to zero) cannot reduce its von Neumann entropy. Hence
\begin{equation}
    S_{L:R}[\ket\psi] = S(\rho_L) \leq H(q_L)
\end{equation}
where $H$ is the Shannon entropy of a classical distribution. Similarly, $S(\rho_R) \leq H(q_R)$. Define expectation values of Hamming weights $\mu_L = \mathbb{E}|\vec r_L|_1, \mu_R = \mathbb{E}|\vec r_R|_1,$ and $\mu=\mu_L+\mu_R = \mathbb{E}|\vec r|_1$. By the subadditivity of Shannon entropy, we have $H(q_L) \leq \sum_{i\in L} H(r_i)$, where $H(r_i)$ is the entropy of the binary variable $r_i \in \{0,1\}$ distributed according to the marginalization of $p(\vec r)$. In other words, writing $p_i = \mathbb{P}[r_i = 1]$, we have $H(r_i) = h_2(p_i)$ with $h_2$ the binary entropy. Next, we use the concavity of entropy to write
\begin{equation}
    H(q_L) \leq \sum_{i\in L} h_2(p_i) \leq |L|h_2\left(\frac{\mu_L}{|L|}\right)
\end{equation}
and similarly $H(q_R) \leq |R|h_2\left(\frac{\mu_R}{|R|}\right)$. Since both expressions bound the same entanglement entropy, adding them and using concavity a second time gives
\begin{equation}
    2S_{L:R}[\ket\psi] \leq |L| h_2\left(\frac{\mu_L}{|L|}\right) + |R|h_2\left(\frac{\mu_R}{|R|}\right)\leq n h_2\left(\frac{\mu}{n}\right).
\end{equation}
Therefore $S_{L:R}[\ket\psi]  \leq \frac{n}{2} h_2\left(\frac{\mu}{n}\right)$. It remains to relate the mean Walsh degree $\mu$ to Walshness. By Jensen’s inequality, we have 
\begin{equation}
    e^{\alpha \mu} = e^{\alpha \mathbb{E}[|\vec r|_1]} \leq \mathbb{E}\left[ e^{\alpha |\vec r|_1}\right]  = e^{\alpha \w[\ket\psi]},
\end{equation}
and hence $\mu \leq \w[\ket\psi]$. If $\w/n\leq1/2$, then $\mu/n\leq\w/n\leq1/2$, and $h_2$ is increasing on this interval, so 
\begin{equation}
\label{eq:SLRketpsileqfracn2}
    S_{L:R}[\ket\psi] \leq \frac{n}{2}h_2\left(\frac{\w[\ket\psi]}{n}\right).
\end{equation}
If instead $\wdd>1/2$, we may use $h_2(x) \leq \log 2$ and combine with the general Hilbert space bound $ S_{L:R}[\ket\psi]\leq \ell \log 2$. Together, this proves Eq.~\eqref{eq:SLRequation}. For the final claim, if $\w=o(n)$, then $h_2\left(\frac{\w}{n}\right)=o(1),$ and hence $S_{L:R}=o(n)$.
\end{proof}
It is useful to give some intuition for the magnitude of the bound in
Eq.~\eqref{eq:SLRketpsileqfracn2}. Consider a balanced bipartition, $\ell=n/2$, and define the
Walshness density $\mathrm w_\alpha=\mathrm W_\alpha/n$. For $\mathrm w_\alpha\le 1/2$,
Eq.~\eqref{eq:SLRketpsileqfracn2} gives
\begin{equation}
\frac{S_{L:R}}{(n/2)\log 2}
\le
\frac{h_2(\mathrm w_\alpha)}{\log 2}.
\end{equation}
Some representative values are shown in Table~\ref{tab:walshness-entanglement}.

\begin{table}[t]
\centering
\begin{tabular}{c c}
\hline\hline
$\wdd=\w/n$
&
Bound on $S_{L:R}/S_{\max}$
\\
\hline
$0.01$ & $8.1\%$ \\
$0.05$ & $28.6\%$ \\
$0.10$ & $46.9\%$ \\
$0.25$ & $81.1\%$ \\
$0.40$ & $97.1\%$ \\
$0.50$ & $100\%$ \\
\hline\hline
\end{tabular}
\caption{
Entanglement bound for a balanced bipartition, where
$S_{\max}=(n/2)\log 2$ is the maximum bipartite entanglement entropy.
The second column is $h_2(\mathrm{W}_\alpha/n)/\log 2$.
}
\label{tab:walshness-entanglement}
\end{table}

For small Walshness density,
%\jessica{For consistent notation, should we have $\mathrm{w}\rightarrow \wdd$?}
\begin{equation}
h_2(\mathrm w)
=
\mathrm w\log\frac{1}{\mathrm w}+\mathrm w+O(\mathrm w^2),
\end{equation}
and therefore Eq.~\eqref{eq:SLRketpsileqfracn2} gives
\begin{equation}
S_{L:R}
\le
\frac{\w}{2}\log\frac{e n}{\w}
+
O\!\left(\frac{\w^2}{n}\right),
\qquad
\w\ll n.
\end{equation}
Thus, for example, $\w=O(1)$ implies
$S_{L:R}=O(\log n)$ across any bipartition.

\begin{fact}[Haar-random states]
For fixed $\alpha>0$ and a Haar-random state $\ket{\psi_{\mathrm{Haar}}}\in\mathbb C^{2^n}$, the fixed-basis
Walshness takes the form
\begin{equation} \wz[\ket{\psi_{\mathrm{Haar}}}]
    =
    \frac{n}{\alpha}\log\frac{1+e^\alpha}{2}
    + \varepsilon_{n,\alpha},
\end{equation}
where the random correction satisfies
\begin{equation}
    \mathbb E\!\left[\varepsilon_{n,\alpha}^2\right]
    \leq
    4\left(n^2+\frac{1}{\alpha^2}\right)
    \left(\frac{1+e^{2\alpha}}{(1+e^\alpha)^2}\right)^n.
    \label{eq:haar-walshness-error}
\end{equation}
In particular, both its root-mean-square size and its variance
decay exponentially in $n$.
\end{fact}

\begin{proof}
Assume $\ket{\psi_{\mathrm{Haar}}}$ is normalized. Note that since $H^{\otimes n}$ is unitary,
$H^{\otimes n}\ket{\psi_{\mathrm{Haar}}}$ is also Haar random.
The Walsh spectrum
$p(\vec r)=|\bra{\vec r}H^{\otimes n}\ket{\psi_{\mathrm{Haar}}}|^2$
therefore has the standard Haar moments $\mathbb E[p(\vec r)]=2^{-n}, \mathbb E[p(\vec r)p(\vec s)]
    =
    \frac{1+\delta_{\vec r,\vec s}}{2^n(2^n+1)}.$ Next, we define $    Z_\alpha
    \coloneqq
    e^{\alpha\wz}
    =
    \sum_{\vec r}e^{\alpha|\vec r|_1}p(\vec r)$ which has mean
\begin{equation}
    \mathbb E[Z_\alpha]
    =
    2^{-n}\sum_{\vec r}e^{\alpha|\vec r|_1}
    =
    2^{-n}\sum_{k=0}^n\binom nk e^{\alpha k}
    =
    \left(\frac{1+e^\alpha}{2}\right)^n,
\end{equation}
and second moment given by
\begin{equation}
\begin{aligned}
    \mathbb E[Z_\alpha^2]
    &=
    \sum_{\vec r,\vec s}
    e^{\alpha(|\vec r|_1+|\vec s|_1)}
    \frac{1+\delta_{\vec r,\vec s}}{2^n(2^n+1)}
    \\
    &=
    \frac{
        \left(\sum_{\vec r}e^{\alpha|\vec r|_1}\right)^2
        +\sum_{\vec r}e^{2\alpha|\vec r|_1}
    }{2^n(2^n+1)}
    \\
    &=
    \frac{(1+e^\alpha)^{2n}+(1+e^{2\alpha})^n}
         {2^n(2^n+1)}.
\end{aligned}
\end{equation}
Consequently,
\begin{equation}
\begin{aligned}
    \frac{\operatorname{Var}(Z_\alpha)}
         {(\mathbb E[Z_\alpha])^2}
    &=
    \frac{\mathbb E[Z_\alpha^2]}
         {(\mathbb E[Z_\alpha])^2}-1
    \\
    &=
    \frac{
        2^n\left(\frac{1+e^{2\alpha}}{(1+e^\alpha)^2}\right)^n-1
    }{2^n+1}
    \\
    &\leq
    \left(\frac{1+e^{2\alpha}}{(1+e^\alpha)^2}\right)^n.
\end{aligned}
\end{equation}
Next, we wish to control the logarithm of $Z_\alpha$. By the expression for
$\mathbb E[Z_\alpha]$, the correction $\varepsilon_{n,\alpha}$ takes the exact form $    \varepsilon_{n,\alpha}
    =
    \frac{1}{\alpha}
    \log\frac{Z_\alpha}{\mathbb E[Z_\alpha]}.
$ Whenever
$\left|Z_\alpha/\mathbb E[Z_\alpha]-1\right|\leq\frac12$,
the inequality $|\log(1+x)|\leq2|x|$ gives $|\varepsilon_{n,\alpha}|
    \leq
    \frac{2}{\alpha}
    \left|\frac{Z_\alpha}{\mathbb E[Z_\alpha]}-1\right|.$ When this is not the case, we may use
$|\varepsilon_{n,\alpha}|\leq n$, since both
$\wz$ and
$\frac n\alpha\log[(1+e^\alpha)/2]$ lie in $[0,n]$.
Splitting the expectation between these two cases and applying
Chebyshev's inequality yields
\begin{equation}
\begin{aligned}
    \mathbb E[\varepsilon_{n,\alpha}^2]
    &\leq
    \frac{4}{\alpha^2}
    \mathbb E\!\left[
        \left(\frac{Z_\alpha}{\mathbb E[Z_\alpha]}-1\right)^2
    \right]
    +
    n^2\,
    \mathbb P\!\left[
        \left|\frac{Z_\alpha}{\mathbb E[Z_\alpha]}-1\right|
        >\frac12
    \right]
    \\
    &\leq
    4\left(\frac{1}{\alpha^2}+n^2\right)
    \frac{\operatorname{Var}(Z_\alpha)}
         {(\mathbb E[Z_\alpha])^2}
    \\
    &\leq
    4\left(\frac{1}{\alpha^2}+n^2\right)
    \left(\frac{1+e^{2\alpha}}{(1+e^\alpha)^2}\right)^n.
\end{aligned}
\end{equation}
This proves Eq.~\eqref{eq:haar-walshness-error}. The exponential
decay follows from $    \frac{1+e^{2\alpha}}{(1+e^\alpha)^2}<1$ and $\operatorname{Var}(\varepsilon_{n,\alpha})
    \leq \mathbb E[\varepsilon_{n,\alpha}^2].$
\end{proof}

We also note the following facts about intrinsic Walshness. Every $n$-qubit pure state satisfies the upper bound $\w[\ket\psi]
    \leq
    \frac{n}{\alpha}\log\frac{1+e^\alpha}{2}.$ For fixed $\alpha>0$, Haar-random states asymptotically saturate
this bound: $\w[\ket{\psi_{\mathrm{Haar}}}]
    =
    \frac{n}{\alpha}\log\frac{1+e^\alpha}{2}
    +\tilde o(1),
$ where the correction is nonpositive and vanishes in probability
as $n\to\infty$. \par

\begin{figure}
    \centering
    \includegraphics[width=0.4\linewidth]{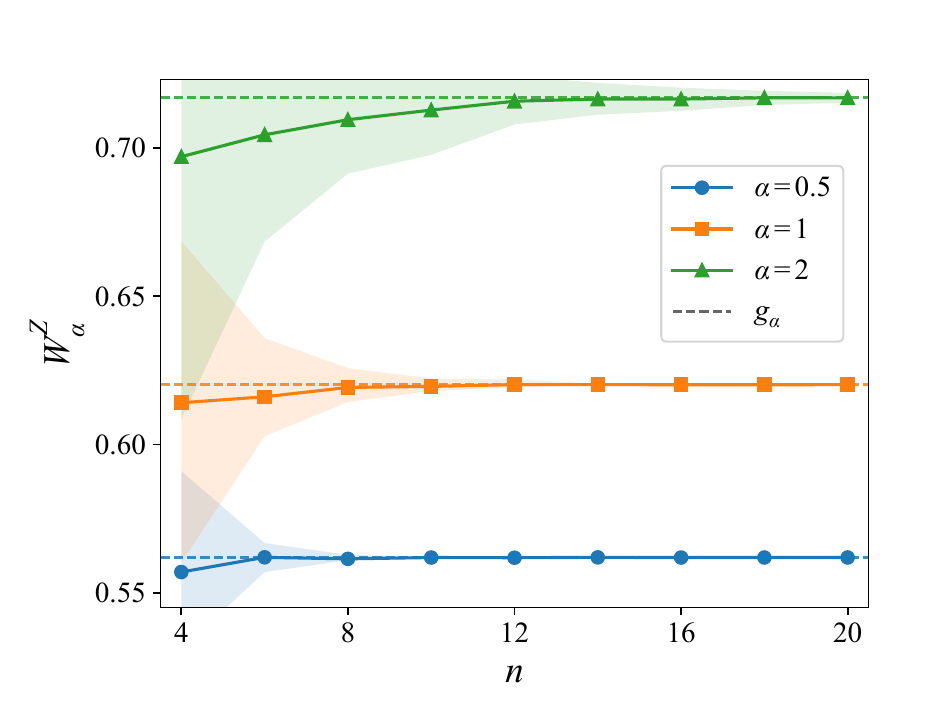}
    \caption{ \textbf{Walshness of random stabilizer states.} We plot the fixed-basis Walshness density $\wzd$ versus system size $n$ for several values of $\alpha$. Markers show the mean over random stabilizer states and shaded regions indicate the sample variation. Dashed lines denote $g_\alpha=\alpha^{-1}\log[(1+e^\alpha)/2]$, the value for a uniform Walsh spectrum.
}
\label{fig:random_stabilizer}
\end{figure}

Next, it is natural to also consider random stabilizer states.
In Fig.~\ref{fig:random_stabilizer}, we plot the fixed-basis Walshness
$\wz/n$ for several values of $\alpha$ for random stabilizer states on $n$ qubits. We find that $\wz/n$ rapidly approaches $g_\alpha\coloneqq\frac{1}{\alpha}\log\frac{1+e^\alpha}{2}$ as in the Haar case. We prove this next. 

\begin{fact}[Random stabilizer states]
For fixed $\alpha>0$ and a state $\ket{\psi_{\mathrm{stab}}}$
chosen uniformly from all pure $n$-qubit stabilizer states,
the fixed-basis Walshness takes the form
\begin{equation}
    \wz[\ket{\psi_{\mathrm{stab}}}]
    =
    \frac{n}{\alpha}\log\frac{1+e^\alpha}{2}
    + \varepsilon_{n,\alpha},
\end{equation}
where the random correction satisfies the same bound
\eqref{eq:haar-walshness-error}.
\label{fact:stabilizer}
\end{fact}
\begin{proof}
The uniform ensemble of $n$-qubit stabilizer states forms a
complex projective $2$-design~\cite{kueng2015qubit}, meaning
that its first two density-matrix moments agree with those
of Haar-random states. Moreover, $H^{\otimes n}$ is a Clifford
unitary, so $H^{\otimes n}\ket{\psi_{\mathrm{stab}}}$ is also
uniformly distributed over stabilizer states. Its Walsh spectrum
$p(\vec r)=|\bra{\vec r}H^{\otimes n}\ket{\psi_{\mathrm{stab}}}|^2$
therefore satisfies $\mathbb E[p(\vec r)]=2^{-n}$ and
$\mathbb E[p(\vec r)p(\vec s)]
=(1+\delta_{\vec r,\vec s})/[2^n(2^n+1)]$,
exactly as in the preceding Haar proof.
Consequently, $Z_\alpha=e^{\alpha\wz}$ has
the same mean and variance. The remainder of that proof
uses only these moments and the bound
$0\leq\wz\leq n$ to control the logarithm,
and therefore gives Eq.~\eqref{eq:haar-walshness-error}
without modification.
\end{proof}

\subsection{Computing Walshness}
\label{sec:computing_walshness}
In this section, we describe how we compute the Walshness by performing the minimization over single-site unitaries. Suppose we have a normalized state 
\begin{equation}
    \ket{\psi} = \sum_{\vec s} \psi(\vec s) \ket{\vec s}.
\end{equation}
We compute its Walsh spectrum by applying the fast Walsh–Hadamard transform,
\begin{equation}
p(\boldsymbol r)
=
\left|
\bra{\boldsymbol r}H^{\otimes n}\ket{\psi}
\right|^2.
\end{equation}
We compute the fixed-basis Walshness, 
\begin{equation}   \wz =\frac{1}{\alpha} \log\left(\sum_{\vec r} e^{\alpha |\vec r|_1} p(\vec r)\right)
\end{equation}
using LogSumExp in SciPy. This is mathematically equivalent but more numerically stable to the possibility of multiplying extremely small $p(\vec r)$ with extremely large $e^{\alpha |\vec r|_1}$. We compute the intrinsic Walshness starting from Eq.~\eqref{eq:Walternate3}, which we reproduce here: 
\begin{equation}
    \w 
=
\frac n2+
\frac{1}{\alpha}
\log
\min_{\hat{\vec n}_j\in S^2}E[\{\hat{\vec n}_j\}],
\end{equation}
with $E[\{\hat{\vec n}_j\}]
=
\left\langle
\prod_{j=1}^n
A_\alpha(\hat{\boldsymbol n}_j)\right\rangle$ and $A_\alpha(\hat{\boldsymbol n})
=
\cosh\left(\frac{\alpha}{2}\right)I
-
\sinh\left(\frac{\alpha}{2}\right)
\hat{\boldsymbol n}\cdot\boldsymbol\sigma
$. We make the basic observation that $E$ is linear in each $\hat{\vec n}_j$ separately in order to define a \textit{single coordinate update}. For this, we fix every vector except $\hat{\vec n}_j$, define the environment state
\begin{equation}
\ket{\phi_{-j}}
=
\prod_{k\neq j}
A_\alpha(\hat{\boldsymbol n}_k)\ket{\psi},
\end{equation}
and also define the three-component real vector $\boldsymbol v_j
=
\bra{\psi}
\boldsymbol\sigma_j
\ket{\phi_{-j}}$. In terms of these, 
\begin{equation}
E[\{\hat{\vec n}_j\}]
=
\cosh(\frac{\alpha}{2}) \langle\psi|\phi_{-j}\rangle -
\sinh(\frac{\alpha}{2})
\hat{\boldsymbol n}_j\cdot\boldsymbol v_j.
\end{equation}
The first term does not depend on $\hat{\boldsymbol n}_j$. Since $\alpha>0$, minimizing $E[\{\hat{\vec n}_j\}]$ is equivalent to maximizing $\hat{\boldsymbol n}_j\cdot\boldsymbol v_j$, which is obtained by 
\begin{equation}
    \hat{\boldsymbol n}_j\leftarrow \boldsymbol v_j / \|\boldsymbol v_j\|. 
\end{equation}
whenever $\boldsymbol v_j\neq0$. If $\boldsymbol v_j=0$, we do nothing. In one ``sweep", we perform this minimization for each of the $n$ spins in linear order. \par 

A direct implementation would reconstruct $\ket{\phi_{-j}}$ from scratch for every site, leading to an unnecessary cost per sweep. Instead, we retain the full product state
\begin{equation}
\ket{\phi}
=
\prod_{k=1}^n
A_\alpha(\hat{\boldsymbol n}_k)\ket{\psi}.
\end{equation}
To update site $j$, we first perform
\begin{equation}
\ket{\phi_{-j}}
=
A_\alpha(\hat{\boldsymbol n}_j)^{-1}\ket{\phi},
\end{equation}
using the fact that the inverse is particularly simple: $A_\alpha(\hat{\boldsymbol n})^{-1}
=
\cosh\left(\frac{\alpha}{2}\right)I
+
\sinh\left(\frac{\alpha}{2}\right)
\hat{\boldsymbol n}\cdot\boldsymbol\sigma$, because $(\hat{\boldsymbol n}\cdot\boldsymbol\sigma)^2=I$. We compute the three components of \(\boldsymbol v_j\), update \(\hat{\boldsymbol n}_j\), and reconstruct the full product state by applying the new local operator,
\begin{equation}
\ket{\phi}
\leftarrow
A_\alpha(\hat{\boldsymbol n}_j^{\mathrm{new}})
\ket{\phi_{-j}}.
\end{equation}
A complete sweep cycles through the sites sequentially. Since applying a single-qubit operator to a state vector requires $O(2^n)$ operations, one sweep costs $O(n2^n)$ operations. Reducing this exponential scaling, e.g. with recently established classical spin model bootstrap techniques~\cite{paul2026bootstrapping}, would be an interesting avenue of future study. Moreover, we note that since each coordinate update is an exact minimization, $E[\{\hat{\vec n}_j\}]$ cannot increase during a sweep.\par 

Although every single-site update is exact, the joint optimization over all $n$ classical spins is nonconvex. The procedure outlined above can therefore converge to a spurious local minimum. We address this by running the optimization from several initial configurations $\{\hat{\vec n}_j^{\mathrm{init.}}\}$. We sample these initial configurations randomly and uniformly from $S^2$. For each state, we retain the smallest value obtained across the initial conditions. We also store the distribution of final values, whose spread across initial conditions diagnoses the presence of competing local minima. We sometimes find that optimization becomes more difficult as $\alpha$ increases. To combat this, we sometimes first optimize at a smaller value of $\alpha$, then use the resulting spins as the initial configuration at larger $\alpha$.\par 

\begin{figure}
    \centering  \includegraphics[width=0.4\linewidth]{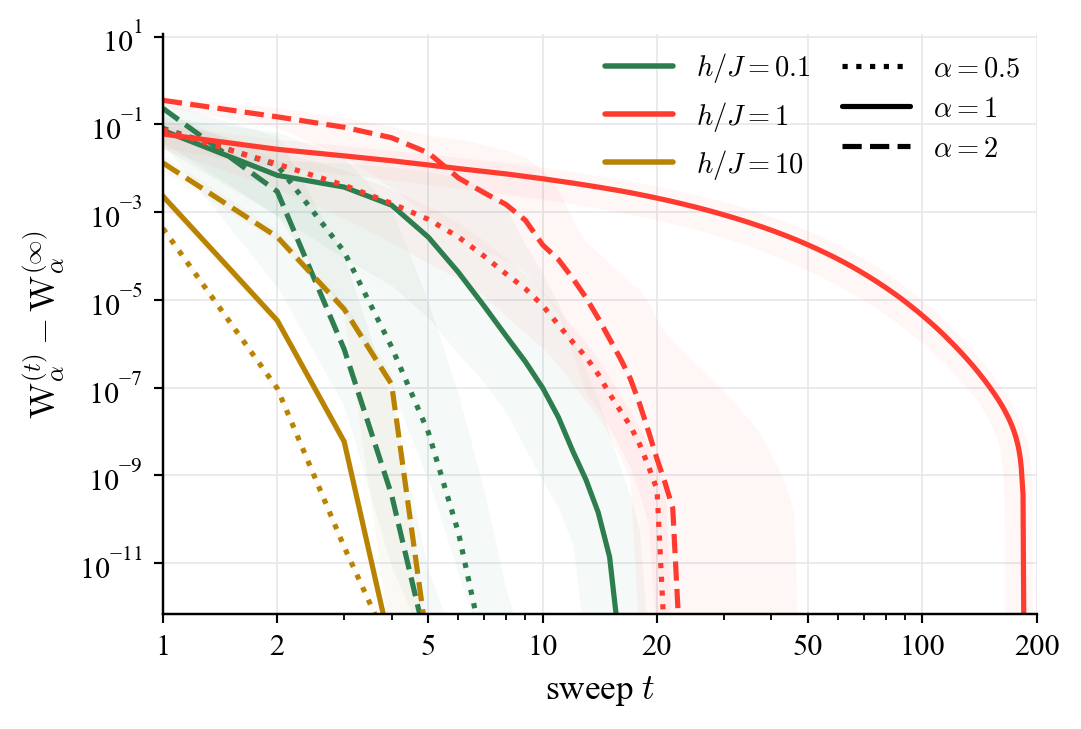}  \caption{\textbf{Convergence}. Analysis of the convergence of Walshness for the ground states of the transverse-field Ising model at system size $n=12$, varying couplings $h/J$ (indicated by the colors), and varying $\alpha$ (indicated by the line style). The x-axis shows the number of sweeps $t$ in the coordinate descent algorithm, while the y-axis shows the difference between the numerically obtained and the converged value of Walshness. Both axes are on a log-scale. Solid lines indicate medians and shaded ribbons indicate the 10th-90th percentile across 16 runs.}   \label{fig:convergence}
\end{figure}

After sweep $t$, letting $E^{(t)}$ denote the intermediate scalar objective, our convergence criterion is
\begin{equation}
E^{(t)}
-
E^{(t+1)}
\leq
\epsilon_{\mathrm{rel}}
\max
\left(
1,
\left|E^{(t)}\right|
\right).
\end{equation}
All calculations in this work use a relative tolerance of $\epsilon_{\mathrm{rel}}=10^{-10}$. 
For the TFIM, we use 32 random initial conditions and at most 150 sweeps per continuation stage; convergence curves for the TFIM ground states are shown in Fig.~\ref{fig:convergence}.

\section{Representability bounds from Walshness}
\label{sec:theorems}

In this section, we will prove a theorem whose high-level idea is:
\begin{equation}
    \text{$\psi$ is a small MLP} \quad \to \quad \text{$\psi$ has low fixed-basis Walshness.}
\end{equation}
For analytic activations under the assumptions below, the corresponding bound concerns approximate Walsh degree. Conversely, we prove a theorem valid for any $\psi$ whose high-level idea is
\begin{equation}
    \text{$\psi$ has low fixed-basis Walshness} \quad \to \quad \text{$\psi$ has a small MLP representation.}
\end{equation}
Together, these results provide complementary upper and lower bounds relating Walshness to representability by small MLPs. For comparison, there are few quantities, if any, in the literature that afford such a characterization. One exception is small conditional mutual information, which is, however, limited to positive wavefunctions~\cite{yang2024can}. Of course, these statements are not yet quantitative and we will carefully lay out what is meant by low Walshness and small MLPs. While we focus on multilayer perceptrons (MLPs) for now, our results will generalize beyond them. \par

The first, or forward, direction rests on the notion of Walsh degree. Define the \textit{\textbf{Walsh degree}} as
\begin{equation}
    \deg(\psi) \coloneqq \max\{|\vec r|_1 \mid \widehat\psi(\vec r)\neq 0\}.
\end{equation}
This is, however, not robust to small perturbations of $\psi$, and therefore it is more useful to define the \textit{\textbf{$\epsilon$-approximate Walsh degree}}
\begin{equation}
    \deg_\epsilon(\psi) \coloneqq \min\Big\{D \;\Big|\; \sum_{|\vec r|_1 > D} p(\vec r)
    \leq \epsilon^2\Big\},
\end{equation}
where $p(\vec r)$ is the Walsh spectrum of $\psi$, which satisfies $\sum_{\vec r} p(\vec r)=1$ and is quadratic in $\psi$. The above is the smallest Walsh degree at which truncating the Walsh expansion of $\psi$ incurs an infidelity at most $\epsilon^2$. In Theorem~\ref{thm:poly-bound} we will prove that an MLP of depth $L$ whose
nonlinearities are polynomials of degrees $d_1,\ldots,d_L$ satisfies $\deg(\psi) \leq D \coloneqq \min(n, \prod_\ell d_\ell)$, and hence
$\wz \leq D$ for every $\alpha$. This bound is independent of the widths,
of the weights, and of $\alpha$, and constrains the architecture only through the degrees
of its nonlinearities. For \textit{nonpolynomial} nonlinearities we bound $\deg_\epsilon$ instead by replacing each nonlinearity with a polynomial
approximant. In Theorem~\ref{thm:analytic-bound} we prove that
\begin{equation}
    \label{eq:Wfromdegeps}
    \wz \leq \deg_\epsilon(\psi) + \frac{1}{\alpha}
    \log\left(1 + \left(e^{\alpha(n - \deg_\epsilon(\psi))}-1\right)\epsilon^2\right).
\end{equation}
The right-hand side can be understood as $\deg_\epsilon(\psi)$ plus ``error". Some analysis is required to bound the error tightly in terms of the size of the MLP architecture, and we carry this out explicitly for some simple examples of nonlinearities such as $\tanh$. At fixed positive $\alpha$, this error term need not be small, so low approximate degree does not by itself imply bounded $\mathrm{W}_\alpha^Z$. We use this relation below to obtain depth lower bounds under the conditions $\mathrm{W}_0^Z=\Omega(n)$ or $\mathrm{W}_\alpha^Z \ge n-O(1)$. \par 

For the converse direction, the conceptual chain is as follows: a small $\wz$ implies an exponentially small high-degree Walsh spectral tail, which in turn implies a small $\mathrm{deg}_\epsilon(\psi)$, which in turn implies a small MLP representation. Consider the random variable $Y = e^{\alpha |\vec r|_1}$, with $\vec r \sim p(\vec r)$, which satisfies $e^{\alpha \wz} = \mathbb{E}[Y]$. Applying Markov's inequality to $Y$ yields $\mathbb{P}[Y> e^{\alpha D}] \leq \frac{\mathbb{E}[Y]}{e^{\alpha D}} = e^{\alpha(\wz-D)}$, for any $D$. At the same time, $\mathbb{P}[Y>e^{\alpha D}] = \sum_{|\vec r|_1>D}p(\vec r)$, which measures the tail of the Walsh spectrum. Thus we have
\begin{equation}
    \label{eq:markovtail}
    \sum_{|\vec r|_1 > D} p(\vec r) \leq e^{\alpha( \wz-D)}.
\end{equation}
Now, consider taking $D = \wz - \frac{1}{\alpha}\log(\epsilon^2)$, so that the right-hand side becomes
$\epsilon^2$. For this choice of $D$, we have $\sum_{|\vec r|_1 > D} p(\vec r) \leq \epsilon^2$. By the definition of $\mathrm{deg}_\epsilon(\psi)$, we have
\begin{equation}
    \label{eq:degepsfromW}
    \deg_\epsilon(\psi) \leq \wz -\frac{1}{\alpha}\log(\epsilon^2).
\end{equation}
This quantifies how small fixed-basis Walshness implies low $\epsilon$-approximate Walsh degree (at fixed $\alpha,\epsilon$). It remains to show that low $\epsilon$-approximate Walsh degree implies an efficient MLP representation, and we provide an explicit construction in the theorem below using nothing more than $x^2$ nonlinearities.

\textbf{\textit{Multilayer perceptrons (MLPs).}} Throughout this section we consider a depth-$L$ network with hidden widths
$m_1,\ldots,m_L$ and inputs $u_i^{(0)}(\vec s) = (-1)^{s_i}$ for $i = 1,\ldots,n$. For
$\ell = 1,\ldots,L$ the preactivations and activations are
\begin{equation}
    z_j^{(\ell)} = b_j^{(\ell)} + \sum_{k} W_{jk}^{(\ell)} u_k^{(\ell-1)},
    \qquad u_j^{(\ell)} = \sigma_\ell(z_j^{(\ell)}),
\end{equation}
and the output is $\psi = c + \sum_j a_j u_j^{(L)}$. We write
$s_\ell \coloneqq \max_j \sum_k |W_{jk}^{(\ell)}|$ for the maximum row $\ell_1$ norm at
layer $\ell$ and $a_{\mathrm{out}} \coloneqq \sum_j |a_j|$ for the $\ell_1$ norm of the
output coefficients. We refer to such a network as an MLP of depth $L$ with
nonlinearities $\sigma_1,\ldots,\sigma_L$.

\subsection{From MLPs to Walshness}
\label{sec:MLP2W}
\begin{figure}[t]
\centering
\resizebox{0.75\textwidth}{!}{%
\begin{tikzpicture}[
  x=1cm,y=1cm,
  leaf/.style={circle,draw=wblue,fill=wblue!20,line width=0.6pt,inner sep=2pt},
  gate/.style={circle,draw=wpurple,fill=wpurple!25,line width=0.6pt,inner sep=2.6pt},
  root/.style={circle,draw=worange,fill=worange!35,line width=0.7pt,inner sep=3pt},
  wire/.style={line width=0.5pt,draw=black!40},
  bar/.style={draw=wblue,fill=wblue!40,line width=0.5pt},
  dbar/.style={draw=black!35,fill=black!8,line width=0.5pt,dashed},
  plab/.style={font=\small}
]
 
% ---------- (a) the network ----------
\foreach \j in {0,1,2}{\node[leaf] (i\j) at (0,{0.7*\j+0.9}) {};}
\node[plab,left=3pt of i2] {$s_1$};
\node[plab,left=3pt of i1] {$\vdots$};
\node[plab,left=3pt of i0] {$s_n$};
\foreach \j in {0,1,2}{\node[gate] (h\j) at (1.5,{0.7*\j+0.9}) {};}
\foreach \j in {0,1,2}{\node[gate] (g\j) at (3.0,{0.7*\j+0.9}) {};}
\node[root] (o) at (4.5,1.6) {};
\node[plab,above=4pt of o] {$\psi(\vec s)$};
\foreach \a in {0,1,2}{\foreach \b in {0,1,2}{
  \draw[wire] (i\a) -- (h\b); \draw[wire] (h\a) -- (g\b);}}
\foreach \a in {0,1,2}{\draw[wire] (g\a) -- (o);}
\draw[decorate,decoration={brace,amplitude=5pt,mirror},draw=black!55,line width=0.5pt]
  (1.5,0.5) -- (3.0,0.5) node[midway,below=7pt,plab] {$L$};
\node[plab] at (-0.85,2.95) {(a)};
 
% ---------- depth compounds the error ----------
\draw[->,>=stealth,line width=0.8pt,draw=black!55] (5.85,1.6) -- (6.95,1.6);
 
% ---------- (b) the Walsh spectrum ----------
\fill[worange!12] (10.44,0.5) rectangle (12.3,2.6);
\foreach \i/\h in {0/0.4,1/1.2,2/1.75,3/1.3,4/0.85}{
  \draw[bar] ({0.42*\i+8.4},0.5) rectangle ({0.42*\i+8.7},{0.5+\h});
}
\foreach \i/\h in {5/0.28,6/0.16,7/0.09,8/0.05}{
  \draw[dbar] ({0.42*\i+8.4},0.5) rectangle ({0.42*\i+8.7},{0.5+\h});
}
\draw[black!70,line width=0.6pt] (8.3,0.5) -- (12.3,0.5);
\draw[black!70,line width=0.6pt] (8.3,0.5) -- (8.3,2.7);
\draw[dashed,line width=0.8pt,draw=worange] (10.44,0.4) -- (10.44,2.6);
\node[plab] at (10.44,2.85) {$D$};
\node[plab,below=3pt] at (9.6,0.5) {$|\vec r|_1$};
\node[plab,below=3pt] at (11.37,0.5) {$\epsilon^2$};
\node[plab,rotate=90] at (7.8,1.6) {$p(\vec r)$};
\node[font=\scriptsize,align=center] at (11.42,1.8)
  {$D \to n$\\[1pt] as $L\to\infty$};
\node[plab] at (7.55,2.95) {(b)};
\end{tikzpicture}}
\caption{\textbf{Walsh spectrum of MLPs.}
(a) A depth-$L$ MLP with analytic nonlinearities can be $\epsilon$-approximated by a polynomial MLP whose Walsh degree grows multiplicatively with depth, $D\lesssim d^L$, where $d$ is the polynomial degree of each nonlinearity in the approximate MLP. (b) Consequently, the Walsh spectrum of the original network has at most $\epsilon^2$ weight above degree $D$. See Sec.~\ref{sec:MLP2W} for details.}
\label{fig:analytic-tail}
\end{figure}
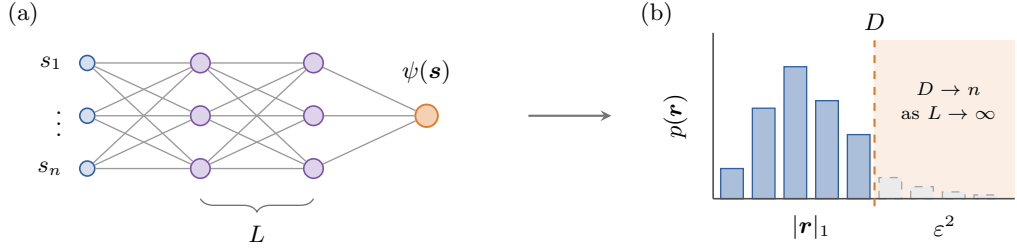

\begin{theorem}[Walsh degree of polynomial MLPs]
\label{thm:poly-bound}
Let $\psi$ be the output of an MLP of depth $L$ whose nonlinearity at layer $\ell$ is a
polynomial of degree $d_\ell$. Then $\deg(\psi) \leq D \coloneqq \min(n,\prod_\ell d_\ell)$,
and consequently $\wz \leq D$ for every $\alpha > 0$.
\end{theorem}
\begin{proof}
To prove this, we track how the Walsh degree grows layer by layer throughout the network. Each input $u_i^{(0)} = (-1)^{s_i}$ has Walsh degree $1$. Affine maps do not increase the degree. To see this, suppose $h_1, \dots, h_m$ each have Walsh degree at most $K$. Then any linear combination $b + \sum_k w_k h_k$ also has Walsh degree at most $K$ since Walsh coefficients $\widehat \psi(r)$ are additive. \par 

Next, we show Walsh degree behaves multiplicatively under polynomial composition. That is, if $h$ has Walsh degree at most $k$ and $\sigma$ is a polynomial of degree $d$, then $\sigma(h)$ has Walsh degree at most $dk$. To see this, note that $\sigma(h) = \sum_{m=0}^d c_m h^m$, so it suffices to show $\deg(h^m) \le m\,\deg(h)$. This follows by induction from the fact that $\deg(fg) \le \deg(f) + \deg(g)$ for any $f, g$, which we now verify. The Walsh coefficients of a product satisfy
\begin{equation}
  \widehat{fg}(\vec r) = \sum_{\vec s} \widehat{f}(\vec s)\,\widehat{g}(\vec r \oplus \vec s).
\end{equation}
A term in this sum is nonzero only if $|\vec s|_1 \le \deg(f)$ (otherwise $\widehat f(\vec s)=0$) and $|\vec r\oplus \vec s|_1 \le \deg(g)$ (otherwise the other factor vanishes). Since $|\vec r|_1 \le |\vec s|_1 + |\vec r\oplus \vec s|_1$, every nonzero term requires $|\vec r|_1 \le \deg(f) + \deg(g)$. \par 

Next, we proceed by induction through the network. After layer $0$, each $u_i^{(0)}$ has Walsh degree $1$. After the affine map at layer $\ell$, each preactivation $z_j^{(\ell)}$ has Walsh degree at most $\prod_{t=1}^{\ell-1} d_t$ (the degree coming in from the previous layers). After applying $\sigma_\ell$ of degree $d_\ell$, each activation $u_j^{(\ell)}$ has Walsh degree at most $\prod_{t=1}^{\ell} d_t$. After all $L$ layers, the output $\psi$ (an affine combination of the $u_j^{(L)}$) has Walsh degree at most $\prod_{\ell=1}^L d_\ell$. Since Walsh degree can never exceed $n$, we get $\deg(\psi) \le D = \min(n, \prod_\ell d_\ell)$.
\end{proof}

As an application, suppose each nonlinearity is a polynomial $z\mapsto z^d$ and we are trying to represent a function using an $L$-layer MLP. Then we have $\wz \leq \min(n,d^L)$ and we need $\wz=n$. Therefore a depth of at least $L = \log_d n$ is necessary. In other words, 
\begin{equation}
    L =\Omega(\log n)
\end{equation}
in this case. \par 

Next, we will establish similar, albeit weaker, upper bounds for more general nonlinearities. Our strategy will be to use a polynomial approximation. Loosely speaking, we will assume the nonlinearities are well-approximated by polynomials of degree $d$ (\textit{e.g.} they are analytic), and then consider a surrogate neural network obtained by replacing each nonlinearity with a polynomial. We first need a way to relate Walshness of  to $\bar M_\alpha(\widetilde{g})$ when $\widetilde{g}$ is a nearby function of bounded Walsh degree. This is the role of the following lemma.\par 

\begin{lemma}[Spectral tail]
\label{lem:spectral-tail}
Let $\ket\psi \neq 0$, let $0 \le D \le n$, and let $\tau_D = \sum_{|\vec r|_1 > D}
p(\vec r)$. Then
\begin{equation}
    \wz \leq D + \frac{1}{\alpha}\log\left(1 +
    \left(e^{\alpha(n-D)}-1\right)\tau_D\right).
\end{equation}
In particular, taking $D = \deg_\epsilon(\psi)$ gives Eq.~\eqref{eq:Wfromdegeps}, and
taking $\epsilon = 0$ recovers $\wz \le \deg(\psi)$.
\end{lemma}
\begin{proof}
    Starting from the definition $e^{\alpha \wz}= \sum_{\vec r} e^{\alpha |\vec r|_1} p(\vec r)$, we split the sum into the two pieces 
\begin{equation}
    e^{\alpha \wz}= \sum_{|\vec r|_1 \leq D } e^{\alpha |\vec r|_1} p(\vec r)+ \sum_{|\vec r|_1 > D } e^{\alpha |\vec r|_1} p(\vec r).
\end{equation}
For the first term, we use $e^{\alpha |\vec r|_1} \leq e^{\alpha D}$, while for the second term we use $e^{\alpha |\vec r|_1 }\leq e^{\alpha n}$ to write 
\begin{equation}
    e^{\alpha \wz} \leq e^{\alpha D} (1-\tau_D) + e^{\alpha n}\tau_D. 
\end{equation}
Factoring out $e^{\alpha D}$, taking the logarithm and dividing by $\alpha$ yields $ \wz \leq D +\frac{1}{\alpha}\log\left[1+ (e^{\alpha(n-D)}-1)\tau_D\right]$, as desired. The remaining claims are then straightforward. 
\end{proof}
\begin{corollary}[Mean Walshness and approximate Walsh degree]
\label{cor:W0-degree}
For any $\ket\psi\neq 0$ and $0<\epsilon<1$,
\begin{equation}
\label{eq:W0-degree}
\mathrm{W}_0^Z[\ket\psi]
\leq
\deg_\epsilon(\psi)
+
\left[n-\deg_\epsilon(\psi)\right]\epsilon^2.
\end{equation}
Equivalently,
\begin{equation}
\label{eq:degree-from-W0}
\deg_\epsilon(\psi)
\geq
\frac{\mathrm{W}_0^Z[\ket\psi]-n\epsilon^2}
{1-\epsilon^2}.
\end{equation}
In particular, if
$\mathrm{W}_0^Z[\ket{\psi_n}]\geq cn$
for some constant $c>0$, then for every fixed $\epsilon$ satisfying
$\epsilon^2<c$,
\begin{equation}
\deg_\epsilon(\psi_n)=\Omega(n).
\end{equation}
\end{corollary}

\begin{proof}
Taking the limit $\alpha\to0^+$ in
Lemma~\ref{lem:spectral-tail} gives
\begin{equation}
\mathrm{W}_0^Z[\ket\psi]
\leq
D+(n-D)\tau_D.
\end{equation}
Setting $D=\deg_\epsilon(\psi)$ gives
$\tau_D\leq\epsilon^2$, and therefore
\begin{equation}
\mathrm{W}_0^Z[\ket\psi]
\leq
\deg_\epsilon(\psi)
+
\left[n-\deg_\epsilon(\psi)\right]\epsilon^2.
\end{equation}
Rearranging yields Eq.~\eqref{eq:degree-from-W0}.
If $\mathrm{W}_0^Z[\ket{\psi_n}]\geq cn$ and
$\epsilon^2<c$, then
\begin{equation}
\deg_\epsilon(\psi_n)
\geq
\frac{c-\epsilon^2}{1-\epsilon^2}\,n
=
\Omega(n),
\end{equation}
which proves the final statement.
\end{proof}

In the following, we will assume the output of an MLP $f(\vec s)$ is normalized so that $\|f\|_2^2 \coloneqq 2^{-n}\sum_{\vec s} |f(\vec s)|^2=1$. This can be achieved by rescaling the final affine layer. We assume the hidden weights and biases are real, while the final affine output may be complex.

\begin{theorem}[Walshness of analytic MLPs]
\label{thm:analytic-bound}
Let $\psi(\vec s)\neq 0$ be the output of a depth$-L$ MLP, normalized as above. Suppose that every pre-activation lies in $[-R,R]$ and that the nonlinearities admit uniformly bounded analytic continuations to a common complex neighborhood of $[-R-1,R+1]$. There is a constant $C$, depending only on $R$ and these analytic bounds, such that the following holds. For $0<\epsilon<1$, define
\begin{equation}
\label{eq:harmlessC}
d= \left\lceil C\left[ \sum_{\ell=1}^{L}\log(2+s_\ell) + \log\frac{2+a_{\mathrm{out}}}{\epsilon} \right] \right\rceil,
\end{equation} 
with $D=\min\{n,d^L\}$. Then $\deg_\epsilon(\psi)\leq D$ and, for every $\alpha>0$, we have $ \wz[\ket\psi] \leq D+ \frac{1}{\alpha} \log\left[ 1+\left(e^{\alpha(n-D)}-1\right)\epsilon^2 \right]. $
\end{theorem}
The corollary below then follows. 
\begin{corollary}
\label{cor:loglogn}
For $0<\epsilon<1$ independent of $n$, consider a family of wavefunctions $\psi_n(\vec s)$ represented by depth-$L$ MLPs satisfying the assumptions of the theorem above. Assume that the preactivation range and analytic continuation bounds are uniform in $n$ and across layers, and that $s_\ell,a_{\mathrm{out}}\le n^c$ for a constant $c$ independent of $n$ and the layer. The output coefficients are measured after the mean-square normalization specified above. If $\mathrm{deg}_\epsilon(\psi_n) = \Omega(n)$, then the depth satisfies
\begin{equation}
    L = \Omega\left(\frac{\log n}{\log\log n}\right).
\end{equation}
In particular, by Corollary~\ref{cor:W0-degree}, this applies whenever
$\mathrm{W}_0^Z[\ket{\psi_n}]\geq cn$
for a constant $c>0$ and $\epsilon^2<c$.
For any fixed $\alpha>0$, it also applies whenever
$\wz[\ket{\psi_n}]\geq n-C_*$
for a constant $C_*$ independent of $n$, provided $\epsilon$ is
sufficiently small depending only on $\alpha$ and $C_*$.
The special case $\wz[\ket{\psi_n}]=n$ applies for every
$0<\epsilon<1$.
\end{corollary}
\begin{proof}[Proof of Thm.~\ref{thm:analytic-bound}] The assumptions on analyticity imply that there exists a common Lipschitz constant $K\geq1$ on $[-R-1,R+1]$ for the MLP. They also imply that each nonlinearity can be uniformly approximated on this interval to error $\delta$ by a polynomial of degree $O(\log(2/\delta))$, with constants uniform across layers. This is a standard exponential convergence bound for polynomial approximation of analytic functions from classical approximation theory~\cite{trefethen}. Choose
\begin{equation}
     \delta= \frac{\epsilon} {(1+a_{\mathrm{out}}) \prod_{\ell=1}^{L}(1+Ks_\ell)}.
\end{equation}
For a sufficiently large $C$, the stated choice of $d$ therefore implies the existence of polynomials $P_\ell$ of degree at most $d$ satisfying
\begin{equation}
     \sup_{|x|\leq R+1}|\sigma_\ell(x)-P_\ell(x)|\leq \delta. 
\end{equation}
We now consider the MLP obtained by replacing each $\sigma_\ell$ by $P_\ell$, keeping the weights and biases unchanged. Let $u_j^{(\ell)}(\vec s)$ denote the neuron $j$ in layer $\ell$ of the original network, and let $\widetilde u_j^{(\ell)}(\vec s)$ denote the corresponding neuron in the polynomialized network. Define the error 
\begin{equation}
    E_\ell = \max_{\vec s} \max_{j}\left|u_j^{(\ell)}(\vec s)-\widetilde u_j^{(\ell)}(\vec s)\right|.
\end{equation}
The original and polynomialized networks have identical inputs, so $E_0=0$. More explicitly, the two networks obey
\begin{equation}
\begin{aligned}
z_j^{(\ell)}(\vec s) = b_j^{(\ell)} + \sum_k W_{jk}^{(\ell)} u_k^{(\ell-1)}(\vec s),\\
\widetilde z_j^{(\ell)}(\vec s) = b_j^{(\ell)} + \sum_k W_{jk}^{(\ell)} \widetilde u_k^{(\ell-1)}(\vec s), 
\end{aligned}
\end{equation}
where $u_j^{(\ell)}(\vec s) = \sigma_\ell\!\left(z_j^{(\ell)}(\vec s)\right)$ and $\widetilde u_j^{(\ell)}(\vec s) = P_\ell\!\left(\widetilde z_j^{(\ell)}(\vec s)\right)$. Thus, since $s_\ell = \max_j \sum_k |W_{jk}^{(\ell)}|$, the preactivation error satisfies $\max_{\vec s,j} \left| z_j^{(\ell)}(\vec s) - \widetilde z_j^{(\ell)}(\vec s) \right| \le s_\ell E_{\ell-1}$. \par 
We claim that
\begin{equation}
    E_\ell \leq \delta\prod_{t=1}^{\ell}(1+Ks_t).
\end{equation}
We prove this by induction. The claim is trivial for $\ell=0$. Suppose it holds through layer $\ell-1$. Then the preactivation error obeys
\begin{equation}
    \max_{\vec s,j} \left| z_j^{(\ell)}(\vec s) - \widetilde z_j^{(\ell)}(\vec s) \right| \leq s_\ell E_{\ell-1}.
\end{equation}
Since $0<\epsilon<1$, the induction hypothesis gives $s_\ell E_{\ell-1} \leq \frac{s_\ell} {(1+Ks_\ell)\prod_{t=\ell+1}^{L}(1+Ks_t)} \leq 1.$ Thus, if the preactivations of the original network lie in $[-R,R]$, those of the polynomialized network lie in $[-R-1,R+1]$, so the polynomial-approximation bound applies. We therefore have
\begin{equation}\begin{aligned} \left| u_j^{(\ell)} - \widetilde u_j^{(\ell)} \right| &= \left| \sigma_\ell(z_j^{(\ell)}) - P_\ell(\widetilde z_j^{(\ell)}) \right|\\ &\leq \left| \sigma_\ell(z_j^{(\ell)}) - \sigma_\ell(\widetilde z_j^{(\ell)}) \right| + \left| \sigma_\ell(\widetilde z_j^{(\ell)}) - P_\ell(\widetilde z_j^{(\ell)}) \right|\\ &\leq Ks_\ell E_{\ell-1}+\delta. \end{aligned}\end{equation}
Hence $E_\ell \leq Ks_\ell E_{\ell-1}+\delta.$ Using the induction hypothesis,
\begin{equation}\begin{aligned} E_\ell &\leq \delta + Ks_\ell\delta \prod_{t=1}^{\ell-1}(1+Ks_t)\\ &\leq \delta \prod_{t=1}^{\ell}(1+Ks_t), \end{aligned}\end{equation}
which proves the claim. At the output layer,
\begin{equation}\begin{aligned} |\psi(\vec s)-\widetilde\psi(\vec s)| &\leq \sum_j |a_j| \left| u_j^{(L)}(\vec s)-\widetilde u_j^{(L)}(\vec s) \right|\\ &\leq a_{\mathrm{out}}E_L\\ &\leq \epsilon. \end{aligned} \end{equation}
Thus $\|\psi-\widetilde\psi\|_\infty\leq\epsilon.$ Since the polynomialized network has polynomial nonlinearities of degrees $d_1,\ldots,d_L$, Thm.~\ref{thm:poly-bound} implies $ \deg(\widetilde\psi) \leq D.$ Therefore $\widehat{\widetilde\psi}(\vec r)=0$ whenever $|\vec r|_1>D$. Parseval's identity then gives
\begin{equation}\begin{aligned} \sum_{|\vec r|_1>D}p(\vec r) &= \sum_{|\vec r|_1>D} |\widehat\psi(\vec r)|^2\\ &= \sum_{|\vec r|_1>D} |\widehat\psi(\vec r)-\widehat{\widetilde\psi}(\vec r)|^2\\ &\leq \sum_{\vec r} |\widehat\psi(\vec r)-\widehat{\widetilde\psi}(\vec r)|^2\\ &= 2^{-n}\sum_{\vec s} |\psi(\vec s)-\widetilde\psi(\vec s)|^2\\ &\leq \epsilon^2. \end{aligned}\end{equation}
Consequently, $\deg_\epsilon(\psi)\leq D.$ Applying the previous Lemma then gives 
\begin{equation}
     \wz \leq D+ \frac{1}{\alpha} \log\left[ 1+ \left(e^{\alpha(n-D)}-1\right)\epsilon^2 \right].
\end{equation}
Finally, analyticity gives polynomial approximants with degree logarithmic in $1/\delta$~\cite{trefethen}, 
 Since $\log\frac1\delta = \log\frac{1+a_{\mathrm{out}}}{\epsilon} + \sum_{\ell=1}^{L}\log(1+Ks_\ell),$
we may choose each $d_\ell$ according to Eq.~\eqref{eq:harmlessC} after rearranging constants. 
\end{proof}
Next we prove Corollary~\ref{cor:loglogn}. 
\begin{proof}[Proof of Cor.~\ref{cor:loglogn}]
By the assumptions above, there is a constant $C_0>0$, independent of $n$ and $L$, such that $d\leq C_0(L+1)\log n$ for all sufficiently large $n$. Therefore the preceding theorem gives $\deg_\epsilon(\psi_n) \leq d^L \leq [C_0(L+1)\log n]^L.$ By assumption, there is a constant $c_0>0$ such that $\deg_\epsilon(\psi_n)\geq c_0 n.$ Combining these bounds gives $c_0 n\leq[C_0(L+1)\log n]^L$. Taking logarithms, $\log n+\log c_0 \leq L\left[\log C_0+\log(L+1)+\log\log n\right].$ If $L\geq\log n$, the claimed bound already follows. Otherwise $L<\log n$, so $\log(L+1)=O(\log\log n)$, and hence $\log n+O(1)\leq O(L\log\log n).$ Therefore $L=\Omega\left(\frac{\log n}{\log\log n}\right).$ For the stated finite-$\alpha$ condition, suppose $\wz[\ket{\psi_n}]\geq n-C_*$. Taking $D=n-\Delta$ in the preceding Lemma with any fixed $\Delta>C_*$, and rearranging gives $\sum_{|\vec r|_1>n-\Delta}p(\vec r) \geq \frac{e^{\alpha(\Delta-C_*)}-1}{e^{\alpha\Delta}-1}.$ The right-hand side is a positive constant independent of $n$. Choosing $\epsilon^2$ smaller than this constant implies $\deg_\epsilon(\psi_n)>n-\Delta=\Omega(n),$ so the preceding depth bound applies.
\end{proof}
The same depth lower bounds apply to approximate representations.
Let $\psi_n$ and the network output $f_n$ have unit mean-square
norm, with $f_n$ satisfying the network assumptions of
Corollary~\ref{cor:loglogn}. Fix $0<\epsilon<1$ independently
of $n$, and suppose that $\deg_\epsilon(\psi_n)=\Omega(n)$
and $\operatorname{F}(\psi_n,f_n)\geq1-\epsilon^2/8$.
After choosing an overall phase of $f_n$,
\begin{equation}
\|\psi_n-f_n\|_2^2
=2\bigl(1-\sqrt{\operatorname{F}(\psi_n,f_n)}\bigr)
\leq2\bigl(1-\operatorname{F}(\psi_n,f_n)\bigr)
\leq\epsilon^2/4.
\end{equation}
Applying Theorem~\ref{thm:analytic-bound} to $f_n$ at tolerance
$\epsilon/2$ gives a polynomial of degree at most $D$ within
distance $\epsilon/2$ of $f_n$. The triangle inequality then
gives $\deg_\epsilon(\psi_n)\leq D$, so the preceding depth
argument is unchanged. For polynomial networks, the same
distance bound gives
$\deg_\epsilon(\psi_n)\leq\deg(f_n)\leq\prod_\ell d_\ell$
without the analytic-network assumptions.\par 

We remark that the $\log\log n$ denominator has a simple origin. By analyticity and the polynomial bounds on the network parameters, each layer can be replaced by a polynomial of degree $d=O((L+1)\log n)$, so a depth-$L$ network has approximate Walsh degree at most $O((L+1)\log n)^L$. Reaching degree $\Omega(n)$ therefore requires $L=\Omega(\log n/\log\log n)$.

\subsection{From Walshness to MLPs}
We now establish the converse: low Walshness implies an efficient MLP representation for the quantum state. 
\begin{lemma}[Approximate degree from Walshness]
\label{lem:degeps-from-W}
Let $\ket\psi\neq 0$, $\alpha>0$, and $0<\epsilon<1$. Then
\begin{equation}
\label{eq:degepspsiminnW}
\deg_\epsilon(\psi)
\leq
\min\left\{
n,
\left\lceil
\wz
+
\frac{2}{\alpha}\log\frac{1}{\epsilon}
\right\rceil
\right\}.
\end{equation}
\end{lemma}
\begin{proof}
Let $\vec r$ be distributed according to the Walsh spectrum $p(\vec r)$ and define the nonnegative random variable $Y=e^{\alpha|\vec r|_1}.$ By definition of the fixed-basis Walshness, $\mathbb E[Y]=e^{\alpha\mathrm W\alpha^Z}.$ For any $D\geq0$, Markov's inequality therefore gives $\sum_{|\vec r|_1>D}p(\vec r)=\mathbb P\left[Y>e^{\alpha D}\right] \leq e^{-\alpha D}\mathbb E[Y]=e^{\alpha(\wz-D)}.$ Taking $D=
\left\lceil
\wz-
\frac{1}{\alpha}\log(\epsilon^2)
\right\rceil$ makes the right-hand side at most $\epsilon^2$. Eq.~\eqref{eq:degepspsiminnW} then follows immediately from the definition of $\deg_\epsilon(\psi)$, together with the trivial bound $\deg_\epsilon(\psi)\leq n$.
\end{proof}

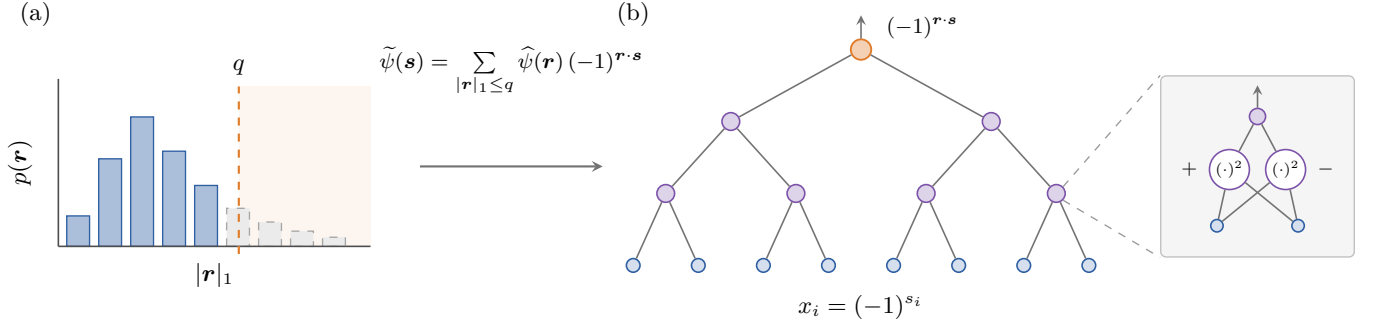
\begin{figure}[t]
\centering
\resizebox{\textwidth}{!}{%
\begin{tikzpicture}[
  x=1cm,y=1cm,
  leaf/.style={circle,draw=wblue,fill=wblue!20,line width=0.6pt,inner sep=2pt},
  gate/.style={circle,draw=wpurple,fill=wpurple!25,line width=0.6pt,inner sep=2.6pt},
  root/.style={circle,draw=worange,fill=worange!35,line width=0.7pt,inner sep=3pt},
  sq/.style={circle,draw=wpurple,fill=white,line width=0.6pt,inner sep=1.2pt,font=\scriptsize},
  bar/.style={draw=wblue,fill=wblue!40,line width=0.5pt},
  dbar/.style={draw=black!35,fill=black!8,line width=0.5pt,dashed},
  wire/.style={line width=0.6pt,draw=black!55},
  plab/.style={font=\small}
]
 
% ---------- (a) truncated Walsh spectrum ----------
\fill[worange!7] (2.32,0.55) rectangle (4.05,2.65);
\foreach \i/\h in {0/0.4,1/1.15,2/1.7,3/1.25,4/0.8}{
  \draw[bar] ({0.42*\i+0.06},0.55) rectangle ({0.42*\i+0.36},{0.55+\h});
}
\foreach \i/\h in {5/0.5,6/0.32,7/0.2,8/0.12}{
  \draw[dbar] ({0.42*\i+0.06},0.55) rectangle ({0.42*\i+0.36},{0.55+\h});
}
\draw[black!70,line width=0.6pt] (-0.05,0.55) -- (4.05,0.55);
\draw[black!70,line width=0.6pt] (-0.05,0.55) -- (-0.05,2.75);
\draw[dashed,line width=0.8pt,draw=worange] (2.32,0.45) -- (2.32,2.65);
\node[plab] at (2.32,2.9) {$q$};
\node[plab,below=3pt] at (2.0,0.55) {$|\vec r|_1$};
\node[plab,rotate=90] at (-0.55,1.65) {$p(\vec r)$};
\node[plab] at (-0.35,3.6) {(a)};
 
% ---------- filtered state, directly above the arrow ----------
\draw[->,>=stealth,line width=0.8pt,draw=black!55] (4.7,1.6) -- (7.1,1.6);
\node[font=\footnotesize] at (5.9,2.85)
  {$\widetilde\psi(\vec s) = \sum\limits_{|\vec r|_1 \leq q}
    \widehat\psi(\vec r)\,(-1)^{\vec r\cdot\vec s}$};
 
% ---------- (b) product tree ----------
\begin{scope}[shift={(7.5,0.3)},scale=0.9,transform shape]
  \foreach \i in {0,...,7}{\node[leaf] (l\i) at ({0.95*\i},0) {};}
  \foreach \i/\a/\b in {0/0/1,1/2/3,2/4/5,3/6/7}{
    \node[gate] (a\i) at ($(l\a)!0.5!(l\b)$) [yshift=1.05cm] {};
    \draw[wire] (l\a) -- (a\i);
    \draw[wire] (l\b) -- (a\i);
  }
  \foreach \i/\a/\b in {0/0/1,1/2/3}{
    \node[gate] (b\i) at ($(a\a)!0.5!(a\b)$) [yshift=1.05cm] {};
    \draw[wire] (a\a) -- (b\i);
    \draw[wire] (a\b) -- (b\i);
  }
  \node[root] (r) at ($(b0)!0.5!(b1)$) [yshift=1.05cm] {};
  \draw[wire] (b0) -- (r);
  \draw[wire] (b1) -- (r);
  \draw[wire,->,>=stealth] (r) -- ++(0,0.5);
  \node[above right=-2pt and 4pt of r,plab] {$(-1)^{\vec r\cdot\vec s}$};
\end{scope}
\node[plab] at (7.5,3.6) {(b)};
\node[plab] at (10.49,-0.25) {$x_i = (-1)^{s_i}$};
 
% ---------- inset, one node opened up ----------
\begin{scope}[shift={(15.7,1.6)},scale=0.82,transform shape]
\node[draw=black!25,fill=black!4,rounded corners=2pt,line width=0.5pt,
      minimum width=3.1cm,minimum height=2.9cm] (box) at (0,0) {};
\node[leaf] (p1) at (-0.65,-0.95) {};
\node[leaf] (p2) at (0.65,-0.95) {};
\node[sq] (s1) at (-0.45,-0.05) {$(\cdot)^2$};
\node[sq] (s2) at (0.45,-0.05) {$(\cdot)^2$};
\node[gate] (m)  at (0,0.8) {};
\draw[wire] (p1) -- (s1);
\draw[wire] (p2) -- (s2);
\draw[wire] (p1) -- (s2);
\draw[wire] (p2) -- (s1);
\draw[wire] (s1) -- (m);
\draw[wire] (s2) -- (m);
\node[left=2pt of s1,font=\small] {$+$};
\node[right=2pt of s2,font=\small] {$-$};
\draw[wire,->,>=stealth] (m) -- ++(0,0.5);
\end{scope}
\draw[dashed,draw=black!40,line width=0.5pt] (a3.north east) -- (box.north west);
\draw[dashed,draw=black!40,line width=0.5pt] (a3.south east) -- (box.south west);
\end{tikzpicture}}
\caption{\textbf{MLP representation from Walsh spectrum.}
(a) We truncate the Walsh spectrum at Hamming weight $q$ and discard
the tail, costing infidelity $\epsilon^2$. (b) Each surviving $(-1)^{\vec r\cdot\vec s}$ has degree at most $q$ and
can be evaluated by a binary tree whose nodes multiply two inputs, so
the tree has depth $O(\log q)$. Summing the trees against the coefficients
$\widehat\psi(\vec r)$ gives $\widetilde\psi(\vec s)$. Inset: multiplication achieved via $xy = [(x+y)^2-(x-y)^2]/4$, with nonlinearity $x\mapsto x^2$. Each node costs two nonlinearities, and each tree costs $q+\frac{q}{2} + \frac{q}{4} + \cdots = O(q)$. We refer to Theorem~\ref{thm:representation} for details. }
\label{fig:construction}
\end{figure}

It remains to show that low $\epsilon$-approximate Walsh degree implies an efficient MLP representation. We do so constructively using only the quadratic nonlinearity $\sigma(x)=x^2$. This is schematically depicted in Fig.~\ref{fig:construction}. 

\begin{theorem}[MLP representation]
\label{thm:representation}
Let $\ket\psi=\sum_{\vec s}\psi(\vec s)\ket{\vec s}\neq 0$, let $\alpha>0$, and define $\omega= \wz[\ket{\psi}].$ For any $0<\epsilon<1$, there exists an MLP $\widetilde{\psi}_{\vec \theta}(\vec s)$ with quadratic nonlinearities $\sigma(x)=x^2$ whose associated normalized state has infidelity at most $\epsilon^2$ with $\ket\psi$. The MLP can be chosen to have width
\begin{equation}
\label{eq:widthscalingMLP}
n^{O\left(\omega+\alpha^{-1}\log(1/\epsilon^2)\right)}
\end{equation}
and depth
\begin{equation}
\label{eq:depthscalingMLP}
O\left(
\log\left[
\omega+\alpha^{-1}\log(1/\epsilon^2)
\right]
\right).
\end{equation}
In particular, for fixed $\alpha$ and $\varepsilon$, $\omega=O(1)$ implies a polynomial-width, constant-depth MLP representation.
\end{theorem}

\begin{proof}
    Define $q=
\min\left\{
n,
\left\lceil
 \omega
+
\frac{2}{\alpha}\log\frac{1}{\epsilon}
\right\rceil
\right\}$. By the preceding lemma, $\deg_\epsilon(\psi)
\leq q$. Define a ``filtered" state 
\begin{equation}\label{eq:psitilderep}
    \widetilde\psi(\vec s) =\sum_{|\vec r|_1\leq q} \widehat\psi(\vec r) (-1)^{\vec r\cdot\vec s}.
\end{equation}
In plain terms, we take the Walsh-Hadamard transform of $\psi(\vec s)$, truncate to $|\vec r|_1\leq q$, and transform back to obtain $\widetilde\psi(\vec s)$. Because $\sum_{|\vec r|_1>q}p(\vec r)\leq \epsilon^2$, the normalized $\ket{\widetilde\psi}$ has infidelity $\leq \epsilon^2$ with $\ket\psi$. In other words, the fidelity satisfies
\begin{equation}
  \mathrm{F}=  \frac{|\langle \psi|\widetilde\psi\rangle|^2}{\langle\psi|\psi\rangle\langle\widetilde\psi|\widetilde\psi\rangle} =1-\sum_{|\vec r|_1>q}p(\vec r) \geq 1-\epsilon^2.
\end{equation}
We now represent $\widetilde\psi(\vec s)$ by an MLP. Write 
\begin{equation}
    x_i = (-1)^{s_i},\qquad (-1)^{\vec r\cdot \vec s} = \prod_{i \mid r_i = 1} x_i. 
\end{equation}
The factor on the right is a monomial of degree at most $q$ in the $x_i$. Using a quadratic nonlinearity $\sigma(x)=x^2$, multiplication of two scalars can be implemented exactly through the handy identity
\begin{equation}
    xy = \frac{(x+y)^2-(x-y)^2}{4}. 
\end{equation}
Hence a product of at most $q$ variables can be evaluated with a balanced binary tree of multiplication gates. In other words, a product $x_1\ldots x_q$ is evaluated by $w_1 = x_1 x_2$, $w_2 = x_3x_4$, $\ldots$ $w_{q/2} = x_{q-1}x_q$, and $t_1 = w_1 w_2$, and so on hierarchically. This requires depth $O(\log q)$. Moreover, it requires $O(q)$ nonlinearities per monomial, since $q + \frac{q}{2} + \frac{q}{4} + \cdots \leq 2q$. We compute each monomial $(-1)^{\vec r\cdot \vec s}$ occurring Eq.~\eqref{eq:psitilderep} in this form. The number of such monomials is 
\begin{equation}
    N_q = \sum_{k=0}^q \binom{n}{k} \leq \sum_{k=0}^q n^k\leq (q+1)n^q.
\end{equation}
The total network width is at most $O(qN_q) = O(q^2 n^q)$. The total depth is $O(\log q)$. Substituting in the value of $q$, we recover Eqs.~\eqref{eq:widthscalingMLP} and \eqref{eq:depthscalingMLP}. 
\end{proof}

\section{Multiplicative Walshness and Restricted Boltzmann Machines}
\label{sec:RBM}

In Sec.~\ref{sec:theorems}, we related fixed-basis Walshness to representability by multilayer perceptrons (MLPs). We now consider neural network ans\"atze whose natural structure is multiplicative. A standard such example is the restricted Boltzmann machine (RBM). An RBM wavefunction involves visible spins (the input spins) and hidden spins in a Boltzmann factor, with a further sum over the hidden spin variables. After summing, the RBM is a product of factors, each depending only on the visible spins and the weights connecting visible and hidden spins, but not the hidden spins. Taking a logarithm turns this product into a sum of local functions. This suggests measuring complexity by an analog of the Walsh degree from before, which is suited to the \textit{logarithm} of a wavefunction amplitude.\par 

In this section, we carry this out by defining \textit{multiplicative Walshness}. In particular, we prove a theorem whose high-level idea is:
\begin{equation}
    \psi \text{ is a small RBM}
    \quad \to  \quad
    \psi \text{ has low multiplicative Walshness.}
\end{equation}
Conversely, we prove a theorem valid for any $\psi$ whose high-level idea is
\begin{equation}
    \psi \text{ has low multiplicative Walshness}
    \quad \to  \quad
    \psi \text{ has a small RBM representation.}
\end{equation}
Of course, these statements are not yet quantitative and we will carefully lay out what is meant by small multiplicative Walshness and small RBMs.

\subsection{Definition of multiplicative Walshness}

Let $\psi(\vec s):\{0,1\}^n\to\mathbb C$ be nowhere vanishing, let $\ket\psi = \sum_{\vec s}\psi(\vec s)\ket{\vec s}$ as usual, and define $K$ by
\begin{equation}
\psi(\vec s)=e^{K(\vec s)}.
\label{eq:log-lift-def}
\end{equation}
The choice of logarithm $K$ is part of the representation and
need not be principal-valued. Different logarithms of the same
wavefunction can have different multiplicative Walshness.
The forward theorem constructs a logarithm from the individual
RBM factors; the converse applies to any chosen logarithm.
In the numerical example below, $\psi>0$ and we use the real
logarithm $K=\log\psi$.\par

Let us write the Walsh expansion of $K$ as
\begin{equation}
K(\vec s)=\sum_{\vec r\in\{0,1\}^n}\widehat K(\vec r)(-1)^{\vec r\cdot \vec s},
\label{eq:K-walsh-expansion}
\end{equation}
and define its degree-$D$ truncation
\begin{equation}
K_{\le D}(\vec s):=\sum_{|\vec r|_1\le D}\widehat K(\vec r)(-1)^{\vec r\cdot \vec s}.
\label{eq:K-truncation}
\end{equation}
For two nonzero wavefunctions $f,g$, define the fidelity
\begin{equation}
\operatorname{F}(f,g):=
\frac{\left|\sum_{\vec s} f(\vec s)^*g(\vec s)\right|^2}
{\left(\sum_{\vec s}|f(\vec s)|^2\right)\left(\sum_{\vec s}|g(\vec s)|^2\right)}.
\label{eq:projective-fidelity}
\end{equation}
We define a notation for the infidelity of $K$ with $K_{\leq D}$, namely $e_D^*(K)\coloneqq
1-\operatorname{F}\!\left(e^K,e^{K_{\le D}}\right)$. For a given $K$, we can consider $e_D^*$ as a sequence in $D$, and it need not be monotone. We therefore define the running minimum
\begin{equation}
E_D^*(K):=\min_{0\le j\le D}e_j^*(K),
\label{eq:multiplicative-running-error}
\end{equation}
along with conventions $E_{-1}^*:=1$ and $E_n^*:=0$ (noting $D\leq n$). We can treat this as a cumulative distribution and define it with an associated probability distribution
\begin{equation}
p_D^*(K):=E_{D-1}^*(K)-E_D^*(K),\label{eq:multiplicative-degree-distribution}
\end{equation}
Since $E_D^*$ is nonincreasing, we have $p_D^*\ge0$, and the telescoping sum $\sum_Dp_D^*=1$, so $p_D^*$ is a classical distribution. It is the multiplicative analog of the Walsh spectrum. We next define the \textit{multiplicative Walshness} by
\begin{equation}
\mathrm{W}_\alpha^*(K):=
\frac{1}{\alpha}
\log \mathbb{E}\left[ e^{\alpha D}\right],\qquad \alpha>0
\label{eq:multiplicative-walshness}
\end{equation}
where the expectation is in $D\sim p_D^*(K)$. Thus $0\le \mathrm W_\alpha^*\le n$. Moreover, adding a constant to $K$ multiplies both the target and every truncated wavefunction by the same nonzero scalar, so $\mathrm W_\alpha^*(K+c)=\mathrm W_\alpha^*(K)$ for every constant $c\in\mathbb C$.\par 

It is useful to note that Eq.~\eqref{eq:multiplicative-walshness} is closely related to the ordinary fixed-basis Walshness introduced earlier. To make this analogy precise, consider instead using the Walsh truncation $\psi_{\le D}(\vec s):=\sum_{|r|_1\le D}\widehat\psi(r)(-1)^{\vec r\cdot \vec s}$ and defining $e_D^{\mathrm{lin}}:=1-\operatorname{F}(\psi,\psi_{\le D})$. Then orthogonality gives us $e_D^{\mathrm{lin}}
=
\frac{\sum_{|\vec r|_1>D}|\widehat\psi(\vec r)|^2}
{\sum_{\vec r}|\widehat\psi(\vec r)|^2}.
$ and hence $e_{D-1}^{\mathrm{lin}}-e_D^{\mathrm{lin}}
=
\sum_{|\vec r|_1=D}p(\vec r)
$, and substituting these into Eq.~\eqref{eq:multiplicative-walshness} recovers $\mathrm{W}_\alpha^Z$. We note the key difference in these constructions is the different truncations used: $\psi_{\le D}$ versus $e^{K_{\le D}}$. In analogy with earlier, we also define an $\epsilon$-approximate multiplicative degree, 
\begin{equation}
\deg_\epsilon^*(K)
:=
\min\left\{D\,\middle|\,E_D^*(K)\le\epsilon^2\right\}.
\label{eq:approx-multiplicative-degree}
\end{equation}
This is the smallest $D$ for which a multiplicative truncation up to degree \(D\) has infidelity at most \(\epsilon^2\).

\subsection{From RBMs to multiplicative Walshness}

We next consider the standard complex RBM wavefunction written in terms of $x_i\coloneqq (-1)^{s_i}=\pm1,$
\begin{equation}
\psi_{\mathrm{RBM}}(\vec s)
=
\exp\!\left(\sum_{i=1}^n a_i x_i\right)
\prod_{j=1}^M
2\cosh\!\left(b_j+\sum_{i=1}^n W_{ji}x_i\right).
\label{eq:rbm-wavefunction}
\end{equation}
For hidden unit $j$, define $S_j:=\{i\,|\,W_{ji}\neq0\}$ and 
$\kappa:=\max_j|S_j|$. We call $\kappa$ the maximum hidden-unit connectivity.

\begin{theorem}[Multiplicative Walshness of sparse RBMs]
Suppose $\psi_{\mathrm{RBM}}$ is nowhere vanishing and has maximum hidden-unit connectivity $\kappa$. Then, for every $\alpha>0$, $\psi_{\mathrm{RBM}}$ admits a logarithm $K$, with $e^K = \psi_{\mathrm{RBM}}$, such that
\begin{equation}
\mathrm{W}_\alpha^*(K)
\le
\max\{1,\kappa\}.
\label{eq:rbm-forward-bound}
\end{equation}
\end{theorem}

\begin{proof}
Because the RBM is nowhere vanishing, every factor in
Eq.~\eqref{eq:rbm-wavefunction} is nonzero on every $\vec s$.
For each hidden unit, define
\begin{equation}
k_j(\vec x_{S_j})
:=
\operatorname{Log}\!\left[
2\cosh\!\left(b_j+\sum_{i\in S_j}W_{ji}x_i\right)
\right],
\end{equation}
where $\operatorname{Log}$ denotes the principal logarithm. Define
\begin{equation}
K(\vec s)
=
\sum_i a_i x_i
+
\sum_{j=1}^M k_j(\vec x_{S_j}).
\label{eq:rbm-log-decomposition}
\end{equation}
Then $e^K=\psi_{\mathrm{RBM}}$, although $K$ need not be the
principal logarithm of the full wavefunction. Each $k_j$
depends only on the variables in $S_j$. We first note that $k_j$ has Walsh degree at most $|S_j|$, since it only depends on at most $|S_j|$ spins. Thus $\deg(k_j)\le |S_j|\le\kappa.$ The term $\sum_i a_ix_i$ has Walsh degree at most one, and addition does not increase the degree beyond the maximum degree of the summands. Hence
\begin{equation}
\deg(K)
\le
D_\kappa:=\max\{1,\kappa\}.
\label{eq:rbm-log-degree}
\end{equation}
For every $D\ge D_\kappa$, we therefore have $K_{\le D}=K$, and consequently $e_D^*=E_D^*=0.$ It follows that $p_D^*=0$ for $D>D_\kappa$. Using Eq.~\eqref{eq:multiplicative-walshness},
\begin{equation}
e^{\alpha \mathrm{W}_\alpha^*}
=
\sum_{D=0}^{D_\kappa}e^{\alpha D}p_D^*
\le
 e^{\alpha D_\kappa}
\sum_{D=0}^{D_\kappa}p_D^*
=
 e^{\alpha D_\kappa},
\end{equation}
which proves Eq.~\eqref{eq:rbm-forward-bound}.
\end{proof}

\subsection{From multiplicative Walshness to RBMs}

We now establish the converse: low multiplicative Walshness implies an approximate RBM representation with controlled maximum hidden-unit connectivity.

\begin{lemma}[Approximate multiplicative degree from multiplicative Walshness]
Let $K$ be as above, let $\alpha>0$, and let $0<\epsilon<1$. Then
\begin{equation}
\deg_\epsilon^*(K)
\le
\min\left\{
 n,
 \left\lceil
 \mathrm W_\alpha^*(K)
 +\frac{2}{\alpha}\log\frac{1}{\epsilon}
 \right\rceil
\right\}.
\label{eq:approx-multiplicative-degree-bound}
\end{equation}
\end{lemma}

\begin{proof}
Let $D$ be distributed according to $p_D^*$, and define the nonnegative random variable $Y=e^{\alpha D}$. By definition, $\mathbb E[Y]=e^{\alpha \mathrm{W}_\alpha^*}.
$ Moreover, telescoping Eq.~\eqref{eq:multiplicative-degree-distribution} gives $E_D^*
=
\sum_{j>D}p_j^*.
$ Markov's inequality therefore gives
\begin{equation}
E_D^*
=
\mathbb P\!\left[Y>e^{\alpha D}\right]
\le
 e^{-\alpha D}\mathbb E[Y]
=
 e^{\alpha(\mathrm{W}_\alpha^*-D)}.
\label{eq:multiplicative-tail-bound}
\end{equation}
Taking
\begin{equation}
D=
\left\lceil
\mathrm{W}_\alpha^*
+\frac{2}{\alpha}\log\frac{1}{\epsilon}
\right\rceil
\end{equation}
makes the right-hand side at most \(\epsilon^2\). Equation~\eqref{eq:approx-multiplicative-degree-bound} follows from the definition of \(\deg_\epsilon^*\), together with the trivial bound \(\deg_\epsilon^*\le n\).
\end{proof}

It remains to show that a low-degree $e^{K_{\leq D}(\vec s)}$ has an efficient RBM representation. We prove this below, with the essential step of characterizing the expressive power of one hidden unit. 

\begin{lemma}[Power of one hidden unit]
Let $A\subseteq[n]$ be nonempty and define $\chi_A(\vec s):=\prod_{i\in A}x_i,
x_i=(-1)^{s_i}.$ For every $c_A\in\mathbb C$, the expression $e^{c_A\chi_A(\vec s)}$ can be represented, up to a nonzero overall scalar, by a complex RBM using at most one hidden unit connected only to the visible units in $A$.
\label{lem:one-hidden-parity}
\end{lemma}
\begin{proof}
Define $S_A:=\sum_{i\in A}x_i$. Since $x_i=\pm1$,
\begin{equation}
e^{-i\pi S_A/2}
=
\prod_{i\in A}e^{-i\pi x_i/2}
=
 e^{-i\pi |A|/2}\chi_A(\vec s).
\label{eq:sum-to-parity}
\end{equation}
Suppose first that \(\sinh c_A\cosh c_A\neq0\). Choose \(b_A\in\mathbb C\) such that
\begin{equation}
e^{-2b_A}e^{-i\pi |A|/2}=\tanh c_A.
\label{eq:hidden-bias-choice}
\end{equation}
Then
\begin{align}
&e^{-i\pi S_A/4}
2\cosh\!\left(b_A+\frac{i\pi}{4}S_A\right)
\nonumber\\
&\qquad =
 e^{b_A}
 +e^{-b_A}e^{-i\pi S_A/2}
\nonumber\\
&\qquad =
 e^{b_A}\left(1+\tanh(c_A)\chi_A\right)
=
\frac{e^{b_A}}{\cosh c_A}e^{c_A\chi_A}.
\label{eq:one-hidden-parity-construction}
\end{align}
The prefactor on the left is a bias contribution, while the \(\cosh\) term is one hidden unit with weights \(W_i=i\pi/4\) for \(i\in A\) and zero otherwise. Thus its connectivity is \(|A|\). If \(\sinh c_A=0\), then \(e^{c_A\chi_A}\) is constant. If \(\cosh c_A=0\), then \(e^{c_A\chi_A}\) is proportional to \(\chi_A\), and Eq.~\eqref{eq:sum-to-parity} shows that \(\chi_A\) can be represented using visible biases alone. This proves the claim in all cases.
\end{proof}

\begin{theorem}[RBM representation from multiplicative Walshness]
Let $K$ be as before, let $\alpha>0$, and define $\omega=\mathrm W_\alpha^*(K)$. For any $0<\epsilon<1$, define
\begin{equation}
q\coloneqq
\min\left\{
 n,
 \left\lceil
 \omega+\frac{2}{\alpha}\log\frac{1}{\epsilon}
 \right\rceil
\right\}.
\label{eq:rbm-converse-q}
\end{equation}
Then there exists a complex RBM $\widetilde\psi_{\mathrm{RBM}}$
satisfying simultaneously $1-\operatorname{F}\!\left(e^K,\widetilde\psi_{\mathrm{RBM}}\right)
\le \epsilon^2$, $\kappa\le q,$ and $M\le\sum_{k=2}^{q}\binom{n}{k}$, 
where we set $\kappa=0$ when $M=0$.
In particular, bounded multiplicative Walshness permits RBM
approximations with constant connectivity and polynomially many
hidden units at fixed $\alpha$ and accuracy.
\end{theorem}
\begin{proof}
Eq.~\eqref{eq:approx-multiplicative-degree-bound} gives
$\deg_\epsilon^*(K)\le q$. Thus we can choose $0\le j\le q$ such that
\begin{equation}
1-\operatorname{F}\!\left(e^K,e^{K_{\le j}}\right)
\le\epsilon^2.
\label{eq:chosen-multiplicative-truncation}
\end{equation}
We expand the truncated logarithmic amplitude as
\begin{equation}
K_{\le j}(\vec s)
=
c_\emptyset+\sum_i c_i x_i
+\sum_{\substack{A\subseteq[n]\\2\le|A|\le j}}
c_A\chi_A(\vec s),
\label{eq:truncated-log-interactions}
\end{equation}
where $c_i=0$ if $j=0$. Exponentiating gives
\begin{equation}
e^{K_{\le j}(\vec s)}
=
e^{c_\emptyset}e^{\sum_i c_i x_i}
\prod_{\substack{A\subseteq[n]\\2\le|A|\le j}}
e^{c_A\chi_A(\vec s)}.
\label{eq:truncated-factorization}
\end{equation}
The constant factor does not affect the normalized state. We represent
the degree-one terms using visible biases. By
Lemma~\ref{lem:one-hidden-parity}, each remaining factor requires at
most one hidden unit, connected only to the spins in $A$.
Collecting these hidden units and adding their visible-bias
contributions gives a single RBM proportional to $e^{K_{\le j}}$.
Its connectivity and number of hidden units satisfy
\begin{equation}
\kappa\le j\le q,
\qquad
M\le\sum_{k=2}^{j}\binom{n}{k}
\le\sum_{k=2}^{q}\binom{n}{k},
\end{equation}
with $\kappa=0$ when no hidden units are needed.
Equation~\eqref{eq:chosen-multiplicative-truncation} then gives the
claimed infidelity bound. For fixed $\alpha$ and $\epsilon$,
$\omega=O(1)$ implies $q=O(1)$, so the connectivity is constant
and the number of hidden units is polynomial in $n$.
\end{proof}

\subsection{RBM learning experiments}
\begin{figure*}[t]
    \centering
    \includegraphics[width=\textwidth]{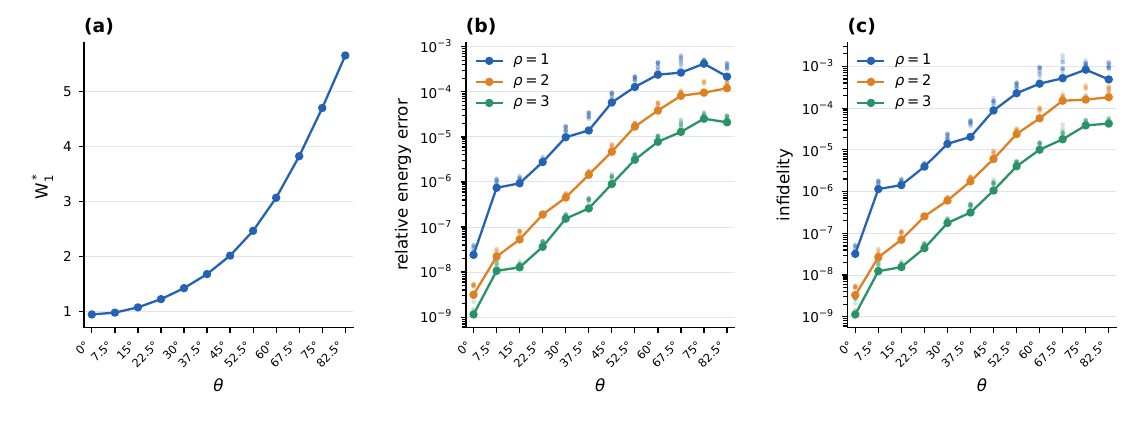}
    \caption{\textbf{Multiplicative Walshness and RBM learning.}
    We consider the $L=10$, $g=1.5$ transverse-field Ising model under
    the global basis rotation of
    Ref.~\cite{cortes2025basis}.
    (a) Multiplicative Walshness $\mathrm{W}^*_{1}$ of the exact ground state.
    (b) Relative energy error
    $|E_{\mathrm{RBM}}-E_{\mathrm{ED}}|/|E_{\mathrm{ED}}|$ and
    (c) infidelity
    $1-|\langle\psi_{\mathrm{ED}}|\psi_{\mathrm{RBM}}\rangle|^2$
    after $1000$ optimization steps for complex RBMs with hidden-unit
    densities $\rho=N_h/N_v=1,2,3$.
    Faint markers show the ten retained trials at each angle, while the solid
    curves follow the trial with the smallest recomputed checkpoint energy;
    the same selected trial is used in panels (b) and (c).
    Panels (b) and (c) revisit the basis dependence studied in Fig.~5 of
    Ref.~\cite{cortes2025basis}, while panel (a) shows the corresponding
    multiplicative Walshness along the same rotation path.
    Over the range shown, all amplitudes of the exact ground state are
    strictly positive, so the principal logarithm is real, and there is no
    logarithmic branch ambiguity in the evaluation of $\mathrm{W}^*_{1}$.
    All $\rho=1$ checkpoints shown here, together with the $\rho=2,3$
    checkpoints at $\theta=7.5^\circ,30^\circ,52.5^\circ,$ and $75^\circ$,
    were taken from the public repository accompanying
    Ref.~\cite{cortes2025basis}. The remaining $\rho=2,3$ checkpoints were
    generated in this work using the same source code and training protocol.}
    \label{fig:multiplicative-rbm-learning}
\end{figure*}

We next test whether multiplicative Walshness is informative for practical
RBM learning. We consider the $L=10$, $g=1.5$ transverse-field Ising model
and vary the local basis through the one-parameter rotation considered in
Ref.~\cite{cortes2025basis}. This is the same setting used to demonstrate
the strong basis dependence of RBM learning in Fig.~5 of
Ref.~\cite{cortes2025basis}, where the relative energy error and infidelity
were found to vary by several orders of magnitude with the rotation angle.
Here we revisit this behavior from the perspective of multiplicative
Walshness. Figure~\ref{fig:multiplicative-rbm-learning}(a) shows
$\mathrm{W}^*_{1}$ of the exact ground state along the same basis rotation.
For all rotation angles shown, the exact wavefunction amplitudes are strictly
positive. Thus the principal logarithm $K=\log\psi$ is real, and the
logarithmic branch ambiguity discussed above does not arise in this example.

We compare this quantity with variational optimization of complex RBMs.
We denote the hidden-unit density by $\rho=N_h/N_v$ in order to distinguish
it from the parameter $\alpha$ entering Walshness, and consider
$\rho=1,2,3$. Following Ref.~\cite{cortes2025basis}, the RBMs are optimized
using stochastic reconfiguration~\cite{sorella1998green}, with sums evaluated
over the full configuration space. Each RBM is optimized for $1000$ steps.
For every pair $(\rho,\theta)$, we retain ten optimization trials. The faint
markers in Figs.~\ref{fig:multiplicative-rbm-learning}(b) and
\ref{fig:multiplicative-rbm-learning}(c) show these trials, while the solid
curve follows the trial with the smallest recomputed checkpoint energy at
each angle. The same selected trial is used to report both the relative
energy error and the infidelity.

Part of the checkpoint data was taken from the public repository
accompanying Ref.~\cite{cortes2025basis}\footnote{Original code and data:
\url{https://github.com/rscortessa/basis-dependence-of-RBM-for-TFIM},
commit \texttt{77c3ab0c31e3bd4229544774c1f1f1389c21eda6}.
The additional checkpoints generated in this work are available at
\url{https://github.com/hello-world2fy/TFIM},
commit \texttt{d96d63b049246f98397b187b1c3fe3814b027b55}.}.
In particular, all $\rho=1$ checkpoints shown here, together with the
$\rho=2,3$ checkpoints at
$\theta=7.5^\circ,30^\circ,52.5^\circ,$ and $75^\circ$, are taken from
the released data. We generated the remaining $\rho=2,3$ checkpoints using
the same source code, physical setup, optimization method, hyperparameter
search ranges, and $1000$-step training budget. For consistency, all energies
and infidelities shown in Fig.~\ref{fig:multiplicative-rbm-learning} were
recomputed from the retained checkpoints using the same post-processing
procedure.

As the basis is rotated, the multiplicative Walshness increases strongly,
and both the relative energy error and the infidelity exhibit a corresponding
overall increase. This reproduces, for the $g=1.5$ case, the strong
basis-dependent deterioration reported in Fig.~5 of
Ref.~\cite{cortes2025basis}. The interpretation here is complementary:
multiplicative Walshness measures the interaction order required after
passing to the logarithmic amplitude, and the results above relate low
multiplicative Walshness to low-connectivity RBM representations. Thus the
same basis dependence observed numerically in Ref.~\cite{cortes2025basis}
is accompanied by a systematic increase in the multiplicative complexity
of the target wavefunction.

Increasing the hidden-unit density improves the variational accuracy across
essentially the full range of angles, but the strong basis dependence remains
visible for all three network sizes. We emphasize that the present experiment
is not a direct numerical test of the connectivity bound derived above.
The RBMs used here are fully connected, so varying $\rho$ changes the number
of hidden units rather than their maximum connectivity. Rather, the comparison
tests whether multiplicative Walshness also correlates with practical learning
difficulty within a standard dense-RBM ansatz. As in the MLP experiments
above, finite-time optimization errors should not be interpreted as exact
representational lower bounds.

\section{Case studies}
\label{sec:casestudies}
In this section of the supplement, we provide detailed analytical and numerical case studies of Walshness in various models, including ferromagnets, stabilizer states, and ground states of the transverse-field Ising model (TFIM) and mixed-field toric code (MFTC), some of which were presented in the main text and End Matter. We provide complete details on our numerical experiments investigating the utility of Walshness in training NQS. We aim to build evidence that the Walshness-minimizing basis may provide an advantage in NQS training, particularly when the reduction in Walshness is large. Separately, we highlight analytical calculations of fixed-basis Walshness in the transverse-field Ising model and toric code, which may be of independent interest. 
\par 
Before we proceed, a few remarks on the scope of our numerical results are in order. 
First, the MLP experiments below use the infidelity with respect to the target state as the loss function, rather than the variational energy. Energy minimization is the more common choice in practice. However, our analytical results concern the representability of low- and high-Walshness states, and infidelity training aligns more directly with that setting. 
Whether a Walshness-minimizing basis also helps in energy-based optimization remains an interesting open question. 
Second, our analytical results are asymptotic in system size and do not directly constrain small systems. For the TFIM, consistent with this, we observe a clear advantage of the Walshness-minimizing basis only for sufficiently large system sizes. For small systems, the computational basis often performs comparably or better. We therefore emphasize that Walshness should not be expected to offer an advantage in the small-system regime, and our conclusions apply to the large-system behavior.
Third, our analytical results concern representability, which is only a meaningful constraint when the ansatz is underparameterized. A sufficiently expressive NQS can, in principle, represent arbitrary states, including those with high Walshness. We therefore focus on NQS ansätze with relatively few parameters, and we likewise do not claim an advantage in the overparameterized regime.
Finally, we train with Adam rather than the more commonly used stochastic reconfiguration (SR), as we found it to converge more reliably, be more stable, and be computationally cheaper in our particular infidelity-minimization setting.

\subsection{Ferromagnets}
\label{sec:ferro}
\begin{figure}
    \centering
\includegraphics[width=\linewidth]{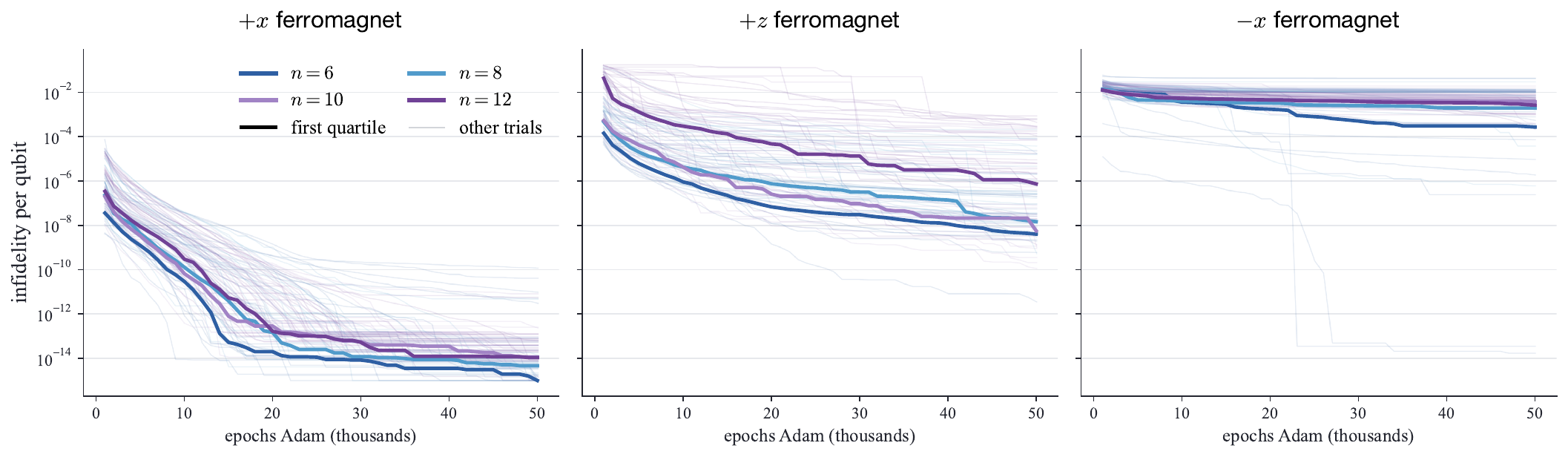}
    \caption{\textbf{Learning ferromagnets.} We train a real-amplitude MLP with two hidden layers, width $n/2$, and $\tanh$ activations on the $2^n$ computational-basis configurations of three states: the $+x$ ferromagnet ($\psi(\vec s)=2^{-n/2}$), the $+z$ ferromagnet ($\psi(\vec s) =\delta_{\vec{s},0^n}$), and the $-x$ ferromagnet ($\psi(\vec s) = 2^{-n/2}(-1)^{|\vec{s}|_1}$). Thick curves plot the first quartile of the infidelities per qubit, $1-\mathrm{F}^{1/n}$, attained up to each training epoch. Thin curves show individual trials across four initializations and six Adam settings, with learning rates $\eta\in\{10^{-3},3\times10^{-3},10^{-2}\}$ and $\beta_1\in\{0.5,0.9\}$. Under this optimization protocol, $+x$ reaches numerical precision, while $+z$ and $-x$ exhibit greater variability.}
    \label{fig:ferromagnetmanytrials}
\end{figure}

\begin{figure}
    \centering
\includegraphics[width=0.4\linewidth]{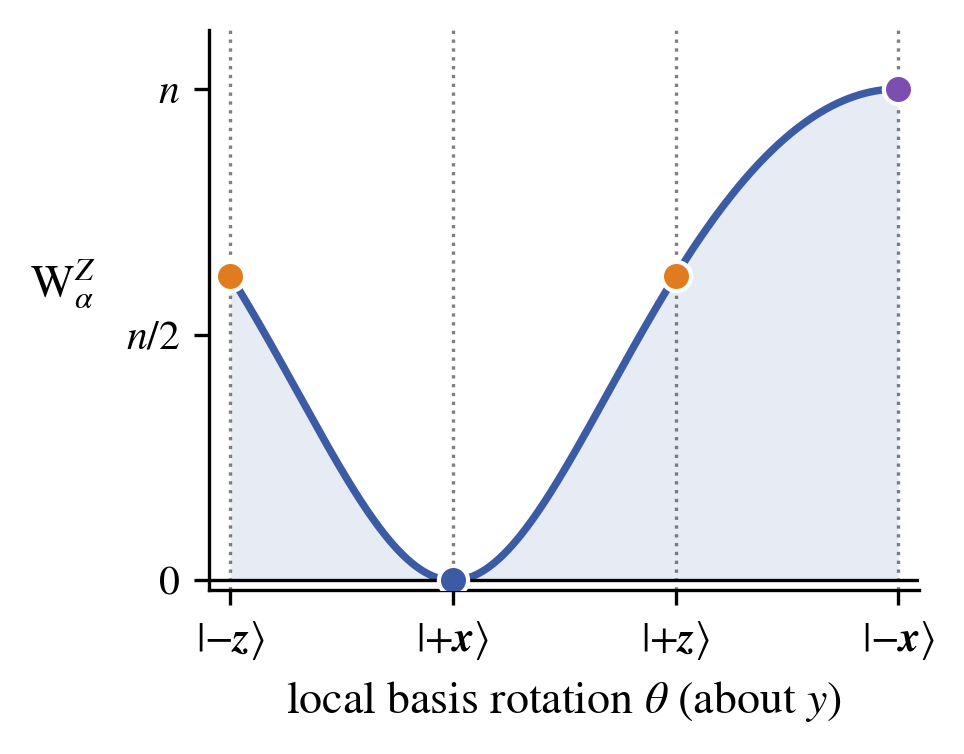}
    \caption{\textbf{Walshness landscape in the ferromagnet.} The y-axis shows the fixed-basis Walshness as a function of the basis on the x-axis, parametrized by the angle $\theta$. The state at every site $i$ is the $+1$ eigenstate of $\cos\theta\,\sigma^x_i + \sin\theta\,\sigma^z_i$, and $\mathrm{W}^Z_\alpha$ is evaluated in the computational ($\sigma^z$) basis. The states relevant for this section and their Walshness, $\ket{-z}, \ket{+x}, \ket{+z},\ket{-x}$ are clearly marked.}    \label{fig:walshness_ferromagnet}
\end{figure}

The three ferromagnetic states discussed in the main text provide a useful starting point to test whether Walshness offers any utility in guiding practical settings in training NQS. Although all three states are product states and therefore satisfy $\w=0$, their fixed-basis Walshness values differ strongly:
\begin{equation}
\wz[\ket{\vec{+x}}]=0,\qquad
\wz[\ket{\vec{+z}}]
=\frac{n}{\alpha}\log\!\left(\frac{1+e^\alpha}{2}\right),
\qquad
\wz[\ket{\vec{-x}}]=n.
\end{equation}
Indeed, the $\ket{\vec{+x}}$ wavefunction is constant, the $\ket{\vec{+z}}$ wavefunction is a delta function whose Walsh spectrum is uniform and hence $|\vec r|_1$ follows a binomial distribution, and the $\ket{\vec{-x}}$ wavefunction is the degree-$n$ parity function. For an illustration of the fixed-basis Walshness of the ferromagnet in the landscape of different basis rotations, see Fig.~\ref{fig:walshness_ferromagnet}. This experiment deliberately trains the amplitudes in the fixed $Z$ basis. Hence, the relevant quantity is $\wz$, rather than the intrinsic Walshness $\w$ obtained after optimizing the local basis.
To reiterate, we expect that the $+x$ target will be the easiest to learn, with the $+z$ target being more difficult, and the $-x$ target being the most difficult to learn.

Figure~\ref{fig:ferromagnetmanytrials} compares these targets using the same depth-two $\tanh$ MLP architecture and full-batch Adam optimization, using the infidelity as the loss function. 
The constant $+x$ target reaches an infidelity of the numerical precision for every $n$ that was tested. 
In contrast, the infidelity in the delta-function $+z$ target converges more slowly, while in general appearing to reach infidelities larger than that of the $+x$ case. 
In stark contrast, the training of the parity $-x$ target reaches the highest infidelity with the same maximum number of $50{,}000$ epochs, while displaying the strongest deterioration in performance with system size. 
Thus, the observed optimization difficulty broadly follows the ordering predicted by fixed-basis Walshness.
\par 
We briefly note that it appears that for $n=6$, there is a trial concerning the $-x$ ferromagnet (with $\eta=10^{-2}$ and $\beta_1=0.9$) which reaches near-perfect precision near epoch $22{,}700$. 
We verified that a second seed with the same Adam parameters reaches comparable precision near epoch $26{,}600$. 
This jump is consistent with Adam entering a narrow parameter basin that reproduces the wavefunction almost perfectly. 
However, no seeds reach comparable precision for $n>6$ within $50{,}000$ epochs, suggesting that this near-perfect optimization is a small-$n$ outlier. The vast majority of the training curves lie around an infidelity per qubit of $\sim 10^{-2}$, suggesting that performance is typically quite poor. 
\par
We emphasize that this experiment does not directly measure the optimal approximation error of the architecture. Each curve reports the smallest error found within a finite training horizon and a finite hyperparameter sweep. It therefore gives an upper bound on the minimum infidelity achievable by the chosen MLP. In particular, continued improvement at late epochs prevents us from interpreting a nonzero residual as a representational obstruction. The experiment instead tests whether fixed-basis Walshness correlates with practical optimization difficulty under a controlled training protocol.

\subsection{Stabilizer states}

A natural next step after the simple ferromagnetic product states is the class of more general stabilizer states~\cite{gottesman1997stabilizer}. 
Although stabilizer states admit efficient classical descriptions, their amplitudes can have very different sign structures in different local product bases. It will be useful to stratify stabilizer states based on the depth $D$ of a brickwork Clifford circuit which prepares them from the trivial state $\ket{0^n}$. We will vary $D$ and repeat the quantum state learning experiments from the last section. \par 

\begin{figure}
    \centering    \includegraphics[width=\linewidth]{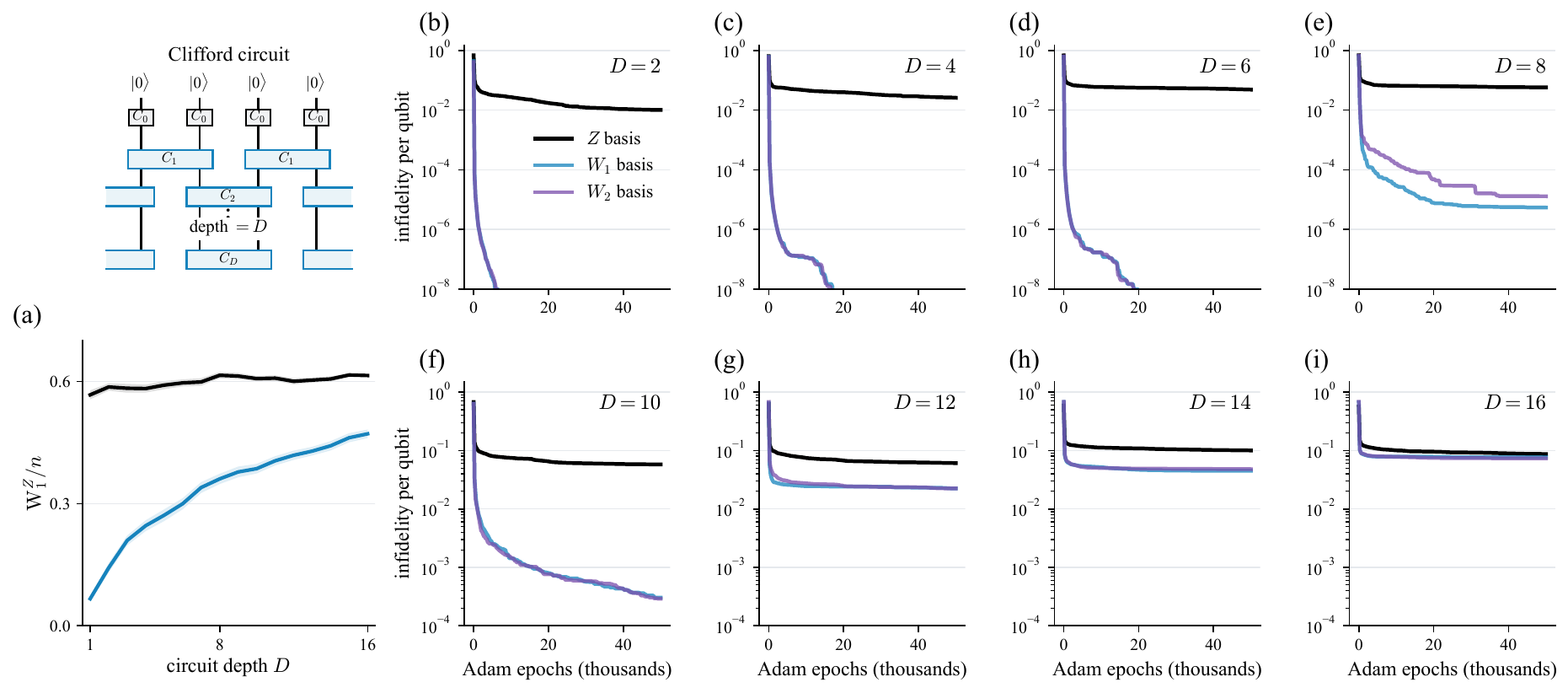}
\caption{\textbf{Learning stabilizer states.} We prepare states on an open chain of $n=12$ qubits by applying independent
random real single-qubit Clifford gates to $|0\rangle^{\otimes n}$, followed by
$D$ alternating brickwork layers of nearest-neighbor real two-qubit Clifford
gates. Each two-qubit gate is the identity with probability $1/2$ and otherwise
is sampled uniformly from the nonidentity real Clifford group. (a) Fixed-basis Walshness density at $\alpha=1$ in the $Z$-basis (black) and numerically optimized $\mathrm{W}_1$ basis (blue), showing the mean over 64 circuit instances, with ribbons showing one standard deviation. Panels
(b)--(i) show depths $D=2,4,\ldots,16$. For every state, we train the same
one-hidden-layer tanh MLP of width $12$ in the computational basis ($Z$, black),
the local basis minimizing $\mathrm{W}_{\alpha=1}$ ($\mathrm{W}_1$ basis, blue), and the local basis
minimizing $\mathrm{W}_2$ ($\mathrm{W}_2$ basis, purple). Each curve is the pointwise median of
the best-so-far infidelity per qubit,
$\epsilon=1-F^{1/n}$, over $16$ circuit instances and six network initializations, with identical initializations across the
three bases. The Walshness-minimizing bases yield orders-of-magnitude smaller
errors at shallow and intermediate depths; the advantage decreases as the
circuits deepen.}  \label{fig:stabilizersupp}
\end{figure}

We generate stabilizer states on an open chain of $n=12$ qubits using random real Clifford circuits. Each circuit begins with an independent random single-qubit Clifford at each site, followed by $D$ alternating layers of nearest-neighbor two-qubit Clifford gates, as shown in Fig.~\ref{fig:stabilizersupp}. Each two-qubit gate is the identity with probability $1/2$ and is otherwise chosen uniformly from the non-identity real two-qubit Clifford set. For every state, we compare learning in the computational $Z$ basis, learning in the basis minimizing $\mathrm{W}_1$, and the basis minimizing $\mathrm{W}_2$. Our choices of the parameter $\alpha = 1,2$ are quite arbitrary, and we do not attempt to optimize over $\alpha$, simply noting that any choice of $\alpha$ is valid (although it would be interesting future work to understand how the optimization depends on $\alpha$). We use identical network initializations and training parameters across these three bases. \par 

\begin{figure}
    \centering
    \includegraphics[width=\linewidth]{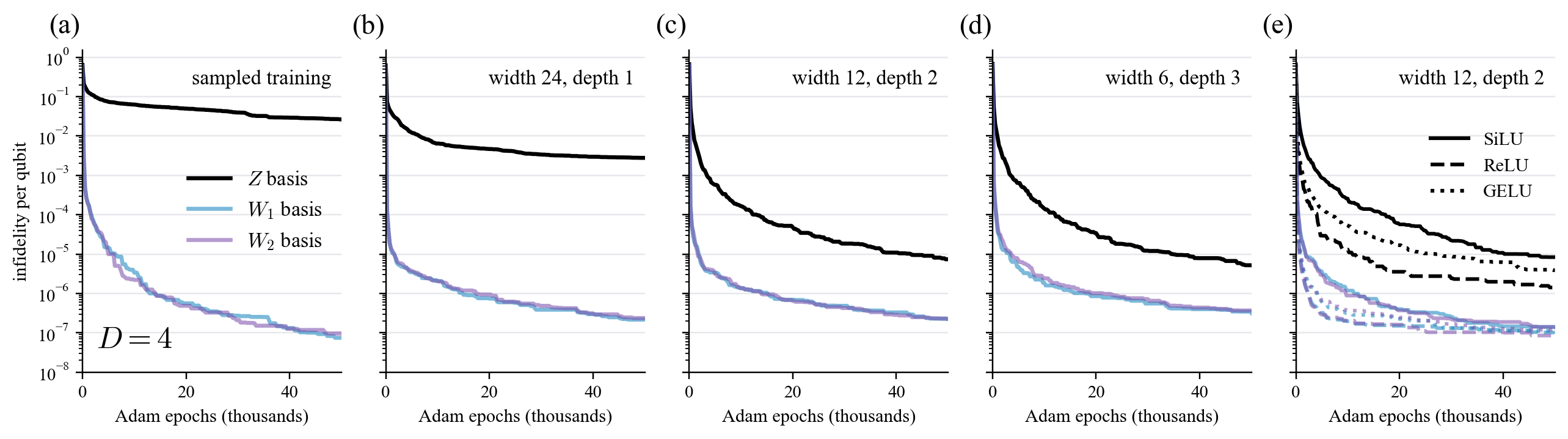}
\caption{\textbf{Robustness of Walshness-minimizing basis advantage.} 
We train neural representations of the $D=4$ stabilizer states from Fig.~\ref{fig:stabilizersupp} in the computational basis
($Z$, black), the local basis minimizing $\mathrm{W}_1$ ($\mathrm{W}_1$ basis, blue), and the
local basis minimizing $\mathrm{W}_2$ ($\mathrm{W}_2$ basis, purple).
(a) A one-hidden-layer width-$12$ tanh MLP trained using $512$ sampled
configurations per Adam step, drawn from
$q(s)=0.9|\psi(s)|^2+0.1/2^{12}$.
(b)--(d) Exact full-batch training with tanh MLPs of network width and hidden
depth $(24,1)$, $(12,2)$, and $(6,3)$, respectively.
(e) Exact full-batch training with width-$12$, depth-$2$ MLPs using SiLU
(solid), ReLU (dashed), and GELU (dotted) nonlinearities.
Each curve is the pointwise median best-so-far infidelity per qubit,
$1-\mathrm{F}^{1/n}$, over $16$ circuit instances and six
network initializations. The Walshness-minimizing bases improve learning across sampling protocols,
network widths, network depths, and nonlinearities.
}  \label{fig:variationonstabilizers}
\end{figure}

We show the results in Fig.~\ref{fig:stabilizersupp}. For shallow circuits, we find that the Walshness-minimizing bases produce a several-orders-of-magnitude improvement in the accuracy reached by the neural network. One can observe that this advantage decreases as $D$ increases. We suspect that this occurs because, as $D$ increases, the random circuit removes the local basis structure and accrues increasing sign complexity, causing the intrinsic Walshness to increase. 
In Fig.~\ref{fig:variationonstabilizers}, we further test the robustness of these results to a number of variations in the neural network ans\"atz. In particular, we vary the sampling method, the network width and depth, and the activation function. While these choices result in modest variations in the training curves, our prior conclusions remain intact for these shallow networks.\par 

\begin{figure}[t]
\centering
\includegraphics[width=\linewidth]{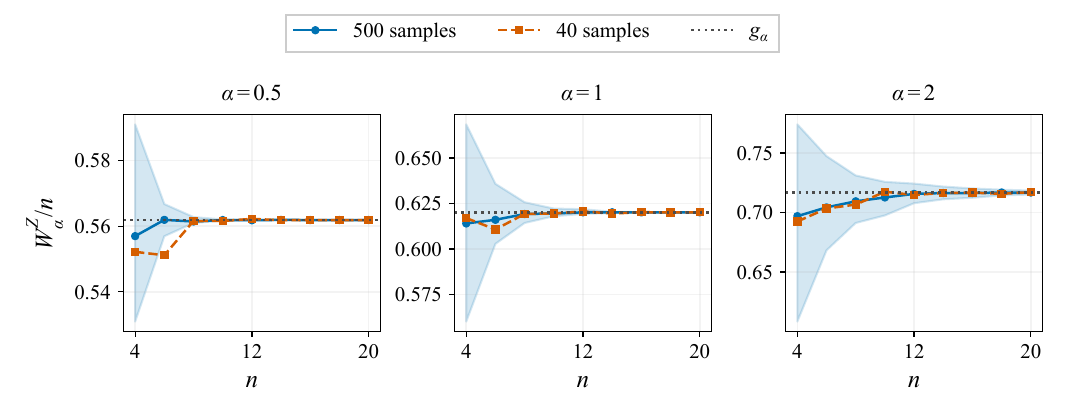}
\caption{\textbf{Fixed-basis Walshness of random stabilizer states.}
$\wz/n$ versus $n$ for $\alpha=0.5$, $1$, and $2$. Markers
show means over $500$ (blue) and $40$ (orange) samples, and shading shows the
$16$th--$84$th percentiles of the $500$-sample run. Dotted lines mark
$g_\alpha$.}
\label{sm:fig_stabilizer}
\end{figure}

\textit{Random-stabilizer scaling.} We next check Fact~\ref{fact:stabilizer}
numerically. For each even $n=4,6,\ldots,20$ and each
$\alpha\in\{0.5,1,2\}$, we sample $500$ random stabilizer states and compute
$\wz/n$ in the computational basis. We show the results in
Fig.~\ref{sm:fig_stabilizer}. For all three values of $\alpha$, the mean
approaches $g_\alpha$ by $n\approx 10$, and the spread across states narrows
rapidly with $n$, consistent with Eq.~\eqref{eq:haar-walshness-error}. We
emphasize that this test concerns only the fixed-basis Walshness.

\subsection{Transverse-field Ising model}

In this section, we present analytical and numerical results on Walshness in the ground state of the transverse-field Ising model
\begin{equation}
\label{eq:tfim}
    H = -\sum_j Z_j Z_{j+1} - g\sum_j X_j,
\end{equation}
which we consider on a periodic chain of $n$ sites. This model has a ferromagnetic phase for $g<1$ and a paramagnetic phase for $g>1$, separated by a quantum critical point of Ising universality at $g=1$ in the thermodynamic limit~\cite{pfeuty1970one}. Throughout this section, we will use $\ket\psi$ to denote the normalized finite-size exact ground state, which has eigenvalue $+1$ under the $\mathbb{Z}_2$ symmetry $\prod_jX_j$. In particular, in the ferromagnetic phase, we consider the GHZ-like state which is an even superposition of states with nonzero $Z$-magnetization. We choose this convention, rather than explicitly symmetry-broken ground states, in order to highlight interesting aspects of Walshness in GHZ-like states. In the symmetry-broken states, like in the paramagnetic regime, intrinsic Walshness goes to zero.\par \par 
In the limit $g\to0$, the ground state is $\ket{\psi_{\mathrm{GHZ}}} = \frac{\ket{0^n}+\ket{1^n}}{\sqrt{2}}$ and, from Eqs.~\eqref{eq:GHZWalshness} and~\eqref{eq:GHZconstant}, the fixed-basis Walshness density is 
\begin{equation}
\label{eq:tfim-ghz}
\left.\wzd\right|_{g=0}
=g_\alpha
+\frac1{\alpha n}\log\!\left[1+\tanh^n\!\left(\frac\alpha2\right)\right].
\end{equation}
where $g_\alpha \coloneqq \frac1\alpha\log\frac{1+e^\alpha}{2}$.
For fixed $\alpha>0$, the subleading term of $\wz$ vanishes exponentially in $n$. Meanwhile, the Walshness density is
\begin{equation}    \left.\wdd\right|_{g=0}=\frac12.
\label{eq:tfimGHZhalf}
\end{equation}
At intermediate $g$, the fixed-basis Walshness can be computed in closed form. From Eq.~\eqref{eq:fbWalternate}, we have
\begin{equation}
    e^{\alpha \wz} = \left\langle \exp\left[\frac{\alpha}{2}\sum_j (1-X_j)\right]\right\rangle = \left\langle \exp\left[\alpha\sum_j c_j^\dagger c_j\right]\right\rangle~,
\end{equation}
where in the second equality, we have used a Jordan-Wigner representation in which $X_j = 1-2c_j^\dagger c_j$ and $c_j$ are fermion operators obeying canonical anticommutation relations $\{c_j^\dagger,c_k\}=\delta_{jk}$. In the $\mathbb{Z}_2$-even sector, fermions obey antiperiodic boundary conditions, $c_{n+1} = -c_1$. Assuming $n$ is even, writing $c_j = n^{-1/2}\sum_k e^{ikj}c_k$, the momenta come in pairs $\pm k$, with $k = (2m+1)\pi/n$, where $m=0,\ldots, \frac{n}{2}-1$ and we have $\sum_jc_j^\dagger c_j = \sum_k c_k^\dagger c_k$. Let $\ket{\Omega}$ denote the fermionic vacuum and let $\ket{0}_k$ denote the vacuum of the modes $\pm k$. From the standard Jordan-Wigner solution we obtain that the ground state is
\begin{equation}
\ket\psi = \prod_{0<k<\pi} \ket{\psi_k}\coloneqq \prod_{0<k<\pi} \left(u_k + iv_k c_k^\dagger c_{-k}^\dagger\right)\ket{\Omega}~,
\end{equation}
where $|u_k|^2+|v_k|^2 = 1$ and $|v_k|^2 =
\frac12
\left(
1-\frac{g-\cos k}{\varepsilon_k}
\right)$, with $\varepsilon_k = \sqrt{1+g^2-2g\cos k}$. Therefore 
\begin{equation}
    \begin{aligned}
        \left\langle \exp \left[\alpha \sum_k c_k^\dagger c_k\right]\right\rangle &= \prod_{0<k<\pi} \left\langle \psi_k\mid e^{\alpha (c_{k}^\dagger c_k + c_{-k}^\dagger c_{-k})}\mid \psi_k\right\rangle\\
        &= \prod_{0<k<\pi} \left(|u_k|^2 + e^{2\alpha}|v_k|^2\right)\\
        &= \prod_{0<k<\pi} \left(1 + (e^{2\alpha}-1)|v_k|^2\right).
    \end{aligned}
\end{equation}
After taking the logarithm and a few simplifications, we obtain the exact finite-size result
\begin{equation}
\label{eq:tfim-fixed-finite}
    \wz
    =\frac n2+\frac1\alpha\sum_{0<k<\pi}
      \log\!\left[
        \cosh\alpha
        -\frac{g-\cos k}{\sqrt{1+g^2-2g\cos k}}\sinh\alpha
      \right].
\end{equation}
Dividing by $n$ and taking the thermodynamic limit yields
\begin{equation}
\label{eq:tfim-fixed-infinite}    \lim_{n\to\infty}\frac{\wz}{n}    =\frac12+\frac1{2\pi\alpha}\int_0^\pi
      \log\!\left[
        \cosh\alpha
        -\frac{g-\cos k}{\sqrt{1+g^2-2g\cos k}}\sinh\alpha
      \right]dk.
\end{equation}
At $g=0$, this expression approaches $\left.{\mathrm{w}}_\alpha^Z\right|_{g=0}=
\frac1\alpha\log\frac{1+e^\alpha}{2},$ which is greater than the intrinsic Walshness density [Eq.~\eqref{eq:tfimGHZhalf}]. Moreover, Eq.~\eqref{eq:tfim-fixed-infinite} vanishes as $g \to \infty$. At the critical point $g=1$, Eq.~\eqref{eq:tfim-fixed-infinite} is continuous with a logarithmic divergence in its derivative with respect to $g$. A straightforward calculation shows that 
\begin{equation}    \dv{g}\left[\lim_{n\to\infty}\frac1n\wz \right] = \frac{\tanh \alpha}{2\pi\alpha}\log|g-1| + O(1)
\end{equation}
near $g\approx 1$. \par
\begin{figure}
    \centering
    \includegraphics[width=0.8\linewidth]{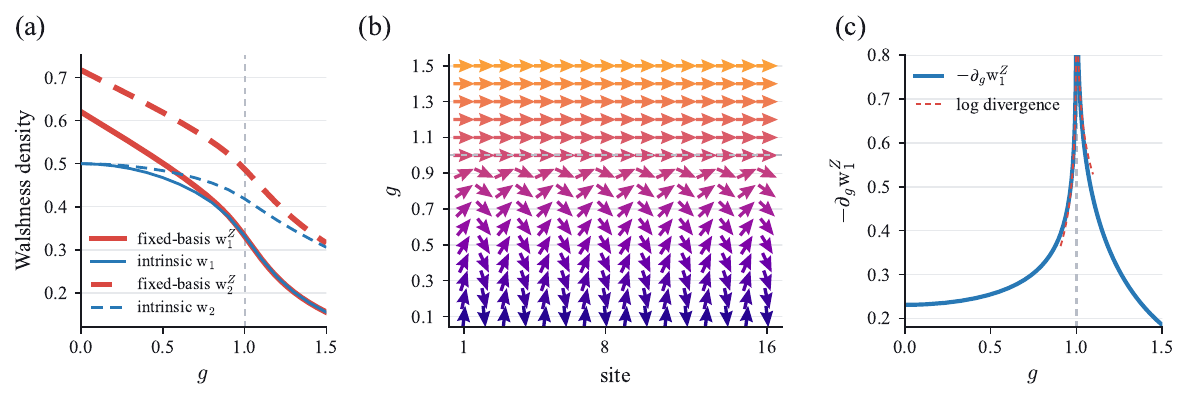}
    \caption{\textbf{Walshness in the transverse-field
Ising model.} We consider the ground state of
$H=-\sum_j Z_jZ_{j+1}-g\sum_jX_j$ with periodic boundary conditions.
(a) Fixed-basis Walshness density $\mathrm{w}_\alpha^Z$ (red)
and intrinsic Walshness density $\mathrm{w}_\alpha$ (blue)
for $n=16$, with $\alpha=1$ (solid) and $\alpha=2$ (dashed).
(b) Walshness-minimizing local axes in the $x$--$z$ plane
for $\alpha=1$. The optimal texture is period-two for small enough $g$.
(c) Thermodynamic-limit derivative $-\partial_g\mathrm{w}_1^Z$,
with red dashed curves showing the predicted logarithmic divergence
as $g\to1^\pm$.}
    \label{fig:tfimwalshnesssupp}
\end{figure}

 
We next examine the intrinsic Walshness numerically. For the
$n=16$ ground states, the optimal axes found are well-described
by the period-two ansatz $ \hat{\boldsymbol n}_j
    =    \bigl(\sin\chi_\alpha,\,0,\,(-1)^j\cos\chi_\alpha\bigr), 0\leq\chi_\alpha\leq\frac{\pi}{2}.$ At $g=0$, the choice $\chi_\alpha=0$ attains the known Walshness density of the GHZ state, namely $\wdd=1/2$. We note that at this point, the minimizing texture is not
unique and any assignment of equal numbers of $+\hat z$ and
$-\hat z$ axes also attain the same value. At nonzero field, however, the
period-two texture becomes the unique choice (up to translation) with axes continuously canting toward $+\hat x$.
We may recall that the axes are defined by
$U_j^\dagger X_jU_j=\hat{\boldsymbol n}_j\cdot\boldsymbol\sigma_j$,
so uniform $+\hat x$ axes correspond to the original
computational basis. For $\alpha=1$, we find that the optimized and
fixed-basis values coincide at all sampled fields $g\geq1$.
However, for $\alpha=2$, the staggered component and the reduction in
Walshness persist through $g=1.5$
[Fig.~\ref{fig:tfimwalshnesssupp}(a),(b)]. We note that the period-two staggering certainly does not indicate that the finite-size ground state breaks translation symmetry. This simple example teaches us that, remarkably, the Walshness-minimizing basis can break symmetries present in the quantum state $\ket\psi$. \par 

\begin{figure}
    \centering
    \includegraphics[width=\linewidth]{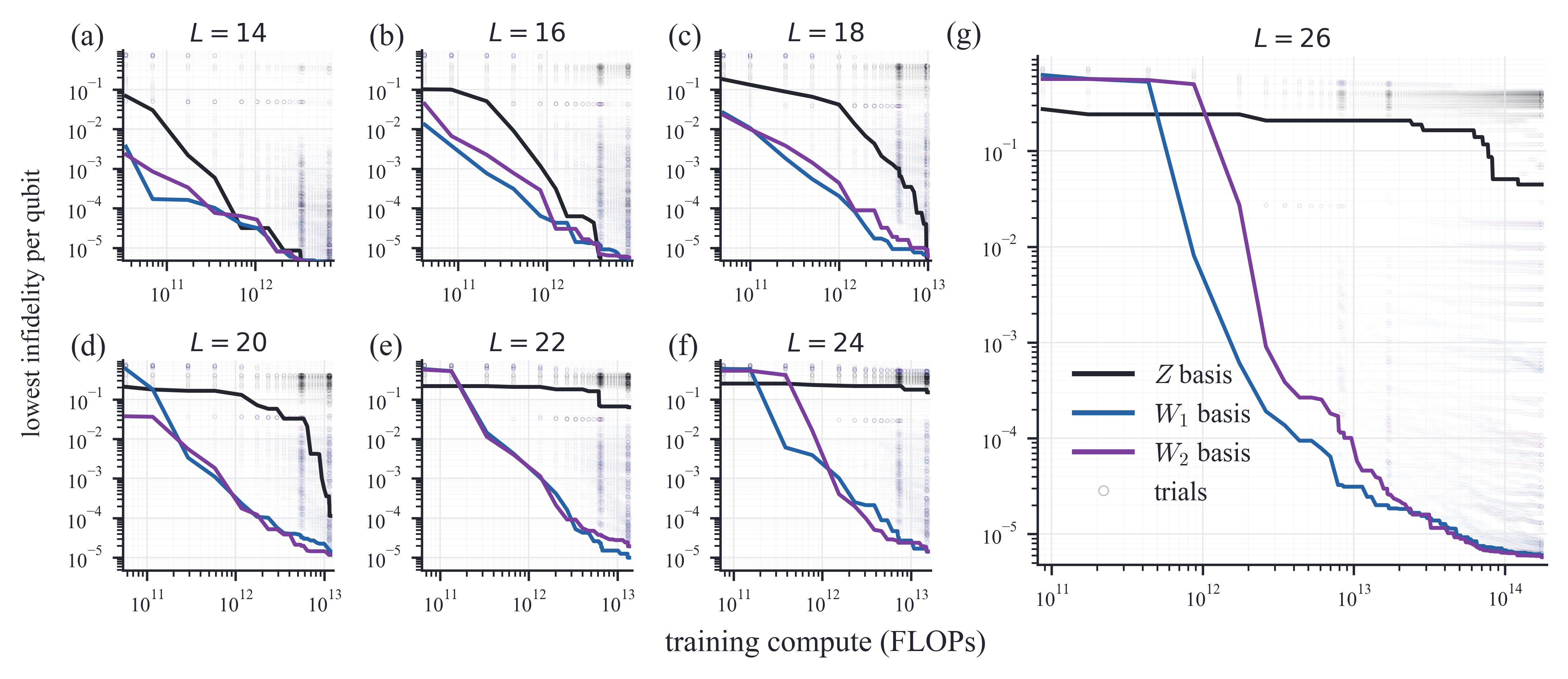}
\caption{\textbf{Learning the transverse-field Ising model ground state.} We learn the ground state of the transverse-field Ising
model (TFIM) at $g=0.2$ using two-hidden-layer $\tanh$ networks of width
$L+8$ for systems of size $L$. We compare the computational $Z$ basis (black) with bases independently
minimizing $\mathrm{W}_1$ (blue) and $\mathrm{W}_2$ (purple). 
The vertical axis is the best-so-far per-qubit infidelity,
$1-F^{1/L}$.
Faint points show individual trials over random initializations, Adam
hyperparameters, and sampling mixtures; solid curves show the pointwise lowest
error attained across these trials.
Panels (a)--(f) show $L=14,16,\ldots,24$ through $2\times10^5$ Adam updates,
using three sampling mixtures and six random initializations. Panel (g) shows
a more thorough $L=26$ study using up to $2\times10^6$ updates and the $\eta=0.25$. Training compute estimates the number of floating point operations (FLOPs) for each trial. Across larger sizes,
Walshness-minimizing bases reach substantially lower infidelity than the
computational basis at comparable large compute.}
\label{fig:tfimsupp}
\end{figure}

We test the consequences of Walshness for learning in the TFIM at $g=0.2$, comparing the
computational basis with the bases minimizing
$\mathrm W_1$ and $\mathrm W_2$.
We use matrix product states to define the quantum states which we are learning at large system sizes where ED becomes infeasible. Each target quantum state (across system sizes) is fitted by a
real-amplitude $\tanh$ MLP with two hidden layers of width
$w=L+8$, with identical initial parameters across all bases ($Z,\mathrm{W}_1,\mathrm{W}_2$). We train using Adam with minibatches of $8192$
configurations. In each basis, configurations are sampled
independently from $ q(\boldsymbol s)
    =
    (1-\eta)|\psi(\boldsymbol s)|^2+\eta\,2^{-n},$ with $
    \eta\in\{0.05,0.10,0.25\},$ where $\psi$ denotes the normalized target amplitudes in
that basis. The uniform component functions to give every
configuration nonzero sampling probability. We use all six combinations of learning rate
$\{10^{-3},3\times10^{-3},10^{-2}\}$ and
$\beta_1\in\{0.5,0.9\}$, with $\beta_2=0.999$,
Adam $\varepsilon=10^{-8}$, and no weight
decay. With six random initializations, we have a total of $3\times 6\times 6=108$ trials per
basis and size. Panels (a)--(f) of
Fig.~\ref{fig:tfimsupp} show $n=14,16,\ldots,24$
through $2\times10^5$ updates. Panel (g) shows a separate
$n=26$ extension through $2\times10^6$ updates, using
$\eta=0.25$ and the same six Adam settings and six
initializations. 
We measure training compute in floating-point operations (FLOPs); for
$t$ updates, batch size $B$, and architecture shape
$[n,w,w,1]$, we define this as $ C(t)=tB\bigl(4nw+6w^2+6w\bigr).$ This counting includes forward and backward matrix products,
counting a multiplication and an addition as two
operations. \par 

We observe that the learning advantage can be substantial, but that it depends on the system size and
training budget. At $n=14,16,18$, favorable
computational-basis trials eventually catch up and
attain lower final infidelities than the best Walshness-minimizing basis trials.
However, at larger sizes a substantial learning advantage appears; for instance at $n=24$ with $2\times10^5$ updates, where the lowest recorded infidelity per qubit is about
$0.15$ in the computational basis and
$1.4\times10^{-5}$ in either Walshness-minimizing basis. In the larger $n=26$ experiment, the corresponding
errors reach about $4.5\times10^{-2}$ and
$6\times10^{-6}$. These experiments suggest that Walshness minimization is a
useful criterion for basis selection in scenarios with limited compute budget. 

\subsection{Walshness of the toric-code ground states}\label{sec:tc_walshness}

In this section, we derive an exact closed form for the fixed-basis Walshness $\wz$ of the toric-code ground states on the $L\times L$ torus that lie in a definite logical sector. The idea of the proof is that the Walsh spectrum of these states is uniform over closed-loop configurations. As a result, $\wz$ reduces to the high-temperature expansion of the two-dimensional Ising model, which Ref.~\cite{kaufman_crystal_1949} solved on the finite torus.

We begin with a few definitions. Let $\Lambda=(V,E,F)$ be the $L\times L$ square lattice with
periodic boundary conditions, with vertex, edge and face sets $V$, $E$ and $F$; thus
$|V|=|F|=L^2$ and $n=|E|=2L^2$. With a qubit on each edge, the toric-code Hamiltonian is
\begin{equation}\label{eq:htc}
H_{\mathrm{TC}}=-\sum_{v\in V}A_v-\sum_{p\in F}B_p,\qquad
A_v=\prod_{e\ni v}X_e,\qquad B_p=\prod_{e\in\partial p}Z_e .
\end{equation}
We identify the $Z$-basis state $\ket{\mathbf s}$, $\mathbf s\in\mathbb F_2^{E}$, with the edge set
$\{e:s_e=1\}$, and write $|\mathbf s|$ for its Hamming weight ($|\mathbf s|_1$ in the definition of
$\wz$) and $\deg_{\mathbf s}(v)$ for the number of edges of $\mathbf s$ incident to $v$. The star
$\partial^*v=\{e:e\ni v\}$ satisfies $A_v\ket{\mathbf s}=\ket{\mathbf s\oplus\partial^*v}$ and
$\mathbf r\cdot\partial^*v\equiv\deg_{\mathbf r}(v)\pmod 2$. Finally, $\gamma_1,\gamma_2\subseteq E$
are non-contractible cycles winding in the two directions of the torus, and
$\bar Z_i=\prod_{e\in\gamma_i}Z_e$.

Now, we define the \emph{closed-loop configurations} on $\Lambda$, which include loops that wind around the torus. This is the set of configurations such that
\begin{equation}\label{eq:C}
\mathcal C=\big\{\mathbf c\in\mathbb F_2^{E}:\ \deg_{\mathbf c}(v)\equiv 0\pmod 2\ \ \forall v\in V\big\}.
\end{equation}
On the dual lattice $\Lambda^*$, whose vertices are the faces of $\Lambda$ and whose edges each
cross one edge of $\Lambda$, the closed-loop configurations are
\begin{equation}\label{eq:Cstar}
\mathcal C^*=\Big\{\mathbf s\in\mathbb F_2^{E}:\ \sum_{e\in\partial p}s_e\equiv 0\pmod 2\ \ \forall p\in F\Big\}.
\end{equation}
For $a,b\in\{0,1\}$, the \emph{winding sector}
\begin{equation}\label{eq:sectors}
\mathcal C^*_{ab}=\big\{\mathbf s\in\mathcal C^*:\ \mathbf s\cdot\gamma_1\equiv a,\ \mathbf s\cdot\gamma_2\equiv b\pmod 2\big\}
\end{equation}
contains the dual loop configurations that cross $\gamma_1$ ($\gamma_2$) an odd number of
times exactly when $a=1$ ($b=1$).

The set $\mathcal C$ is the cycle space of $\Lambda$, of dimension $|E|-|V|+1=L^2+1$, so
$|\mathcal C|=2^{L^2+1}$. The homologically trivial sector $\mathcal C^*_{00}$ is spanned by the
stars $\{\partial^*v\}_{v\in V}$, and the other sectors are its cosets,
$\mathcal C^*_{ab}=\mathbf w_{ab}\oplus\mathcal C^*_{00}$ for any representative
$\mathbf w_{ab}\in\mathcal C^*_{ab}$; each contains $2^{L^2-1}$ configurations. It is well
known that the ground state of $H_{\mathrm{TC}}$ with
$\bar Z_1=(-1)^a$ and $\bar Z_2=(-1)^b$ is ~\cite{kitaev2003fault}
\begin{equation}\label{eq:psi-ab}
\ket{\psi_{ab}}=2^{-(L^2-1)/2}\sum_{\mathbf s\in\mathcal C^*_{ab}}\ket{\mathbf s}~.
\end{equation}

With these definitions, we can now state and prove our result for the Walshness of the
toric-code ground states.

\begin{proposition}\label{prop:tc-walsh}
For each $a,b\in\{0,1\}$, the Walsh spectrum of $\ket{\psi_{ab}}$ is uniform on closed-loop
configurations,
\begin{equation}\label{eq:uniform}
p(\mathbf r)=
\begin{cases}
|\mathcal C|^{-1}, & \mathbf r\in\mathcal C,\\
0, & \mathbf r\notin\mathcal C.
\end{cases}
\end{equation}
Consequently, independently of the logical sector,
\begin{equation}\label{eq:tc-walsh}
\wz[\ket{\psi_{ab}}]
=\frac{1}{\alpha}\log\frac{G(e^{\alpha})}{2^{L^2+1}}
=2L^2+\frac{1}{\alpha}\log\frac{G(e^{-\alpha})}{2^{L^2+1}},
\qquad
G(x)=\sum_{\mathbf c\in\mathcal C}x^{|\mathbf c|}.
\end{equation}
\end{proposition}

We note that the closed form of $G$ that we will derive below holds only for $0<x<1$, and so the expression from the last equality is the most useful in practice. The proof of the proposition rests on two lemmas.

\begin{lemma}\label{lem:open}
Let $\psi$ satisfy $A_v\ket{\psi}=\ket{\psi}$ for all $v\in V$, and suppose that
$\mathbf{r}\notin \mathcal{C}$, i.e., $\mathbf{r}$ has a vertex $v'$ of odd degree. Then
$\widehat\psi(\mathbf r)=0$.
\end{lemma}

\begin{proof}
Define $f(\mathbf s)=\mathbf s\oplus\partial^*v'$, so that $A_{v'}\ket{\mathbf s}=\ket{f(\mathbf s)}$.
Since $f$ is its own inverse, it is a bijection of $\mathbb F_2^{E}$. The condition
$A_{v'}\ket{\psi}=\ket{\psi}$ gives $\psi(f(\mathbf s))=\psi(\mathbf s)$, while
$(-1)^{\mathbf r\cdot f(\mathbf s)}=-(-1)^{\mathbf r\cdot\mathbf s}$ because $v'$ has odd degree in
$\mathbf r$. Reindexing the sum in $\widehat\psi(\mathbf r)=2^{-n}\sum_{\mathbf s}(-1)^{\mathbf r\cdot\mathbf s}\psi(\mathbf s)$
by $f$,
\[
\sum_{\mathbf s}(-1)^{\mathbf r\cdot\mathbf s}\psi(\mathbf s)=\sum_{\mathbf s}(-1)^{\mathbf r\cdot f(\mathbf s)}\psi(f(\mathbf s))
=-\sum_{\mathbf s}(-1)^{\mathbf r\cdot\mathbf s}\psi(\mathbf s),
\]
so the sum vanishes.
\end{proof}

\begin{lemma}\label{lem:closed}
For $\mathbf r\in\mathcal C$ and any representative $\mathbf w_{ab}\in\mathcal C^*_{ab}$,
\begin{equation}\label{eq:psihat}
\widehat\psi_{ab}(\mathbf r)=(-1)^{\mathbf r\cdot\mathbf w_{ab}}\,2^{-2L^2}\,2^{(L^2-1)/2}.
\end{equation}
\end{lemma}

\begin{proof}
By Eq.~\eqref{eq:psi-ab}, $\psi_{ab}(\mathbf s)=2^{-(L^2-1)/2}$ for $\mathbf s\in\mathcal C^{*}_{ab}$
and zero otherwise, and every $\mathbf s\in\mathcal C^{*}_{ab}$ can be written uniquely as
$\mathbf s=\mathbf w_{ab}\oplus\mathbf t$ with $\mathbf t\in\mathcal C^{*}_{00}$. Each such
$\mathbf t$ is a sum of stars $\partial^{*}v$, and since $\mathbf r$ has even degree at every vertex,
$\mathbf r\cdot\partial^{*}v=\deg_{\mathbf r}(v)\equiv0\pmod 2$; hence $\mathbf r\cdot\mathbf t\equiv 0$
and $(-1)^{\mathbf r\cdot\mathbf s}=(-1)^{\mathbf r\cdot\mathbf w_{ab}}$ for all
$\mathbf s\in\mathcal C^{*}_{ab}$. Since $|\mathcal C^{*}_{ab}|=2^{L^2-1}$,
\[
\widehat\psi_{ab}(\mathbf r)=\frac{1}{2^{2L^2}}\sum_{\mathbf s\in\mathcal C^{*}_{ab}}\frac{(-1)^{\mathbf r\cdot\mathbf s}}{2^{(L^2-1)/2}}
=\frac{(-1)^{\mathbf r\cdot\mathbf w_{ab}}}{2^{2L^2}}\,2^{(L^2-1)/2},
\]
and therefore $|\widehat\psi_{ab}(\mathbf r)|^2=2^{L^2-1}/2^{4L^2}=2^{-(3L^2+1)}$.
\end{proof}

Together, these lemmas prove Proposition~\ref{prop:tc-walsh}:

\begin{proof}[Proof of Proposition~\ref{prop:tc-walsh}]
By Lemmas~\ref{lem:open} and~\ref{lem:closed}, $|\widehat\psi_{ab}|^2$ vanishes off $\mathcal C$ and
is constant on $\mathcal C$; normalizing gives Eq.~\eqref{eq:uniform}. Hence
$\mathbb E_p\big[e^{\alpha|\mathbf r|}\big]=|\mathcal C|^{-1}\sum_{\mathbf c\in\mathcal C}e^{\alpha|\mathbf c|}=G(e^{\alpha})/|\mathcal C|$,
which is the first equality of Eq.~\eqref{eq:tc-walsh}. For the second equality, the vector of all ones
$\mathbf 1$ lies in $\mathcal C$ because every vertex has degree four. Thus
$\mathbf c\mapsto\mathbf c\oplus\mathbf 1$ is a bijection of $\mathcal C$ with
$|\mathbf c\oplus\mathbf 1|=2L^2-|\mathbf c|$, so $G(x)=x^{2L^2}G(1/x)$. This proves the second equality.
\end{proof}

We emphasize that the formula above only applies to the ground states that lie in one of the four definite logical sectors $\psi_{ab}, ab\in\{00, 01, 10, 11\}$.
If one picks a generic state lying within the span of the ground state subspace, then the formula above no longer holds.

We now derive a closed form of $\wz$ by establishing a connection to the exact solution of the 2D Ising model.
\begin{lemma}\label{lem:ising}
Let
\[
Z(K)=\sum_{\sigma\in\{\pm1\}^V}\exp\Big(K\sum_{\langle ij\rangle\in E}\sigma_i\sigma_j\Big)
\]
be the partition function of the Ising model $H_{\mathrm{Ising}}=-J\sum_{\langle ij\rangle}\sigma_i\sigma_j$
on $\Lambda$ at inverse temperature $\beta$, with $K\equiv \beta J$. Then, for $0<x<1$,
\begin{equation}\label{eq:G-ising}
G(x)=\left(\frac{1-x^2}{2}\right)^{L^2} Z(\operatorname{arctanh}x)~.
\end{equation}
\end{lemma}
This directly follows from the high-temperature expansion of the 2D Ising model. 
The partition function $Z$ was
obtained in the thermodynamic limit by Onsager~\cite{onsager_crystal_1944} and on the finite
$L\times L$ torus by Kaufman~\cite{kaufman_crystal_1949}. For completeness, we present the closed-form solution here. Substituting Kaufman's result, in the
form of Eqs.~(4)--(5) of Ref.~\cite{ivashkevich_kroneckeraposs_2002}, into Lemma~\ref{lem:ising}
yields, for $0<x<1$,
\begin{equation}\label{eq:G-closed}
\begin{split}
G(x)=\frac12\big[x(1-x^2)\big]^{L^2/2}\sum_{\tau\in\{0,\frac12\}}
\Bigg[&\prod_{k=0}^{L-1}2\cosh\!\big(L\,\omega^{\tau}_{k}(x)\big)\\
&+\prod_{k=0}^{L-1}2\sinh\!\big(L\,\omega^{\tau}_{k}(x)\big)\Bigg],
\end{split}
\end{equation}
where $\omega^{\tau}_{k}(x)=\operatorname{arcsinh}s^{\tau}_{k}(x)$ with
\begin{align}
s^{\tau}_{k}(x)&=
\begin{cases}
m(x), & (k,\tau)=(0,0),\\[2pt]
\sqrt{\sin^2\dfrac{\pi(k+\tau)}{L}+m(x)^2}, & \text{otherwise},
\end{cases}\label{eq:s}\\[4pt]
m(x)&=\frac{x^2+2x-1}{2\sqrt{x(1-x^2)}} .\label{eq:m}
\end{align}
Together with Eq.~\eqref{eq:tc-walsh}, this gives $\wz[\ket{\psi_{ab}}]$ in a closed form, as desired.

Lemma~\ref{lem:ising} also gives the Walshness density in the thermodynamic limit.
\begin{proposition}\label{lem:thermodynamic_limit_tc}
For every $\alpha>0$ and in every logical sector,
\begin{equation}\label{eq:density}
\mathrm{w}^z_\infty(\alpha)\equiv\lim_{n\rightarrow\infty}\wzd[\ket{\psi_{ab}}]
=\frac12+\frac{\Phi(\alpha/2)-\ln 2}{2\alpha}~.
\end{equation}
Here, $\Phi(K)=\lim_{L\to\infty}L^{-2}\ln Z(K)$ is the free energy per site of the 2D Ising model
in the thermodynamic limit (normalized by $\beta$)~\cite{onsager_crystal_1944},
\begin{equation}\label{eq:onsager}
\Phi(K)=\ln(2\cosh 2K)+\frac{1}{2\pi}\int_0^{\pi}\ln\frac{1+\sqrt{1-\kappa^2\sin^2\theta}}{2}\,d\theta~,
\end{equation}
with $\kappa=2\sinh 2K/\cosh^2 2K$.
\end{proposition}
\begin{proof}[Proof of Prop.~\ref{lem:thermodynamic_limit_tc}]
Let $x=e^{-\alpha}$ and $K=\operatorname{arctanh}x$. Then, Eq.~\eqref{eq:tc-walsh} and
Lemma~\ref{lem:ising} give
\[
\begin{aligned}
\lim_{L\to\infty}\frac{\wz}{n}&=\lim_{L\to\infty}\left\{1+\frac{1}{2\alpha}\left[\ln(1-x^2)-2\ln 2+\frac{\ln Z(K)}{L^2}-\frac{\ln 2}{L^2}\right]\right\}\\
&=1+\frac{\ln(1-x^2)-2\ln 2+\Phi(K)}{2\alpha}.
\end{aligned}
\]
To put this in a cleaner form and to express $\Phi$ as a direct function of $\alpha$, note that $\sinh 2K=2x/(1-x^2)=1/\sinh\alpha$ and
$\cosh 2K=\coth\alpha$. Since $\kappa$ is
invariant under $\sinh 2K\mapsto 1/\sinh 2K$, Eq.~\eqref{eq:onsager} gives
\[
\Phi(K)=\Phi(\alpha/2)-\ln\sinh\alpha~.
\]
Equation~\eqref{eq:density} then follows from $\ln(1-x^2)=\ln 2-\alpha+\ln\sinh\alpha$.
\end{proof}

\begin{figure}
    \centering
    \includegraphics[width=\linewidth]{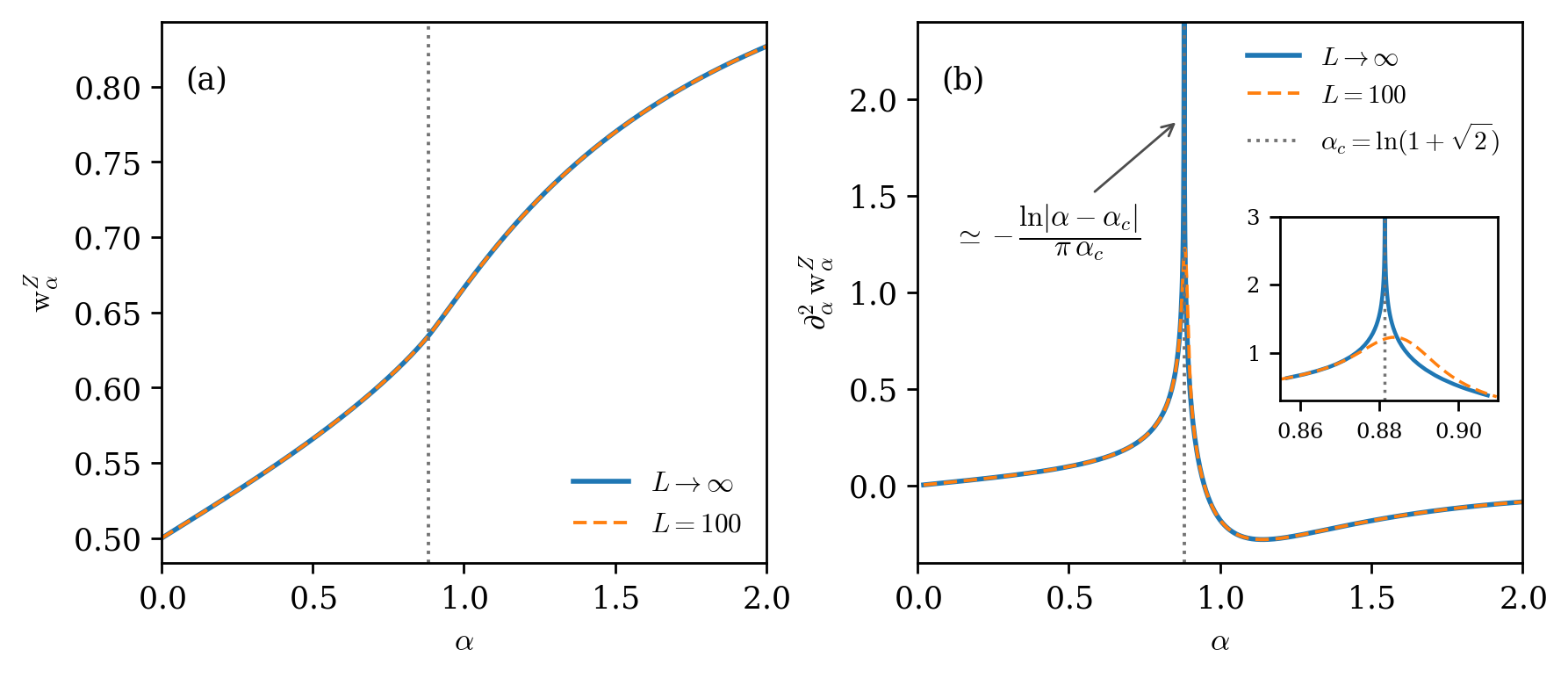}
    \caption{\textbf{Toric code.} Fixed-basis Walshness density of the toric-code ground states in the thermodynamic limit and at system size $L=100$. (a) $\wzd$ as a
    function of $\alpha$ in the thermodynamic limit (solid), compared with $\wzd$ at $L=100$
    (dashed). (b) Second derivative $\partial_\alpha^2\wzd$, which diverges logarithmically at
    $\alpha_c=\ln(1+\sqrt2)\approx0.881$ (dotted line). At $L=100$ the divergence is rounded into a
    finite peak at $\alpha\approx0.884$ (inset).}
    \label{fig:walshness_tc_exact}
\end{figure}

The 2D Ising model has a critical point at $K_c=\ln(1+\sqrt{2})/2$. Since
Eq.~\eqref{eq:density} involves $\Phi$ evaluated at $\alpha/2$, the Walshness density is
non-analytic at $\alpha_c=2K_c=\ln(1+\sqrt{2})\approx 0.881$. Indeed, one can verify that, as
$\alpha\to\alpha_c$,
\begin{equation}\label{eq:log-div}
\frac{\partial^2}{\partial\alpha^2}[\mathrm{w}^z_\infty(\alpha)]\simeq-\frac{\ln|\alpha-\alpha_c|}{\pi\,\alpha_c}.
\end{equation}
To visualize these results, Fig.~\ref{fig:walshness_tc_exact}(a) shows the Walshness density of
the toric-code ground states in the thermodynamic limit as a function of $\alpha$, compared with
the exact result at $L=100$, and Fig.~\ref{fig:walshness_tc_exact}(b) shows $\partial_\alpha^2\wzd$.
Remarkably, the Walshness density of the toric-code ground states encodes the Onsager transition
of the 2D Ising model: its second derivative with respect to $\alpha$ diverges logarithmically at $\alpha_c$. It would be
interesting to determine whether such singularities are a general feature of Walshness in
topologically ordered states, and whether they could be useful in diagnosing or searching for
topological order.

\subsection{Mixed-field toric code}
\label{app:toric-mixed-field}

\begin{figure}
    \centering
    \includegraphics[width=\linewidth]{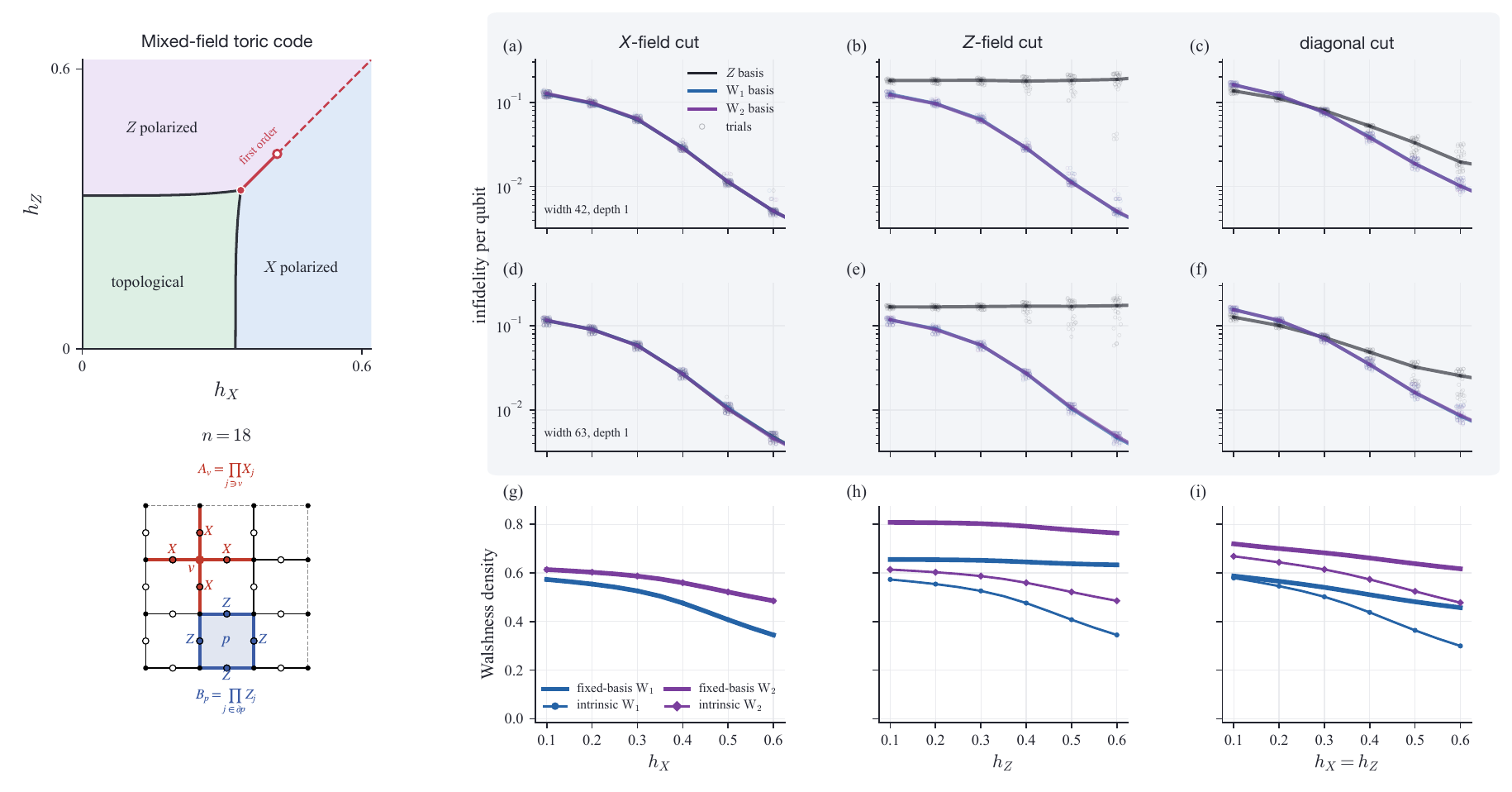}
    \caption{\textbf{Learning the mixed-field toric code ground state.} Left: schematic phase diagram in the $(h_X,h_Z)$ plane and the periodic $L=3$ lattice used for the numerical calculations~\cite{wu2012phase}. Qubits reside on the edges, giving $n=2L^2=18$ qubits. 
(a)--(f) Infidelity per qubit obtained by learning the exact ground state along the $X$-field, $Z$-field, and diagonal cuts. The upper and lower rows use one-hidden-layer tanh networks of widths $42$ and $63$, respectively. We compare the computational $Z$ basis with the local bases minimizing $\mathrm{W}_1$ and $\mathrm{W}_2$. Each curve is the median of the smallest exact infidelity reached during $20,000$ Adam updates over $36$ trials, comprising three importance-sampling mixtures, six Adam settings, and two paired initializations. Faint circles show the individual trials, with a small horizontal displacement for visibility.
(g)--(i) Fixed-basis and intrinsic Walshness densities of the same ground states. Thick curves show the fixed-basis quantities $\mathrm{w}_1^Z$ and $\mathrm{w}_2^Z$ in the computational basis. Thin curves with markers show the corresponding intrinsic quantities $\mathrm{w}_1$ and $\mathrm{w}_2$. Along the $X$-field cut, the computational basis is already Walshness minimizing, and the three learning curves coincide. Along the $Z$-field cut, minimizing Walshness substantially improves learnability. The diagonal cut exhibits a crossover: the rotated bases become favorable only once the reduction in Walshness is sufficiently large. 
}
\label{fig:mftcsupp}
\end{figure}

We now present a numerical study on the relation between Walshness and learnability in a two-dimensional topologically ordered system. We consider the mixed-field toric code Hamiltonian, which is the toric code with applied uniform $X$ and $Z$ fields~\cite{kitaev2003fault}:
\begin{equation}
    H_{\mathrm{TC}}
    =
    -\sum_v A_v
    -\sum_p B_p
    -h_X\sum_e X_e
    -h_Z\sum_e Z_e,
    \label{eq:mixed-field-toric-code}
\end{equation}
where $A_v\coloneqq\prod_{e\ni v}X_e$ and $B_p\coloneqq\prod_{e\in\partial p}Z_e$ are the vertex and plaquette operators, respectively. 
Here, $v$ indicates vertices, or the set of 4 spins surrounding a lattice point, while $\partial p$ denotes the spins surrounding a plaquette in the lattice.
The qubits reside on the edges of an $L\times L$ square lattice with periodic boundary conditions, with total number of qubits $n=2L^2$. A sketch of the lattice, the plaquette and vertex operators, 
and a schematic of the phase diagram are illustrated in the left-hand side of Fig.~\ref{fig:mftcsupp}.
The phase diagram contains a topologically ordered phase at small $h_X,h_Z$ and $X$- and $Z$-polarized phases at large $h_X$ and $h_Z$, respectively. We focus on an $L=3$ torus ($n=18$) with periodic boundary conditions so that near $h_X=h_Z=0$, the ground state manifold is fourfold quasidegenerate. We will focus on the ground state obtained using ED, i.e. the state with logical-$X$ eigenvalues $(+1,+1)$ in the topological phase. When the ground state is truly degenerate, we project to this logical-$X$ subspace to deterministically specify a state.

\begin{figure}
    \centering
    \includegraphics[width=\linewidth]{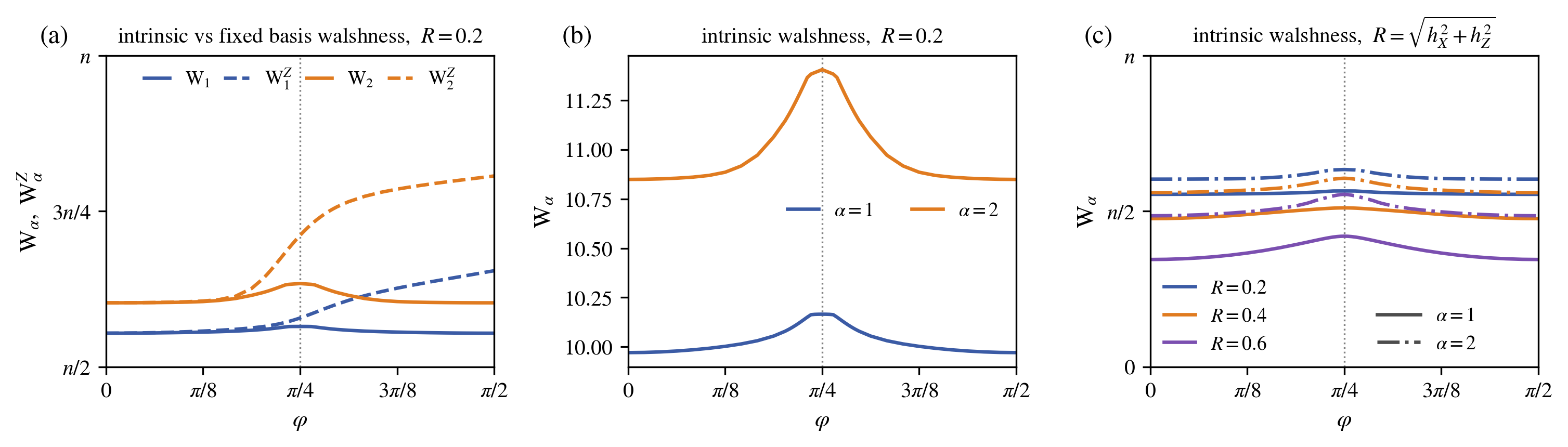}
    \caption{\textbf{Walshness of the mixed-field toric code ground states.} We present calculations of the fixed-computational basis Walshness and the intrinsic Walshness of the ground states of the mixed-field toric code, with $L=3$ and $n=18$. For all plots, we perform a sweep along the line parametrized by the variable $\varphi$, with $(h_X, h_Z) = (R \cos{(\varphi)}, R\sin({\varphi}))$ where $R$ is a constant. (a) Walshness as a function of $\varphi$ with $R=0.2$. Such a parameter regime is deep in the topological phase. The dashed lines indicate the fixed-computational basis Walshness, while the solid lines indicate the intrinsic Walshness. The blue curves indicate $\alpha=1$, while the orange curves correspond to $\alpha=2$. (b) A zoomed-in version of (a), with a focus on the intrinsic Walshness curves. We note a distinct peak at the self-duality point $\varphi=\pi/4$. (c) The intrinsic Walshness at various values of $R$, indicated by the different colored curves, and thus different regions of the phase diagram of $H_{\mathrm{TC}}$. Solid (dashed) lines indicate $\alpha=1$ ($\alpha=2$).
    }
\label{fig:toric_code_gs_walshness}
\end{figure}

To get a sense of the intrinsic and fixed-basis Walshness of the MFTC states, we show the intrinsic and the fixed-computational basis Walshness across various cuts in the phase diagram, for both the $\alpha=1$ and $\alpha=2$ Walshness. 
In the bottom row of Fig.~\ref{fig:mftcsupp}, we compare the fixed-basis and intrinsic Walshness densities along three cuts: increasing $h_X$ with $h_Z=0$, increasing $h_Z$ with $h_X=0$, and increasing $h_X=h_Z$. As expected by the $X\leftrightarrow Z$ self-duality combined with a half-translation to the dual lattice, the intrinsic Walshness is identical for the first two curves. However, only in the former case does intrinsic Walshness agree with the fixed-basis Walshness; in the second case, intrinsic Walshness density is lower by $\sim 0.2-0.3$. 
The $h_X=h_Z$ line cut shows a mixture of these two behaviors, as expected. We do not observe any notable features across the putative topological phase transition points of $h_{X,Z}\sim 0.3$, but we do not rule out insufficient system size to resolve such features. 
\par 

In Fig.~\ref{fig:toric_code_gs_walshness}, we also show additional calculations of the Walshness across the cut $h_X^2+ h_Z^2 = R^2$, where the strength of the transverse and longitudinal fields applied to the regular toric code is kept constant across this cut. 
We parametrize $h_X, h_Z$ with the angle $\varphi$, such that $(h_X, h_Z) = (R\cos[\varphi], R \sin[\varphi])$. 
In panels (a) and (b), we focus on the $R=0.2$ cut, while in panel (c), we compare the intrinsic Walshness across different values of $R$.
From panel (a), the reduction in Walshness is zero when $\varphi\lesssim\pi/4$, and the reduction becomes less when $\varphi\gtrsim\pi/4$.
This aligns with our expectations. At small $\varphi$, the $h_X$ field is relatively large compared to $h_Z$. 
This pushes the system to be closer to a $+x$ paramagnet, whose Walshness-minimizing basis is already the computational basis. 
On the other hand, when $h_Z$ is larger than $h_X$, the system is closer to the $+z$ ferromagnet, which we have shown earlier can experience a reduction in Walshness with the appropriate basis rotation.
These two observations are consistent with what we see here.
Interestingly, one can see clearly that the self-duality point, $h_X=h_Z$, has a larger intrinsic Walshness at these system sizes compared to points away from the self-duality point.
We cannot rule out that this may be a finite-size effect, but a more careful analytical understanding of the Walshness at the self-duality point may be an interesting avenue to provide further understanding of the topological phase diagram.

\par

For every ground state across varying $(h_X,h_Z)$, we minimize fixed-basis Walshness over single-qubit rotations using the procedure described in Sec.~\ref{sec:computing_walshness}. We optimize separately at $\alpha=1$ and $\alpha=2$. We refer to the resulting local product bases as the $\mathrm{W}_1$ and $\mathrm{W}_2$ bases, respectively. Thus, the three bases in which we perform the following machine learning experiments are (1) the original computational $Z$ basis, (2) the $\mathrm{W}_1$ basis, and (3) the $\mathrm{W}_2$ basis. We explicitly rotate the state into each optimized basis and verify that its fixed-basis Walshness reproduces the value obtained from the Walshness minimization.
\par

Next, we test the connection between Walshness and neural-network learnability in this model. We adopt a variational wavefunction which is a real neural network with one hidden tanh layer with input configurations naturally represented by spins $s_j=\pm 1$ in any of the chosen bases. We adopt hidden-layer widths $42$ and $63$, corresponding to $841$ and $1261$ trainable parameters. For each architecture, field point, and basis, we minimize the infidelity using stochastic gradient descent (Adam, $2\times 10^4$ epochs). 
During training, we sample spin configurations for the given target state $\psi_0$ according to $(1-\eta)|\psi_0(\vec s)|^2+\eta\,2^{-n},
$ where $\eta\in\{0.05,0.10,0.25\}$ is a prefactor for a uniform component of this distribution. 
We use minibatches of $8192$ configurations and six Adam settings with learning rates $10^{-3}, 3\times10^{-3}, 10^{-2}$ with $\beta_1=0.5$ or $0.9$, while $\beta_2=0.999$. 
For each setting, we use two random network initializations. Initial parameters are reused across the three bases. 
At predetermined checkpoints, we evaluate the fidelity exactly.
For each trial, we retain the smallest infidelity per qubit
$1-F^{1/n}$ across its checkpoints.
We plot the resulting curves in Figs.~\ref{fig:mftcsupp}(a)--(f), showing the median over all $3\times6\times2=36$ trials, with faint circles showing individual trials. \par 

Remarkably, the neural network performance mirrors the trends seen in previous experiments with the TFIM ground states. First, for the cut along $h_Z$, we observe a strong advantage in both the $\mathrm{W}_{1}$ and $\mathrm{W}_2$ bases over the computational basis in the range $0.1\leq h_Z\leq 0.6$ (we avoid small $h_Z$ due to numerical instabilities and do not claim an advantage there), mirroring the reduction in Fig.~\ref{fig:mftcsupp}h. 

Along the diagonal cut, we observe an advantage for the Walshness-minimizing basis only above $h_X=h_Z\approx 0.3$, whereas for $h_X=h_Z<0.3$, the Walshness-minimizing basis performs worse than the standard computational basis.
These results add a level of nuance to the story, and we discuss below one possible explanation for this observation.
First, we remark that $h_X=h_Z\approx 0.3$ is in the vicinity of the multicritical point. The reduction in Walshness by rotating to the Walshness-minimizing basis is notably smaller in the regime of topological order ($h_X=h_Z\lesssim 0.3$) than in the trivial phase ($h_X=h_Z\gtrsim 0.3$), as seen in panel (i) of Fig.~\ref{fig:mftcsupp}.
Thus, going to the Walshness-minimizing basis in the regime of topological order may not offer a pronounced advantage.
No advantage is seen or expected, of course, in the cut along $h_X$, due to the coinciding of fixed-basis and intrinsic Walshness. 
These trends indicate strong performance of the Walshness-minimizing basis at large $h_Z$; note that large $h_X$ polarizes the ground state towards the $+x$ ferromagnet, which already has vanishing Walshness. \par 
\par 
Taken together, this supports our general thesis that the clearest learning advantages occur when the basis rotation produces a substantial reduction in Walshness.
Moreover, these results extend the connection between Walshness and learnability beyond one-dimensional spin chains. 
In particular, we have investigated, but not fully resolved, the interplay of Walshness with topological order. 
It would be an interesting future direction to establish a link between the two.
\bibliography{bib, nqs}